\makeatletter
\@ifundefined{@parse@version@dash}{%
\def\@parse@version#1{\@parse@version@0#1}
\def\@parse@version@#1/#2/#3#4#5\@nil{%
\@parse@version@dash#1-#2-#3#4\@nil}
\def\@parse@version@dash#1-#2-#3#4#5\@nil{%
  \if\relax#2\relax\else#1\fi#2#3#4 }
}{}
\makeatother

\documentclass[%
twocolumn,
superscriptaddress,
nofootinbib,
 amsmath,amssymb,
 aps,
prd
]{revtex4-1}

\usepackage{graphicx}
\usepackage{dcolumn}
\usepackage{bm}
\usepackage{hyperref}
\usepackage{xprintlen}
\usepackage{float}
\usepackage{amsmath}

\usepackage{xcolor}

\newcommand{\aj}{Astron. J.}
\newcommand{\mnras}{Mon. Not. R. Astron. Soc.}
\newcommand{\jcap}{J. Cosmol. Astropart. Phys.}

\newcommand{\aap}{Astron. Astrophys.}
\newcommand{\apjl}{Astrophys. J. Lett.}

\newcommand{\pasp}{PASP}

\def\drm{\mathrm{d}}
\def\los{{\hat{\bm{n}}}}
\def\vvec{{\bm v}}

\begin{document}

\title{Joint constraints on reionization: a framework for combining the global 21cm signal and the kinetic Sunyaev-Zel'dovich effect}

\author{Jo\"elle-Marie B\'egin}
  \email{joelle-marie.beginmiolan@mail.mcgill.ca}
 \author{Adrian Liu}%
 \author{Ad\'elie Gorce}
 \affiliation{%
Department of Physics and McGill Space Institute, McGill University, Montreal, QC, Canada H3A 2T8
}%

\date{\today}

\begin{abstract}
Recent measurements from the CMB and from high-redshift galaxy observations have placed rough constraints on the midpoint and duration of the Epoch of Reionization. Detailed measurements of the ionization history remain elusive, although two proposed probes show great promise for this purpose: the 21cm global signal and the kinetic Sunyaev-Zel'dovich (kSZ) effect. We formally confirm the  common assumption that these two probes are highly complementary, with the kSZ being more sensitive to extended ionization histories and the global signal to rapidly evolving ones. We do so by performing a Karhunen-Lo\`{e}ve (KL) transformation, which casts the data in a basis designed to emphasize the information content of each probe. We find that reconstructing the ionization history using both probes gives significantly more precise results than individual constraints, although carefully chosen, physically motivated priors play a crucial part in obtaining a bias-free reconstruction.
Additionally, in the KL basis, measurements from one probe can be used to detect the presence of residual systematics in the other, providing a safeguard against systematics that would go undetected when data from each probe is analyzed in isolation. Once detected, the modes contaminated by systematics can be discarded from the data analysis to avoid biases in reconstruction.

\end{abstract}

\maketitle


\section{\label{sec:intro} Introduction}

The cosmological and astrophysical processes that governed the transition of hydrogen from neutral to ionized during cosmic reionization are poorly understood, and explaining the formation of the first luminous sources as well as their role in this large-scale phase transition remains a crucial task in modern cosmology. Although we have a qualitative understanding of the physics governing this Epoch of Reionization (EoR), even basic quantitative constraints such as the ionized fraction of hydrogen as a function of redshift have yet to be made with high precision. Such constraints would serve as incisive tests of EoR models.

As an example of the uncertainty in our theoretical understanding, consider the beginning of reionization. Recent large-scale observations of the CMB seem to favor late reionization scenarios, starting at $z \lesssim 15$ \cite{Planck2018,Gorce2018,Qin2020}, before which the intergalactic medium (IGM) was fully neutral. However, models including the effect of different star populations and self-regulated feedback can result in long tails with low ionized fraction extending to $z \sim 30$ \cite{Miranda2017,Heinrich2018}.

The end of reionization is more tightly constrained, for example via observations of high-redshift quasars and their spectra. Lyman-alpha absorption in quasar spectra due to neutral hydrogen gives rise to the Gunn-Peterson trough, which provides a convenient marker of the end of reionization \cite{Gunn_1965}. From such observations, the nominal redshift for the end of reionization is often taken to be $z\sim6$ \cite{Fan_2006} (see Ref.~\cite{Fan_2006_review} for a review), although recent studies have questioned the robustness of this conclusion \cite{Mesinger2010,Wise2019}.

Another important property of reionization is the redshift of its midpoint, when half the hydrogen in the IGM was ionized. This quantity is often estimated by using the CMB optical depth, $\tau$. Since the optical depth is obtained by integrating along the line of sight, it provides just one number to characterize reionization. However, this can be converted into a constraint for the midpoint under the assumption of a parametric form for the ionization history. For example, \emph{Planck} 2015 data \cite{Planck15} constrain the midpoint of reionization $z_{\text{re}}$ to be $z_{\text{re}} = 8.8^{+1.0}_{-1.1}$ under a redshift-symmetric parametrization of the ionization history \cite{Planck2016}.

Moving beyond an optical depth measurement, another CMB probe of reionization is the kinetic Sunyaev-Zel'dovich (kSZ) effect \citep{Zeldovich_Sunyaev_1969,Sunyaev_Zeldovich_1980}. The kSZ signal is sensitive to reionization since it arises from interactions of CMB photons with energetic free electrons produced during the EoR; see Refs.~\cite{Battaglia2013,Mesinger2012,Park2013,ChoudhuryMukherjee_2021} for examples on how the kSZ is sensitive to and can be used as a probe for reionization. The South Pole Telescope (SPT) collaboration has recently reported the first $\geq 3\sigma$ measurement of the kSZ angular power spectrum at an angular multipole of $\ell = 3000 $ \cite{Reichardt2021}. From this measurement, the authors have reported a constraint for the redshift interval between 25\% and 75\% ionization of $\Delta z = 1.1^{+1.6}_{-0.7}$. Such a measurement is possible because the patchiness of reionization partially sources the amplitude of the kSZ power spectrum, and so more extended and patchy reionization scenarios will lead to a larger contribution to the power. In this regard, the kSZ will be most sensitive to the gradual evolution of the ionization history over a long range of redshifts. Future kSZ measurements utilizing the full shape of the power spectrum rather than just the amplitude at $\ell = 3000$, enabled by observatories such as the Simons Observatory \citep{AdeAguirre_2019} and the CMB-Stage~4 experiments \citep{CMB-S4}, will unlock crucial information regarding the details of the ionization history \cite{Park2013,Alvarez2020,Gorce2020}. 

Perhaps a more direct tracer of the ionization history is the global $21\,\textrm{cm}$ signal. This sky-averaged brightness temperature of neutral hydrogen's hyperfine transition closely tracks the state of hydrogen, and precision measurements of the signal would allow for constraints of the ionized fraction as a function of redshift \cite{Shaver1999, Furlanetto2006globalsig, Pritchard2010, 2012MNRAS.424.2551M}. There are a number of experiments geared toward measuring the global $21\,\textrm{cm}$ signal such as the Experiment to Detect the Global EoR Signature (EDGES, \cite{Bowman2018}), the Probing Radio Intensity at High-Z from Marion (PRI$^{\text{Z}}$M, \cite{Philip2019}) experiment, the Large-aperture Experiment to Detect the Dark Age (LEDA, \cite{Price2018LEDA}), the Radio Experiment for the Analysis of Cosmic Hydrogen (REACH, \cite{2021arXiv210910098C}), and the Shaped Antenna measurement of the background RAdio Spectrum (SARAS, \cite{Singh2018}), with the EDGES measurement at 78~MHz being the only claimed detection of the cosmological signal to date \cite{Bowman2018}. The amplitude of the global signal can be up to four orders of magnitude dimmer than foreground emission such as galactic synchrotron radiation or extragalactic point sources \cite{ChapmanJelic_2019}. Mitigating foregrounds is made particularly difficult by the large dynamic range of the problem, which requires exquisite control of instrumental systematics. Even if instrumental systematics can be characterized and removed, foreground emissions continue to pose a challenge to global $21\,\textrm{cm}$ signal measurements because both the foregrounds and the cosmological signal are expected to be monotonically decreasing smooth functions of frequency. Consequently, abrupt reionization scenarios which give rise to sharp changes in the $21\,\textrm{cm}$ spectrum are easier to distinguish from smooth foreground contaminants and will be more easily detected with the global $21\,\textrm{cm}$ signal.

The global $21\,\textrm{cm}$ signal and the kSZ effect are therefore complementary probes of the EoR, with the former better at detecting abrupt features in the ionization history and the latter better at characterizing its smooth evolution. Previous works have looked at the potential of cross-correlating the kSZ signal with spatial $21\,\textrm{cm}$ fluctuations stemming from reionization \citep{JelicZaroubi_2010,TashiroAghanim_2011,PaulMukherjee_2021,LaPlante_2020,Ma2018,Alvarez_2006}. In this paper, we instead focus on a joint analysis with the global signal, which can lead to improved constraints on the ionization history. We first confirm the intuition that the $21\,\textrm{cm}$ and kSZ are complementary by using the Karhunen-Lo\`eve (henceforth KL) eigenvalue basis to describe the ionization history. This basis decomposes reionization into modes that are ordered by relative information content: the first few modes correspond to shapes in the ionization history best measured by the $21\,\textrm{cm}$ line, and the least well measured by the kSZ; the last few modes correspond to the opposite. Intermediate modes that are reasonably well measured by both can then be used as a consistency check between datasets and guard against systematics.

The rest of the paper is as follows: in Section~\ref{sec:theory}, we outline the theoretical background behind the $21\,\textrm{cm}$ global signal and the kSZ, and specify the types of experiments we are considering. In Section~\ref{sec:math}, we introduce the mathematical formalism for the KL transform in the context of jointly constraining reionization with the $21\,\textrm{cm}$ signal and the kSZ, and use the KL transform to concretely illustrate the complementarity of the two probes. In Section~\ref{sec:combining}, we explore how we can use the KL basis to combine data from $21\,\textrm{cm}$ and kSZ  measurements into a single joint constraint of the ionization history, assuming no experimental systematics. This assumption is discarded in Section~\ref{sec:systematics}, where we show how our KL formalism can be wielded to detect the presence of some common systematics. In Section~\ref{sec:systematics_remain} we go beyond the mere detection of these systematics and present methods for their removal. We summarize our conclusions in Section~\ref{sec:conclusions}.

\section{Theoretical Background and Experimental Assumptions \label{sec:theory}}

In this section, we provide the necessary theoretical background for our two signals of interest. We first present the relevant equations governing them, and outline how each depends on the ionization history. We then present what a typical experiment and error covariances might look like for each of the probes.

\subsection{The global 21\,cm signal \label{sec:theory_21cm}}
\label{sec:21cmintro}
One of the most promising cosmological probes is hydrogen, and characterizing it throughout cosmic time can inform us about the large-scale processes of the Universe. Hydrogen is a particularly good probe of the Dark Ages (the period of time before the first stars and galaxies formed) and the subsequent EoR due to its hyperfine transition line with an emission wavelength of $21\,\textrm{cm}$. There are a number of review papers \cite{Furlanetto2006,Morales2010,Pritchard2012} which comprehensively discuss the physics behind the cosmological $21\,\textrm{cm}$ signal and how it can be used to constrain reionization. Here, we will outline the basic relevant concepts, as well as describe what a typical global signal experiment might look like.  

The $21\,\textrm{cm}$ brightness temperature is seen in contrast to a background radiation, usually (but not always) assumed to be the CMB. Thus we define the differential $21\,\textrm{cm}$ brightness temperature $ \delta T_b$, which is given by \cite{Pritchard2012}:
\begin{align}
    \delta T_b = 27\, &x_\mathrm{HI} \bigg(\frac{T_s - T_{\gamma}}{T_s}\bigg) \bigg(\frac{1+z}{10}\bigg)^{1/2}(1 + \delta_b) \nonumber \\
    &\times\bigg[\frac{\partial v_r /dr}{(1+z)H(z)}\bigg]^{-1} \text{mK}, \label{eq:globalsignal}
\end{align}
where $x_\mathrm{HI}$ is the fraction of neutral hydrogen, ranging from 0 to 1, $T_s$ is the $21\,\textrm{cm}$ spin temperature, $T_{\gamma}$ is the CMB temperature, $\delta_b$ is the baryon overdensity, and $H(z)$ is the Hubble parameter. The last term tracks contributions to the brightness temperature from peculiar velocities $v_r$ along the line of sight, with $r$ denoting the line of sight distance. Here, we focus on the global signal, that is the spatial average of the differential brightness temperature. Since averaging the signal over the sky means the measurement is not sensitive to fluctuations, the $\delta_b$ term as well as the contribution from peculiar velocities are negligible. Furthermore, we make the simplifying assumption \citep{MirochaFurlanetto_2017,HenekaMesinger_2020} that the $21\,\textrm{cm}$ spin temperature is much larger than the CMB temperature during reionization, so that the dependence on $T_s$ drops out. Therefore, in this scenario we may approximate the sky-averaged differential brightness temperature at redshift $z$ by
\begin{equation}
    \delta  T_b (z) \approx 27\,  \overline{x}_\mathrm{HI} (z) \bigg(\frac{1+z}{10}\bigg)^{1/2}\,{\rm mK},\label{eq:globalsignal_reduced}
    \end{equation}
where $\overline{x}_\mathrm{HI}$ is the mean fraction of neutral hydrogen. An experiment aiming to measure the global signal during reionization might target frequencies in the $100$ to $200~\mathrm{MHz}$ range, since this corresponds to redshifted $21\,\textrm{cm}$ signals originating from $z \sim 13$ to $z \sim 6$, and current empirical constraints suggest that reionization occurred roughly in the $z\sim5$ to $15$ range \citep{Planck2016,Gorce2018}. Suppose such an experiment measures the sky at a number of frequency channels; the variance $\sigma_i^2$ in the $i$th frequency channel is given by
\begin{equation}
    \sigma_i^2 = \frac{T_{\text{sky}}^2(\nu_i)}{b\, t_{\text{int}}},
\end{equation}
assuming a sky noise dominated instrument, where $b$ is the channel width and $t_{\text{int}}$ is the integration time. The sky temperature $T_{\text{sky}}$ is in principle the sum of the cosmological brightness temperature $\delta T_b$ and the brightness temperature of the foregrounds $T_{\rm fg}$. However, in practice the foregrounds are so much brighter than the signal that $T_{\text{sky}} \approx T_{\rm fg}$. We model the foregrounds as a polynomial of degree $N$ logarithmic in frequency, 
\begin{equation}
\label{eq:actualTskyfgcontribution}
    \ln T_\mathrm{fg}  = \sum_i^{N} a_i \left[ \ln \bigg( \frac{\nu}{\nu_*}\bigg)\right]^{i},
\end{equation}
where $\nu$ is the frequency, $\nu_* \equiv 150 \,\textrm{MHz}$ is an arbitrary pivot frequency, and following Ref. \cite{Pritchard2010} we take $N=3$ and adopt their best fit $a_i$ values (which were based on data from Ref. \cite{2008AJ....136..641R}). In order to construct a noise covariance $\boldsymbol{\Pi}_{\text{noise}}$ for the experiment, we assume uncorrelated noise between frequency bins:
\begin{equation}
    (\boldsymbol{\Pi}_{\text{noise}})_{ij} = \delta_{ij} \sigma_i^2,
\end{equation}
where $\delta_{ij}$ is the Kronecker delta function.

In this paper, we do not explicitly model the foreground subtraction process. Instead, we include a foreground contribution to the covariance. This will have the effect of severely downweighting the foregrounds in one's downstream analysis. In fact, this can be seen as a conservative way to model the effect of foregrounds, since in practice the covariance ought to represent foreground residuals after one has subtracted off a best-guess estimate (rather than that of the foregrounds themselves, which is presumably larger) \cite{Liu2013globalsig}. We assume that at the relevant frequencies, Galactic synchrotron and free-free emission dominate. Following Ref. \cite{Liu2012}, we compute the foreground covariance $\boldsymbol{\Pi}_\mathrm{fg}$ between frequencies $\nu$ and $\nu^\prime$ for each of these components as
\begin{equation}
    (\boldsymbol{\Pi}_\mathrm{fg})_{\nu\nu'} = A^2 \bigg( \frac{\nu \nu'}{\nu_{*}^2}\bigg)^{-\alpha + \frac{1}{2}\Delta \alpha^2\ln(\nu \nu' / \nu_{*}^2)} - m(\nu)\, m(\nu'),
\end{equation}
with\footnote{Note that although Equations~\eqref{eq:actualTskyfgcontribution} and \eqref{eq:meanfgmodel} give roughly similar results, they are in principle different models. We use Equation~\eqref{eq:actualTskyfgcontribution} for our mean sky temperature because it is based on fits to empirical data. However, at the relevant frequencies the data are generally not sensitive enough to produce empirical covariances that are not artificially rank-deficient \cite{Liu2013globalsig}. It is for this reason that we adopt the semi-empirical approach from Ref.~\cite{Liu2012} for the covariance only.}
\begin{equation}
\label{eq:meanfgmodel}
    m(\nu) = A\bigg(\frac{\nu}{\nu_{*}} \bigg)^{-\alpha + \frac{1}{2}\Delta \alpha^2\ln(\nu / \nu_{*}^2)},
\end{equation}
where $A$ controls the foreground amplitude of the component, $\alpha$ is its spectral index, $\Delta \alpha$ is the running of the spectral index. The values we take for each of these parameters are outlined in Table~\ref{tab:fgnd_params} for the two emission mechanisms, whose covariance we sum to give a total foreground covariance. We then take the total covariance of our hypothetical global signal experiment to be the sum of the instrument and total foreground covariances:
\begin{equation}
    \boldsymbol{\Pi}_{21} = \boldsymbol{\Pi}_{\text{noise}} +\boldsymbol{\Pi}_\mathrm{fg}.\\
\end{equation}
We will use this theoretical experiment and noise covariance in order to compute the Fisher matrix for the global $21\,\textrm{cm}$ signal, which we will then use to define the KL basis and explore the complementarity between the $21\,\textrm{cm}$ signal and the kSZ.

\begin{table}
	\centering
	\caption{Parameters used in the computation of the foreground covariance.}
	\label{tab:fgnd_params}
	\begin{tabular}{lll} 
		\hline
		Parameter \qquad\qquad & Synchrotron emission \qquad\quad & Free-free emission\\
		\hline
		A & 335.4 K & 33.5 K \\
		$\alpha$ & 2.8 & 2.15 \\
		$\Delta \alpha$ & 0.1 & 0.01\\
		\hline
	\end{tabular}
\end{table}

\subsection{The kinetic Sunyaev-Zel'dovich effect}

The scattering of CMB photons off inhomogeneities in the electron density along the line of sight creates anisotropies in the CMB temperature. One such anisotropy is the kinetic Sunyaev-Zel'dovich effect (kSZ) \citep{Zeldovich_Sunyaev_1969,Sunyaev_Zeldovich_1980}, which results from the interaction of low-energy CMB photons with electrons that have a bulk velocity relative to the CMB rest-frame. The temperature fluctuations sourced by the kSZ effect can be written as
\begin{equation}
\label{eq:delta_TkSZ}
  \delta T_{\rm kSZ}(\los) =\frac{\sigma_T}{c} \int \frac{\drm \eta}{\drm z}\frac{\drm z}{(1+z)}\, \mathrm{e}^{-\tau(z)}\, \overline{n}_e(z) \, \vvec\cdot\los ,
\end{equation}
where $\overline{n}_e(z)$ is the free electron number density, $\tau$ the CMB optical depth, both averaged over the sky, $\sigma_T$ the Thomson cross-section, $c$ the speed of light, $\eta$ the comoving distance to redshift $z$, and $\vvec \cdot \los$ the component of the peculiar velocity of the electrons along the line of sight. Note that, in contrast to the 21\,cm signal, the kSZ effect is sensitive to the density of electrons along the line-of-sight, whether they are coming from hydrogen or helium reionization. 
It is comprised of two components: the patchy kSZ, due to the scattering of CMB photons off ionised bubbles during the EoR, and the late-time component, sourced by the large-scale distribution of matter once the IGM has been fully ionized. In this work, we will focus on the former.

Similarly to the primary CMB temperature fluctuations, the kSZ is currently measured in terms of its angular power spectrum $C_\ell^{k \rm SZ} \equiv T^2_\mathrm{CMB} \vert \tilde{\delta T}_\mathrm{kSZ}(k)\vert^2$ where $k \equiv \ell / \eta$ is the Limber wave-vector and $\ell$ is the multipole moment, which can be related to an angular scale in the sky. In Ref. \cite{Gorce2020}, we introduced a way of deriving the $C_\ell^{k \rm SZ}$ in terms of a few cosmological and physically-motivated parameters, including the reionization history $x_i(z)$, which we will use to derive the Fisher matrix of the kSZ signal in the next section. 

Foregrounds dominate the primary CMB signal on small angular scales (smaller than 1~arcmin), making a measurement of the kSZ power challenging \cite{Reichardt2021}. However, forecasts indicate that the new generation of CMB observatories such as CMB-Stage~4 (CMB-S4) \citep{AdeAguirre_2019} will provide a definite measurement of the signal \citep{CalabreseHlozek_2014}. Taking this perspective, to derive the kSZ error covariance matrix defined in Section~\ref{sec:fisher} we will assume the instrumental specifications of CMB-S4. Since unlike any other foreground, the kSZ signal is not frequency-dependent, we can assume a perfect multi-frequency cleaning of CMB-S4 data from other foregrounds and use the noise power spectrum from Ref.~\cite{Alvarez2020} as our diagonal noise covariance. Additionally, we consider cosmic variance errors $\Delta C_\ell$ as a fraction of the observed $C_\ell$, including the primary and the kSZ signal after foreground cleaning:
\begin{equation}
\frac{\Delta C_\ell}{C_\ell} = \sqrt{\frac{2}{f_\mathrm{sky}(2\ell+1)}},
\end{equation}
where $f_\mathrm{sky}=0.45$ is the effective fractional sky area covered by CMB-S4 \citep{Alvarez2020}. Our resulting error covariance is a diagonal matrix\footnote{The assumption of a diagonal error covariance matrix is justified since we will employ rather large bins in $\ell$ in Section~\ref{sec:math}, and in that limit, current experiments appear to be well-modelled by diagonal covariance matrices \cite{Reichardt2021}.} made of noise and cosmic variance contributions.

Of course, even with a proper assessment of errors, some systematics will in practice affect measurements of the kSZ signal. In this paper we will consider a potential residual contribution from the primary CMB to the measured kSZ signal, which could arise from a biased cosmological model. In addition, despite plans for multi-frequency cleanings of foregrounds, some foregrounds will certainly remain at a low level. For example, there exists the cross-spectrum between the Cosmic Infrared Background (CIB) and the thermal SZ effect (tSZ) \citep{AddisonDunkley_2012}, whose improper modelling has prevented a precise measurement of the kSZ amplitude in the past \citep{ReichardtShaw_2012,SieversHlozek_2013,GeorgeReichardt_2015,Reichardt2021}. 

\section{Mathematical Formalism} \label{sec:math}

Having established our model for $21\,\textrm{cm}$ and kSZ measurements in Section~\ref{sec:theory}, we now focus on how each probe contains information about the ionization history. We will first quantify their information content using their respective Fisher information matrices in Section~\ref{sec:fisher} before illustrating their complementarity in Section~\ref{sec:KLformalism} with the KL transform. 

\subsection{The Fisher information matrix \label{sec:fisher}}

Our goal is to provide a set of basis modes that can be used to delineate shapes in the ionization history that the kSZ and global $21\,\textrm{cm}$ signal are sensitive to. A natural starting point is to use the Fisher information matrix to quantify the information contained about the ionization history in each of these probes. Recall that the Fisher matrix $\mathbf{F}$ is defined as
\begin{equation}
    \mathbf{F}_{\alpha\beta} \equiv -\biggl< \frac{\partial^2 \ln \mathcal{L}}{\partial \theta_{\alpha} \partial \theta_{\beta}} \biggr> ,
\end{equation}
where $\mathcal{L}$ is the likelihood function and $\boldsymbol \theta\equiv(\theta_1, \theta_2, \dots)$ are parameters that we wish to constrain using our measurements. The Fisher matrix characterizes the width of the likelihood (as a function of the parameters) under the approximation that it is close to Gaussian; intuitively, it therefore characterizes the information content of a measurement since narrower likelihoods are tantamount to tighter constraints on parameters. For a measurement consisting of observables that have a mean vector $\boldsymbol \mu$ that is measured with covariance $\boldsymbol \Pi$, the Fisher matrix reduces to
\begin{equation}
 \mathbf{F}_{\alpha\beta} = \frac{\partial \boldsymbol \mu^T}{\partial \theta_\alpha} \boldsymbol \Pi^{-1} \frac{\partial \boldsymbol \mu}{\partial \theta_\beta},
\end{equation}
which is the form that we use here. Under ideal conditions, the covariance $\mathbf{C}$ of the final measured parameters stored in $\boldsymbol \theta$ is then given by $\mathbf{F}^{-1}$.

In this paper, our focus is on constraining the ionization history, which we can parametrize by specifying the value of the ionized fraction of hydrogen $x_i$ in a set of predefined redshift bins. These values, which include hydrogen and helium reionization, then form our parameter set, i.e., $\theta_{\alpha} = x_i(z_{\alpha})$. We take 20 bins uniformly spaced in redshift, between $z=6$ and $z=13$, although the formalism can be generalized to any other choice of bins. For the global $21\,\textrm{cm}$ signal, our observable is a series of $21\,\textrm{cm}$ brightness temperatures at different frequencies, $\mathbf{T}_i \equiv T(\nu_i)$. This gives a Fisher matrix $\mathbf{F}_{\alpha\beta}^{21}$ for the $21\,\textrm{cm}$ global signal of the form
\begin{equation}
    \mathbf{F}_{\alpha\beta}^{21} = \frac{\partial \mathbf{T}^T}{\partial x_i(z_{\alpha})} \boldsymbol{\Pi}_{21}^{-1} \frac{\partial \mathbf{T}}{\partial x_i(z_{\beta})},\label{eq:GS_fisher}
\end{equation}
where $\boldsymbol{\Pi}_{21}$ is the covariance matrix of our measurements introduced in Section~\ref{sec:theory_21cm}. For the kSZ, we consider measurements of its dimensionless patchy angular power spectrum at different angular multipoles $\ell$, i.e., $D^p_{\ell} = \ell(\ell +1)C^p_{\ell}/2\pi$, gathered in a vector $\mathbf{D}^p$, with covariance $\boldsymbol{\Pi}_{\rm kSZ}$, such that
\begin{equation}
    \mathbf{F}_{\alpha\beta}^{\rm kSZ} =  \frac{\partial (\mathbf{D}^p)^T}{\partial x_{i}(z_{\alpha})} \boldsymbol{\Pi}_{\rm kSZ}^{-1} \frac{\partial \mathbf{D}^p}{\partial x_{i}(z_{\beta})}.\label{eq:ksz_fisher}
\end{equation}
Here, we consider twenty-two multipole bins of identical width ranging from $\ell = 690$ to $\ell = 7900$, similar to the range of multipoles covered by the SPT \citep{KeislerReichardt_2011,Reichardt2021}.

The two Fisher matrices above then quantify the sensitivity of the global $21\,\textrm{cm}$ signal and the kSZ effect to changes in the ionized fraction in a given redshift bin. The fact that the former is most sensitive to rapidly evolving ionization histories and the latter to extended histories is thus a complementarity that should be reflected in their Fisher matrices. If these matrices were diagonal, the complementarity would be easy to see by simply taking the ratio of their elements. Since this is generally not the case, we can simultaneously diagonalize the two Fisher matrices using a Karhunen-Lo\`eve (KL) transformation which will cast our ionization history into a new basis where the complementarity is clear.

\subsection{The Karhunen-Lo\`eve transform\label{sec:kl}}
\label{sec:KLformalism}

The KL transform casts a measurement of two signals into a basis whose modes and eigenvalues effectively describe the ratio between the two. This method is often used to form a series of modes rank-ordered by their signal-to-noise, which provides a convenient basis for data compression (e.g., \cite{Tegmark2006,Zablocki2016,Szalay2003}). Here, we extend the formalism beyond a signal-to-noise analysis and instead use the KL transform to obtain kSZ-to-21cm modes. In this case, the eigenvalues of the transformation and their corresponding eigenvectors inform us about which modes of the ionization history have a higher kSZ-to-21cm information content and thus are better measured by the kSZ effect. Conversely, modes with lower kSZ-to-21cm information are those that are better measured by the 21cm signal. Therefore, as intended, our KL basis will highlight the complementarity between the two probes.

We begin by solving the generalized eigenvalue problem
\begin{equation}
    \mathbf{F}_{\rm kSZ} \mathbf{v} = \lambda \mathbf{F}_{21} \mathbf{v}, \label{eq:generalized_eigval}
\end{equation}
where $\mathbf{F}_{\rm  kSZ}$ and $\mathbf{F}_{21}$ are the Fisher matrices of the kSZ and 21cm global signal, respectively. Performing a Cholesky decomposition on the $21\,\textrm{cm}$ covariance $\mathbf{C}_{21} = \mathbf{F}^{-1}_{21}$ allows us to write
\begin{equation}
    \mathbf{F}_{21} = \mathbf{C}_{21}^{-1} = \mathbf{L}_{21}^{-T}\mathbf{L}_{21}^{-1},
\end{equation}
where $\mathbf{L}_{21}$ is a lower triangular matrix. Equation~\ref{eq:generalized_eigval} then becomes
\begin{equation}
    \mathbf{L}_{21}^{T} \mathbf{F}_{\rm kSZ} \mathbf{L}_{21} \mathbf{L}_{21}^{-1} \mathbf{v} = \lambda \mathbf{L}^{-1}_{21} \mathbf{v}, 
\end{equation}
which reduces to an eigenvalue problem 
\begin{equation}
    \mathbf{G} \mathbf{w} = \lambda \mathbf{w}, \label{eq:eigenval_problem}
\end{equation}
with $\mathbf{G} \equiv \mathbf{L}_{21}^{T} \mathbf{F}_{\rm kSZ} \mathbf{L}_{21}$ and $\mathbf{w} \equiv \mathbf{L}_{21}^{-1}\mathbf{v}$. With these definitions, we define the KL transformation matrix as 
\begin{equation}
    \mathbf{R} \equiv \mathbf{L}_{21}\mathbf{\Psi},
\end{equation}
where the columns of $\mathbf{\Psi}$ are the eigenvectors $\mathbf{w}$ satisfying Equation~\eqref{eq:eigenval_problem}. If we have a measurement of the ionization history $\mathbf{x} = (x_i(z_1), x_i(z_2),\dots, x_i(z_n))$, its representation $\mathbf{y}$ in the KL basis is given by 
\begin{equation}
    \mathbf{y} = \mathbf{R}^{-1} \mathbf{x}, \label{eq:transformation}
\end{equation}
and the inverse relation is
\begin{equation}
    \mathbf{x} = \mathbf{R} \mathbf{y}. \label{eq:modes}
\end{equation}

In the KL basis, the information content (as expressed by the Fisher information matrices) is diagonal for both global $21\,\textrm{cm}$ and kSZ measurements. Transforming their respective Fisher matrices via appropriate Jacobian factors, we obtain
 \begin{equation}
 \overline{\mathbf{F}}_{21} = \mathbf{\Psi}^{T} \mathbf{L}^{T}_{21} \mathbf{F}_{21} \mathbf{L}_{21} \mathbf{\Psi} = \mathbf{\Psi}^T \mathbf{\Psi} = \mathbf{I}, \label{eq:cov_21_kl}
\end{equation}
for the $21\,\textrm{cm}$ Fisher matrix in the KL basis and 
\begin{equation}
 \overline{\mathbf{F}}_{\rm kSZ} = \mathbf{\Psi}^{T} \mathbf{L}^{T}_{21} \mathbf{F}_{\rm kSZ} \mathbf{L}_{21} \mathbf{\Psi}= \mathbf{\Psi}^T \mathbf{G} \mathbf{\Psi} = \mathbf{\Lambda} \label{eq:cov_ksz_kl}
 \end{equation}
for the kSZ Fisher matrix in the KL basis, where $\boldsymbol \Lambda$ is a diagonal matrix where $\mathbf{\Lambda}_{\alpha \alpha} = \lambda_\alpha$ for $\lambda_\alpha$ satisfying Equation~\eqref{eq:eigenval_problem}. Both Fisher matrices are diagonal

Since the KL transformation defines a basis, it also defines a set of basis vectors, or modes, which we can linearly combine to reconstruct an ionization history. The amplitudes of the linear combination are given by Equation~\eqref{eq:transformation}, which we can write as
\begin{equation}
    \mathbf{x} = \sum_\alpha  \boldsymbol{\varphi}_\alpha y_\alpha,
\end{equation}
where $\{ \boldsymbol{\varphi}_\alpha \}$ are the columns of $\mathbf{R}$ and the modes of our transformation. If we order the elements of our matrices and vectors such that the eigenvalues in $\boldsymbol \Lambda$ are ordered from largest to smallest, then $\boldsymbol \varphi_1$ will be the mode that is comparatively best constrained by kSZ measurements and worst constrained by global $21\,\textrm{cm}$ signal measurements. On the other extreme, the final mode $\boldsymbol \varphi_{n}$ (where $n$ is the number of redshift bins used to describe the ionization history) is best measured by the global $21\,\textrm{cm}$ signal.

\begin{figure}[!ht]
    \centering
    \includegraphics[width = \columnwidth]{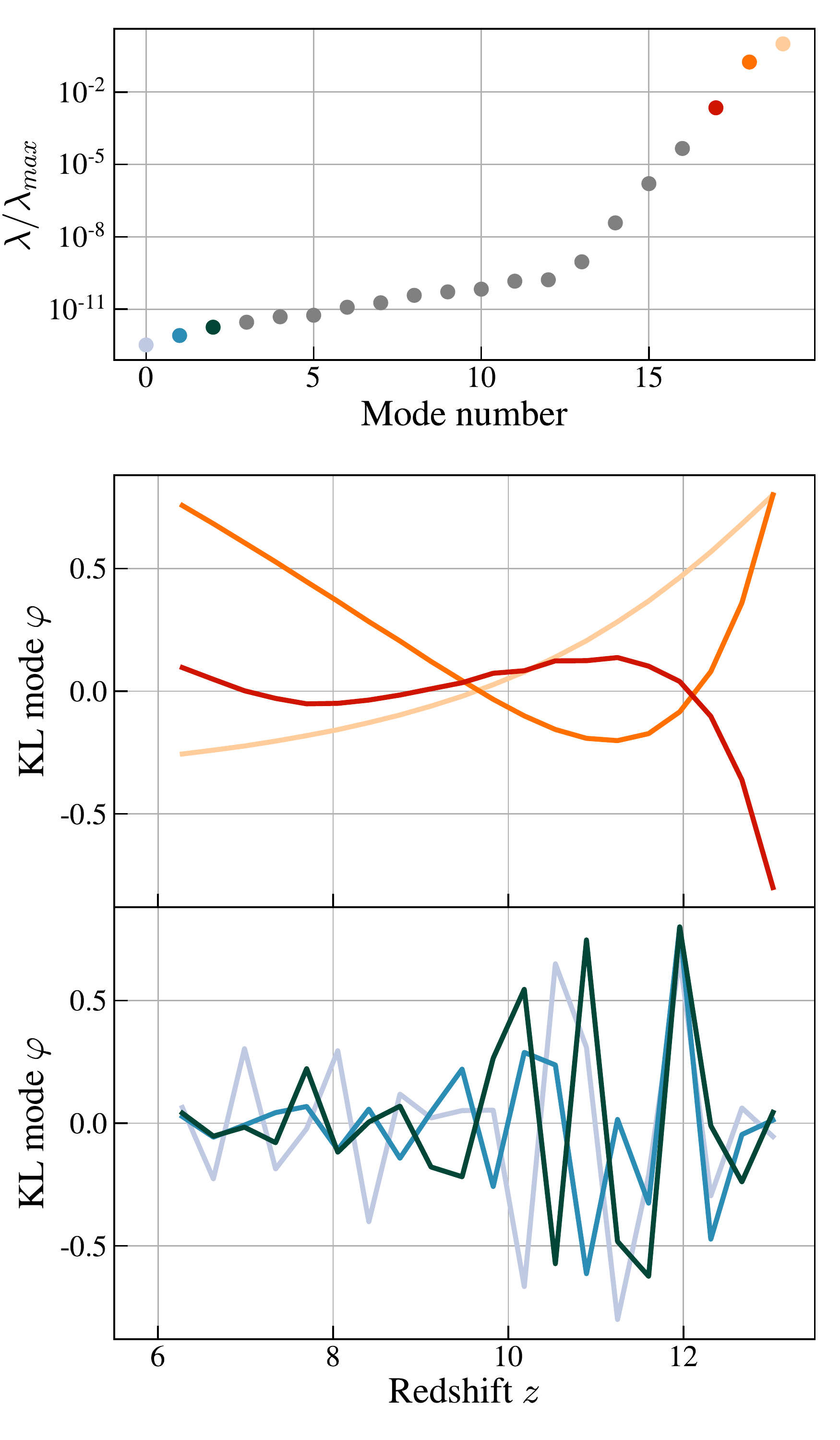}
    \caption{Normalized eigenvalues of the KL transform for an asymmetric reionization history (see text and Fig.~\ref{fig:errorbars}). A larger value corresponds to a mode best measured by the patchy kSZ signal whilst a lower value corresponds to a mode best measured by the global 21\,cm signal. The KL modes associated with the three largest and the three smallest eigenvalues are represented as a function of redshift in, respectively, the middle and the lower panel.}
    \label{fig:modes}
\end{figure}

\subsection{Worked examples of kSZ-to-$21\,\textrm{cm}$ modes}

\begin{figure*}
    \centering 
    \includegraphics[width = \linewidth]{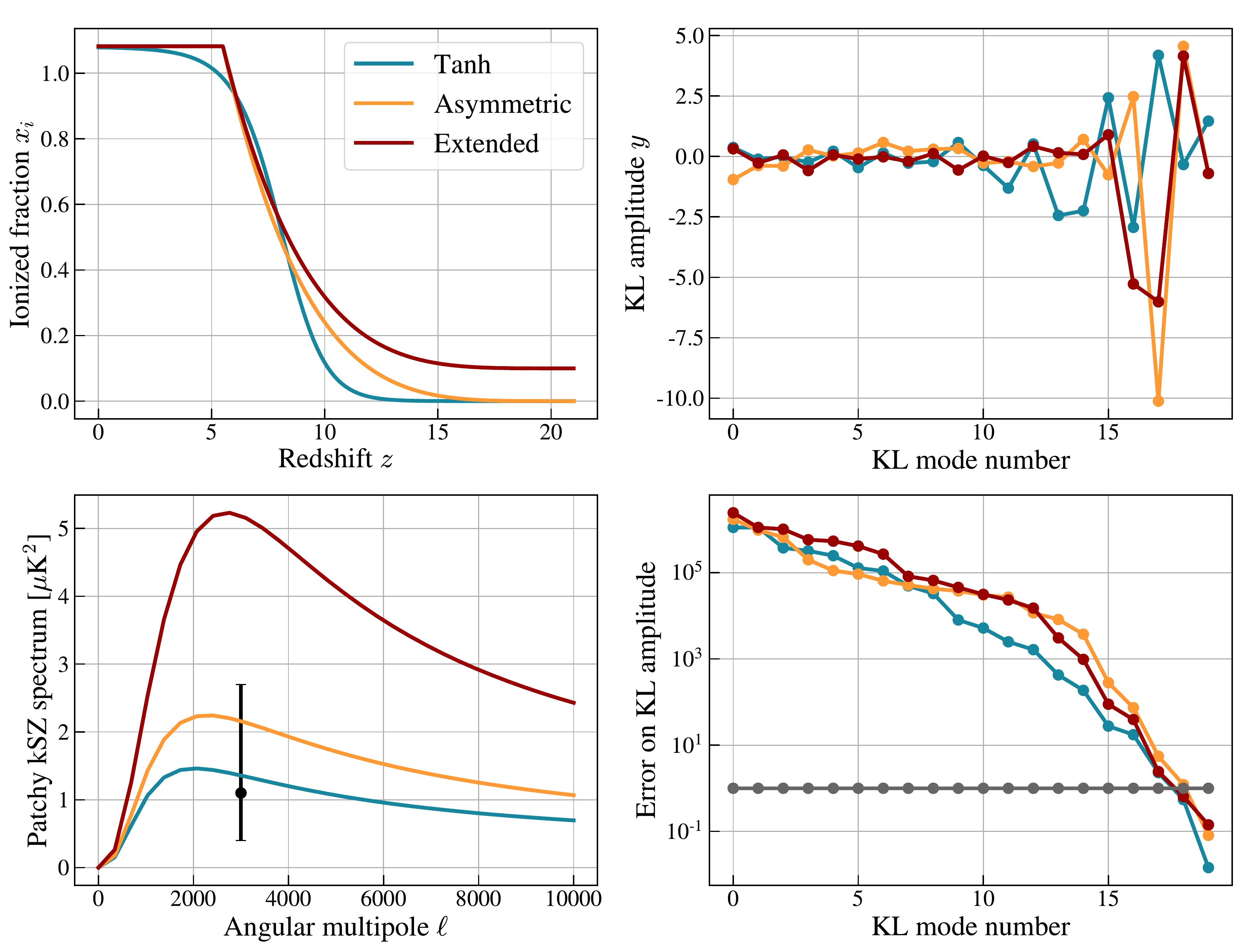}
    \caption{Models considered in the analysis and their KL counterparts. \textit{Left:} Evolution of the IGM ionized fraction with redshift for three EoR models and their corresponding patchy kSZ angular power spectra, compared with the only measurement of the patchy kSZ amplitude to date \citep{Reichardt2021}. Our fiducial results are obtained with the orange (``asymmetric") reionization history. \textit{Right:} Amplitude and expected error bars on the KL modes associated with each model. Errors obtained with the 21\,cm global signal are one by construction.}
    \label{fig:errorbars}
\end{figure*}

In the top panel Figure~\ref{fig:modes}, we plot the eigenvalues of the KL transform for a fiducial ionization history (we take the ``asymmetric" reionization, shown in the top left panel of Figure~\ref{fig:errorbars}), obtained with the parametrization presented in \cite{Douspis2015}. By construction, the transformation is such that the $21\,\textrm{cm}$ Fisher information is the identity matrix in the KL basis, so these eigenvalues quantify the kSZ-to-21cm information ratio. The largest eigenvalues correspond to the modes best measured by the kSZ, the highest three of which are plotted in the middle panel of Figure~\ref{fig:modes}. The smallest eigenvalues correspond to the modes best measured by the global signal, the lowest of which are plotted in the bottom panel of the figure. The shapes of these modes are consistent with our intuition: since the kSZ will best constrain extended ionization histories, the modes that correspond to the highest kSZ-to-21cm signal are relatively smooth and extended. The 21cm global signal, on the other hand, is most sensitive to rapidly evolving ionization histories so the modes with lowest kSZ-to-21cm signal fluctuate more rapidly with redshift.

Although we picked a fiducial asymmetric ionization history for Figure~\ref{fig:modes}, the qualitative intuition remains similar for other models. This is illustrated in Figure~\ref{fig:errorbars}. The left column shows three example ionization histories and their corresponding patchy kSZ power spectra: one is similar to our fiducial reionization history, but presents an extended high-redshift tail, corresponding to sources reionizing the Universe as early as $z=20$. Another is the step-like redshift-symmetric ``instantaneous" reionization model often considered in CMB data analyses \citep{Lewis2008}. We repeated our analysis for a number of non-standard ionization histories resulting, for example, from simulations including feedback in Pop~III stars \cite{Miranda2017,Heinrich2018}. We find that for all these models, results are qualitatively similar.Note that we omit an explicit illustration of the global $21\,\textrm{cm}$ signal since the approximations in Section~\ref{sec:21cmintro} render it directly proportional to the neutral fraction. The top-right panel shows the ionization histories in the KL basis, whilst the bottom-right panel gives the expected errors on the amplitude of each KL mode. By construction, the $21\,\textrm{cm}$ measurements have unit errors in the KL basis. The errors on the kSZ measurements are given by the inverse square root of eigenvalues in Equation~\eqref{eq:cov_ksz_kl}. By comparing the errors of the global signal and the kSZ measurements, we can determine which modes are best measured by which signal. Note that although a cursory glance at the errors suggests that there are vastly more modes that can be measured using the $21\,\textrm{cm}$ line compared to kSZ, one can see that many of these modes contribute very little to the typical ionization history. Thus, in a practical scenario, one is likely to see a less lopsided contribution of information, and indeed, in Section~\ref{sec:combining} we will see that both probes are necessary for accurate reconstructions of the ionization history.

\section{Combining probes when they are consistent \label{sec:combining}}

In an ideal situation, one might find that all datasets at one's disposal are free of systematic effects. Under such circumstances, the next step in one's analysis is to combine data from our two probes into a single consistent ionization history. In this section we establish a formalism for doing so, deferring a discussion of potential systematics (and how to detect them) to Section~\ref{sec:systematics}.

\subsection{A least squares approach}
\label{sec:leastsq}
One method for combining data into a single ionization history is to employ a least-squares estimator, where we wish to construct an estimator $\widehat{\mathbf{x}}$ for the true ionization history $\mathbf{x}$, given a collection of measurements $\mathbf{y}$. In our case, we take $\mathbf{y}_{\rm conc} = (\mathbf{y}_{21},\mathbf{y}_{\rm kSZ})$ to be a concatenation of the KL coefficients. Relating the measurement to what we want to constrain is the linear equation
\begin{equation}
    \mathbf{y}_{\rm conc} = \mathbf{A}\mathbf{x} + \mathbf{n}_{\rm conc},\label{eq:measurement}
\end{equation}
where $\mathbf{A}$ is the design matrix (in this case is a vertical stack of two $\mathbf{R}^{-1}$ matrices), and $\mathbf{n}_{\rm conc}$ is the concatenated noise vector, i.e., $\mathbf{n}_{\rm conc} = (\mathbf{n}_{\rm 21}, \mathbf{n}_{\rm kSZ})$. Note that this noise is not the instrumental noise contribution from the original $21\,\textrm{cm}$ or kSZ measurements, but instead, the ``noise" in our determination of $\mathbf{y}_{\rm 21}$ and $\mathbf{y}_{\rm kSZ}$. It therefore has a covariance matrix $\mathbf{N}$ given by
\begin{equation}
\mathbf{N}=  \begin{pmatrix}
\overline{\mathbf{C}}_{\rm 21} & \mathbf{0} \\
\mathbf{0} & \overline{\mathbf{C}}_{\rm kSZ}
 \end{pmatrix}.
\end{equation}
With these definitions, the least squares estimator for the ionization history takes the form\footnote{Optionally, one may choose to group the constraints on the ionization history into coarser bins than the native bin size used in our Fisher matrices in Equations ~\eqref{eq:GS_fisher} and \eqref{eq:ksz_fisher}. In other words, there is no requirement that $\mathbf{x}$ and $\mathbf{y}_{\rm conc}$ be of the same length; the former can be shorter than the latter. To enable such a setup, one makes the substitution $\mathbf{A} \rightarrow \mathbf{A B}$, where $\mathbf{B}$ is a rectangular matrix of $1$s and $0$s that duplicates entries in a shorter, binned version of $\mathbf{x}$ to transform it into its original full-length equivalent.} \cite{T97maps}
\begin{eqnarray}
    \widehat{\mathbf{x}} &=& (\mathbf{A}^T\mathbf{N}^{-1}\mathbf{A})^{-1}\mathbf{A}^T \mathbf{N}^{-1} \mathbf{y} \label{eq:estimator} \nonumber \\
    &=& \mathbf{R}^T(\overline{\mathbf{F}}_{\rm 21} + \overline{\mathbf{F}}_{\rm kSZ})^{-1}\mathbf{R} \mathbf{R}^{-T}(\overline{\mathbf{F}}_{\rm 21}\mathbf{y}_{\rm 21} + \overline{\mathbf{F}}_{\rm kSZ}\mathbf{y}_{\rm kSZ}) \nonumber  \\
    &=&  \mathbf{C}_{\rm tot} \mathbf{R}^{-T}(\overline{\mathbf{F}}_{\rm 21}\mathbf{y}_{\rm 21} + \overline{\mathbf{F}}_{\rm kSZ}\mathbf{y}_{\rm kSZ}) \label{eq:comboest},
\end{eqnarray}
where we have defined
\begin{equation}
\mathbf{C}_{\rm tot} \equiv (\mathbf{F}_{\rm 21} + \mathbf{F}_{\rm kSZ})^{-1}, \label{eq:cov_tot}
\end{equation}
since this can be shown to be the covariance of our final estimator $\widehat{\mathbf{x}}$. Unsurprisingly, all traces of the KL transform vanish from the final covariance. Moving to KL space is simply a convenient intermediate step that elucidates the complementarity of the global $21\,\textrm{cm}$ signal and (as we shall see in Section~\ref{sec:combining}) allows for the detection of residual systematics. The information content of the datasets is unchanged by our choice of intermediate basis, and thus it is expected that the final covariance in a joint determination of the ionization history is the inverse of
 the sum of the constituent Fisher matrices.

\begin{figure}
    \centering
    \includegraphics[width=0.5\textwidth,trim=0.0cm 2cm 1.75cm 1.90cm,clip]{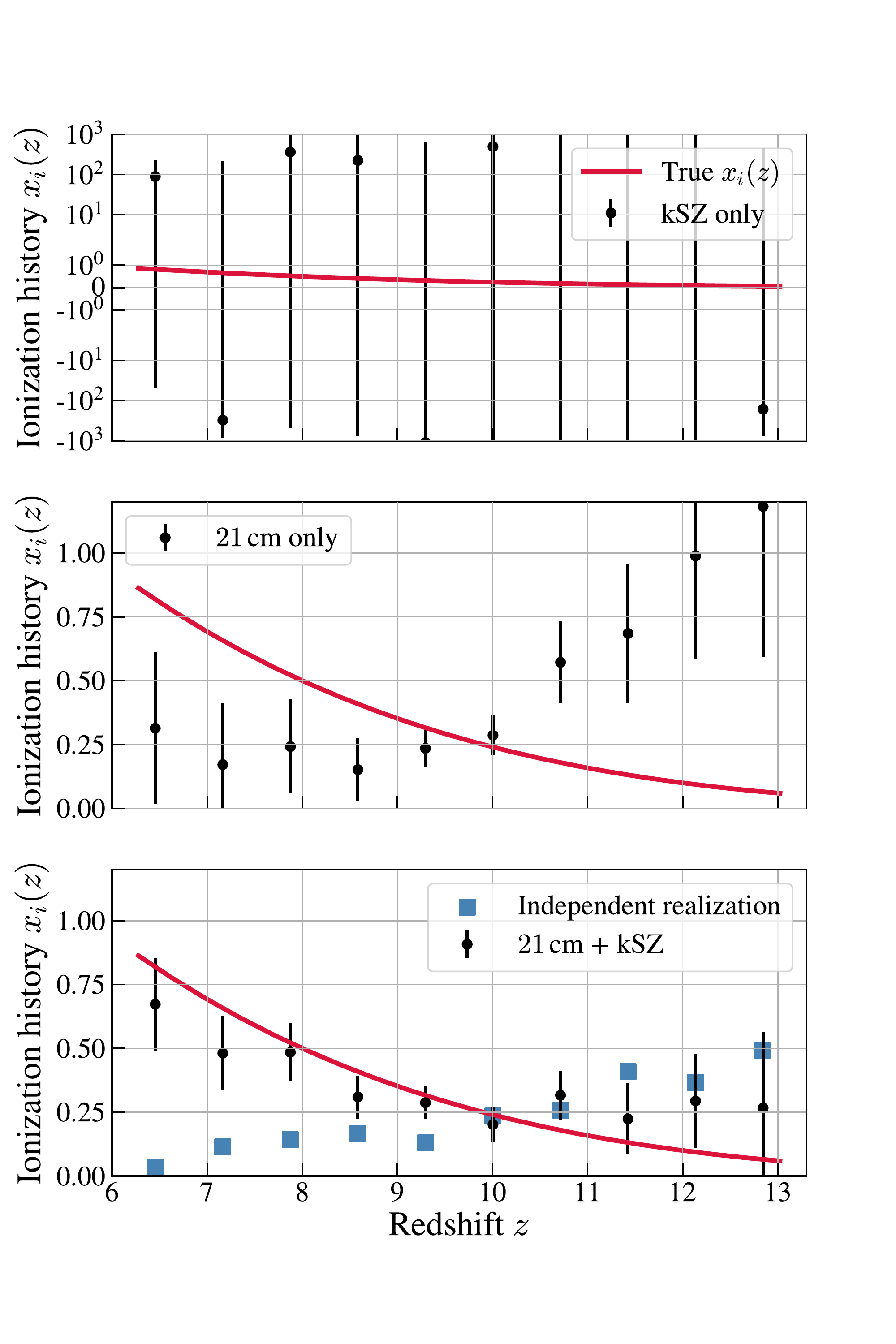}
    \caption{Simulated recoveries (black points with error bars) of the ionization history (red curve) from kSZ measurements alone (top panel), global $21\,\textrm{cm}$ measurements alone (middle panel), and combined constraints (bottom panel). One sees that combining kSZ and $21\,\textrm{cm}$ measurements can significantly improve constraints on the ionization history. However, the correlated nature of the errors mean that the fits can be difficult to evaluate in this space, as illustrated by the blue squares in the bottom panel, where a particularly unlucky noise realization can give results that appear to be egregiously inconsistent.}
    \label{fig:constraint_errs}
\end{figure}

With Equation~\eqref{eq:comboest}, we see that the optimal way to constrain the ionization history is to measure each KL coefficient individually using each of our two probes, and then to average the results together with weights given by each Fisher information matrix. Since the Fisher information is by construction diagonal in this basis for both types of measurement, this is equivalent to an inverse variance weighting. Essentially, we are taking advantage of the complementarity of the global $21\,\textrm{cm}$ signal and the kSZ signal to rely on each probe to deliver the information for each mode when it is available. This is illustrated in Figure~\ref{fig:constraint_errs}, where we show an example reconstructed ionization history with both probes in the bottom panel. Crucially, the top two plots show that the reconstruction fails when we only have one out of our two probes. This establishes that it is fruitful to combine global $21\,\textrm{cm}$ and kSZ measurements directly to constrain the ionization history, and move beyond previous treatments in the literature such as those in Ref. \cite{Monsalve2017reionconstraints}. In those works, constraints were placed on model parameters (such as the midpoint of reionization assuming some parametrized form for the ionization history) rather than on the ionization history itself.

One striking feature of Figure~\ref{fig:constraint_errs} is the trend of smaller error bars in the middle of the redshift range in the $21\,\textrm{cm}$ measurements. In general, this need not be the case, and can be traced back to the specifics of our foreground model from Section~\ref{sec:21cmintro}. A different foreground model may give results that are quantitatively slight different, but we expect the general qualitative complementarity of $21\,\textrm{cm}$ and kSZ to hold.

Although the black points in the bottom panel of Figure~\ref{fig:constraint_errs} visually appear to track the true ionization history quite well, it is important to recall the dangers of a chi-by-eye approach in this space (as opposed to KL space). This is because the errors in the different points are highly correlated. For example, the middle panel does \emph{not} indicate that the $21\,\textrm{cm}$ measurements favor an alternate ionization history at high significance. Instead, it is simply indicative that the $21\,\textrm{cm}$ measurements are unable to measure particular coherent modes (as emphasized by our KL analysis). Similarly, a quick glance at the top panel gives the impression that the kSZ measurements are so large that they provide essentially no information. However, this is again a matter of some crucial missing modes that can be provided by the $21\,\textrm{cm}$ measurements. Finally, we note that while many noise realizations give similar results to the black points shown in the combined constraints, some unlucky draws of noise can occasionally yield ionization histories that visually seem to be egregiously incorrect if one forgets that the errors are correlated. The blue squares in Figure~\ref{fig:constraint_errs} give an example of such a draw, underscoring the need to quote full covariance information when one compares ionization history constraints to theory models. Alternatively, we will see in Section \ref{sec:BayesApproach} that the imposition of suitable priors allows one to leverage the subtle information present here in a faithful reconstruction of the true ionization history.

\subsection{A Bayesian approach}
\label{sec:BayesApproach}
While intuitive to interpret, the least squares approach has a shortcoming in that the incorporation of prior information is more difficult than with a fully Bayesian approach. There are a variety of sensible priors that one might hope to incorporate. Some of these are of the common sense variety, such as requiring that $x_i(z)$ be between $0$ and $1.08$.\footnote{Recall again that our convention for the ionization history includes helium reionization, hence the upper limit at $1.08$ rather than $1$.}
Other priors might be connected to complementary observations, such as that CMB optical depth $\tau$, given by
\begin{equation}
\label{eq:taudef}
\tau = \sigma_T \int \overline{n}_e (z)\, \mathrm{d}l,
\end{equation}
where $\sigma_T$ is the Thomson cross-section, $\overline{n}_e (z)$ is the free electron number density (with the overline denoting a global sky average), and $\mathrm{d}l/\mathrm{d}z$ is the line-of-sight proper distance per unit redshift. Since $\overline{n}_e (z)$ is proportional to the ionization history\footnote{In principle, there is the subtlety that the optical depth is proportional to integral of the ionization history multiplied by the density, rather than just the ionization history alone. Since the ionization field is correlated with the density field, neglecting this can cause a roughly $\sim 10\%$ shift in $\tau$ \cite{Liu2013globalsig}. Here we neglect this effect for simplicity, in light of the fact that our final errors on the ionization history are rather large.}, imposing a prior on $\tau$ from CMB experiments is equivalent to imposing an integral constraint on our recovered ionization histories.
Phrased in the language of a Bayesian analysis, our goal is to constrain the posterior probability distribution $p(\mathbf{x} | \mathbf{y}_{\rm 21}, \mathbf{y}_{\rm kSZ})$ for the ionization history $\mathbf{x}$ given our global $21\,\textrm{cm}$ and kSZ measurements, $\mathbf{y}_{\rm 21}$ and $\mathbf{y}_{\rm kSZ}$. Bayes' theorem states that
\begin{equation}
p(\mathbf{x} | \mathbf{y}_{\rm 21}, \mathbf{y}_{\rm kSZ}) \propto \mathcal{L} ( \mathbf{y}_{\rm 21}, \mathbf{y}_{\rm kSZ} | \mathbf{x}) p(\mathbf{x}),
\end{equation}
where $p(\mathbf{x})$ is the prior and $\mathcal{L} ( \mathbf{y}_{\rm 21}, \mathbf{y}_{\rm kSZ} | \mathbf{x})$ is the likelihood function. The latter can be written as
\begin{eqnarray}
\mathcal{L} ( \mathbf{y}_{\rm 21}, \mathbf{y}_{\rm kSZ} | \mathbf{x}) = \frac{e^{-\frac{1}{2} (\mathbf{y} - \mathbf{A}\mathbf{x}  )^T \mathbf{N}^{-1} (\mathbf{y} - \mathbf{A}\mathbf{x}  )}}{{\rm det}[(2 \pi)^{1/2} \mathbf{N}]},
\end{eqnarray}
assuming Gaussian-distributed measurement errors, whereas the former might take the form
\begin{equation}
\label{eq:tauplus01prior}
p(\mathbf{x}) \propto \exp \left[ \frac{(\tau_{\rm pr} - \mathbf{e}^T \mathbf{x})^2 }{\varepsilon^2_\tau}\right] \prod_{i} \textrm{rect} (x_i) ,
\end{equation}
if we choose to impose a $\tau$ prior as well as a restriction on the possible range of ionization history values. Here, ${\rm rect}(\dots)$ denotes the rectangular function (which is unity between $0$ and $1.08$, and zero otherwise), $\tau_{\rm pr}$ is one's prior value on $\tau$ (assumed to be measured with Gaussian errors to an uncertainty of $\varepsilon_\tau$), and $\mathbf{e}$ is a vector containing all constant factors necessary to convert an ionization history into a $\tau$ value using a discretized version of Equation~\eqref{eq:taudef}.

\begin{figure*}
    \centering
    \includegraphics[width=1.0\textwidth,trim=2.5cm 2.5cm 3.5cm 2.5cm,clip]{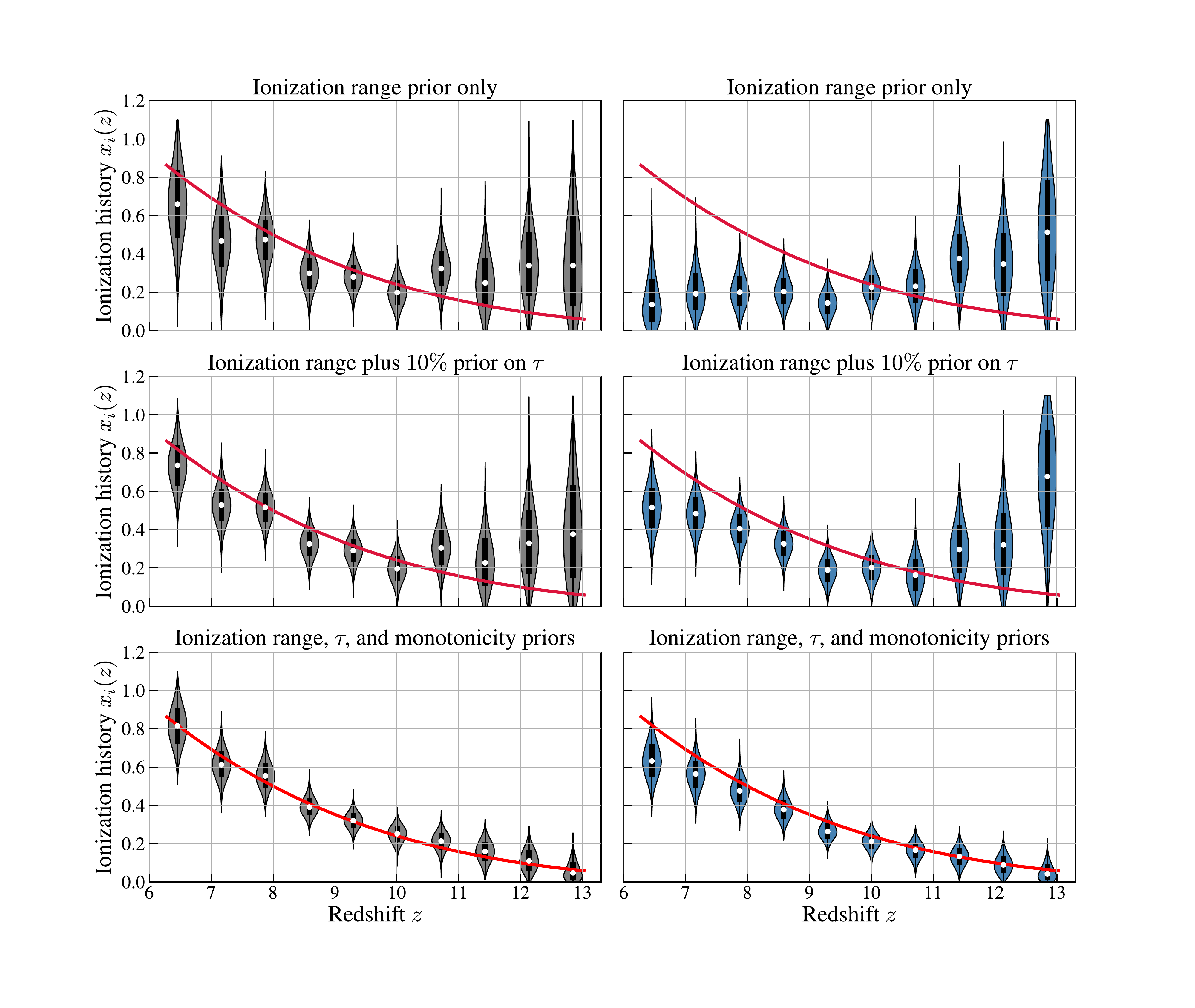}
    \caption{Bayesian reconstructions of the ionization history (red curve) using both kSZ and global $21\,\textrm{cm}$ measurements, imposing a hard prior on the range on possible ionization values (top row), the hard prior plus a $10\%$ prior on $\tau$ (middle row), and both priors plus a monotonicity prior (bottom row). The left column shows the inferred ionization histories from the mock data that was used to generate the black data points in Figure~\ref{fig:constraint_errs}, while the right column is the equivalent for the unlucky noise realization given by the blue squares in Figure~\ref{fig:constraint_errs}. In each plot, the bulges show the marginalized posterior distribution in each redshift bin, the median value is denoted by a white dot, and the thick black bars demarcate $68\%$ credibility regions.}
    \label{fig:bayesianreconstruction}
\end{figure*}

In Figure~\ref{fig:bayesianreconstruction} we show the result of some Bayesian reconstructions of the ionization history. The left column uses the same mock data that was used for the black data points in Figure~\ref{fig:constraint_errs}. The top row places a prior on the range of possible ionization levels. There, we see qualitatively similar results to what we obtained with the least squares approach, except here we have full posterior distribution information and the prior cuts off some of the distributions at low and high ionization levels.  The reconstruction is a reasonable one, which again emphasizes the point that kSZ and $21\,\textrm{cm}$ measurements better work in concert to measure the ionization history. The middle row shows the results of additionally imposing a $10\%$ prior on $\tau$, i.e., using Equation~\eqref{eq:tauplus01prior}. The $\tau$ prior does shrink the errors slightly, as expected, but not to the extent that the constraints are qualitatively better. Admittedly, in neither case is the reconstruction perfect, and one might view some of the features in our reconstruction as undesirable ones. For instance, although models do exist for non-monotonic ionization histories \citep{Wyithe:2002qu,Cen2003,Furlanetto:2004nt}, recent theoretical preferences tend to favor monotonic histories. The bottom row shows the result of adding a prior on monotonicity to the constraint, which produces the best reconstruction yet.

Imposing appropriate priors can be particularly powerful when dealing with unlucky noise realizations. In the right column of Figure~\ref{fig:bayesianreconstruction}, we show the same sequence of Bayesian inferences for the mock data used for the blue squares in Figure~\ref{fig:constraint_errs}. With just the ionization range prior (top row), one again sees a reconstruction that has a much lower ionization level than the truth. The added $\tau$ prior (middle row) improves the situation considerably, lifting overall ionization levels so that the integral of the true ionization history (red curve) is approximately the same as the integral of the reconstructed history. However, with $\tau$ being an overall integral constraint, our inference machinery places some of the extra ionization at unwanted redshifts (i.e., at high redshifts). This is remedied by the addition of our monotonicity prior (bottom row), demonstrating the importance of imposing appropriate (ideally physically motivated) priors.

While in this case the monotonicity prior works quite well, we caution against its use if one is parametrizing the ionization history with a large number of redshift bins. Dividing the redshift axis into a very fine set of bins means that the constraint on each bin is a low signal-to-noise constraint. In this regime, we find from our numerical experiments using twenty bins (rather than the ten shown in Figure~\ref{fig:constraint_errs}) that the monotonicity prior can make it difficult to recover extremely rapid reionization histories where reionization happens much more abruptly than the fiducial histories we have shown. With the measurements being relatively unconstraining when there are many redshift bins, the large volume of accessible parameter space becomes large. Over such a large prior volume, there are simply many more possible ionization histories where ionization happens very gradually over the entire redshift range. In contrast, a model where the ionization remains very low for a long period of time before increasing relatively rapidly is a scenario that is represented by a fairly fine-tuned corner of parameter space that will generally not be explored unless it is strongly demanded by the data.

\begin{figure}
    \centering
    \includegraphics[width=0.5\textwidth,trim=0.0cm 1.5cm 1.cm 1cm,clip]{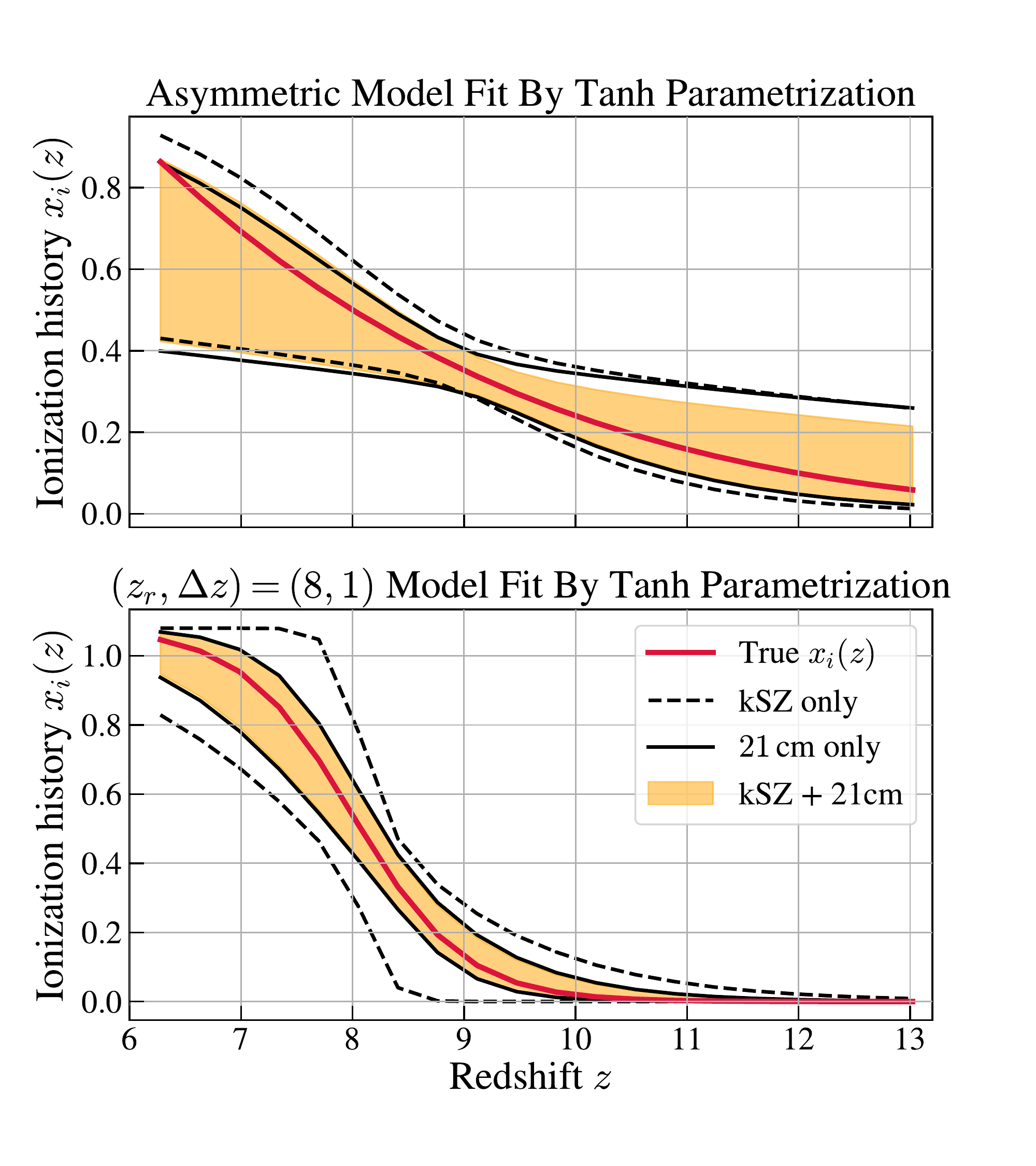}
    \caption{Simulated recoveries of the ionization history (red curve), assuming that the true ionization is given by our fiducial asymmetric model (top plot) and a $\tanh$ model with $z_r = 8$ and $\Delta z = 1$ (bottom plot). The $95\%$ credibility regions coming out of $\tanh$-parametrization fits to mock data are shown for kSZ measurements alone (dashed lines), $21\,\textrm{cm}$ alone (solid lines), and a combined fit (peach region).}
    \label{fig:tanh_param}
\end{figure}

Imposing priors is necessary to include physicality in one's ionization histories when reconstructing it from redshift bins \citep{HuHolder_2003,MilleaBouchet_2018,Planck2018_cosmo_params}.
An alternative method is to give up on a model-independent bin-by-bin reconstruction in favor of a model-dependent parametrization. Examples of this might include the Weibull parametrization suggested in Ref. \cite{Trac2021}, or the popular $\tanh$ parametrization \citep{Lewis2008} where
\begin{equation}
x_i (z) = \frac{1.08}{2} \left[ 1-\tanh \left( \frac{z-z_r}{\Delta z}\right) \right],
\end{equation}
with $z_r$ giving the midpoint of reionization and $\Delta z$ encoding the duration of reionization. Fits using the $\tanh$ form are shown in Figure~\ref{fig:tanh_param}, again using the mock data corresponding to the black points in Figure~\ref{fig:constraint_errs}. Note that we find qualitatively similar behavior with the Weibull parametrization. The top plot assumes that the true ionization history is given by the asymmetric model (i.e., the same fiducial model as we have used for all of our other inference exercises). The dashed lines, solid lines, and peach shaded region demarcate the $95\%$ credibility regions for constraints from kSZ, global $21\,\textrm{cm}$, and the two probes combined, respectively. All assume a $10\%$ prior on $\tau$. Immediately striking is the fact that even the kSZ-only constraints do reasonably well, in contrast to the more model-independent, bin-by-bin inferences. This highlights the fact that using a parametrized form that has just a few parameters is equivalent to making rather strong prior assumptions. Of course, the precise behavior will depend on the true ionization model. Consider the bottom plot of Figure~\ref{fig:tanh_param}, where the true ionization history is given by a $\tanh$ model with $(z_r, \Delta z) = (8,1)$. This is a more rapid reionization scenario than before, and here we see that while kSZ constraints are still powerful, the improvement from introducing $21\,\textrm{cm}$ information is greater. This is in accordance with our intuition from Section~\ref{sec:math}, where we argued that sudden changes in ionization are more easily detected by the $21\,\textrm{cm}$ line and less easily detected by kSZ measurements. 

\section{KL modes as a diagnostic for systematics\label{sec:systematics}}

\begin{figure*}
    \centering 
    \includegraphics[width = \linewidth]{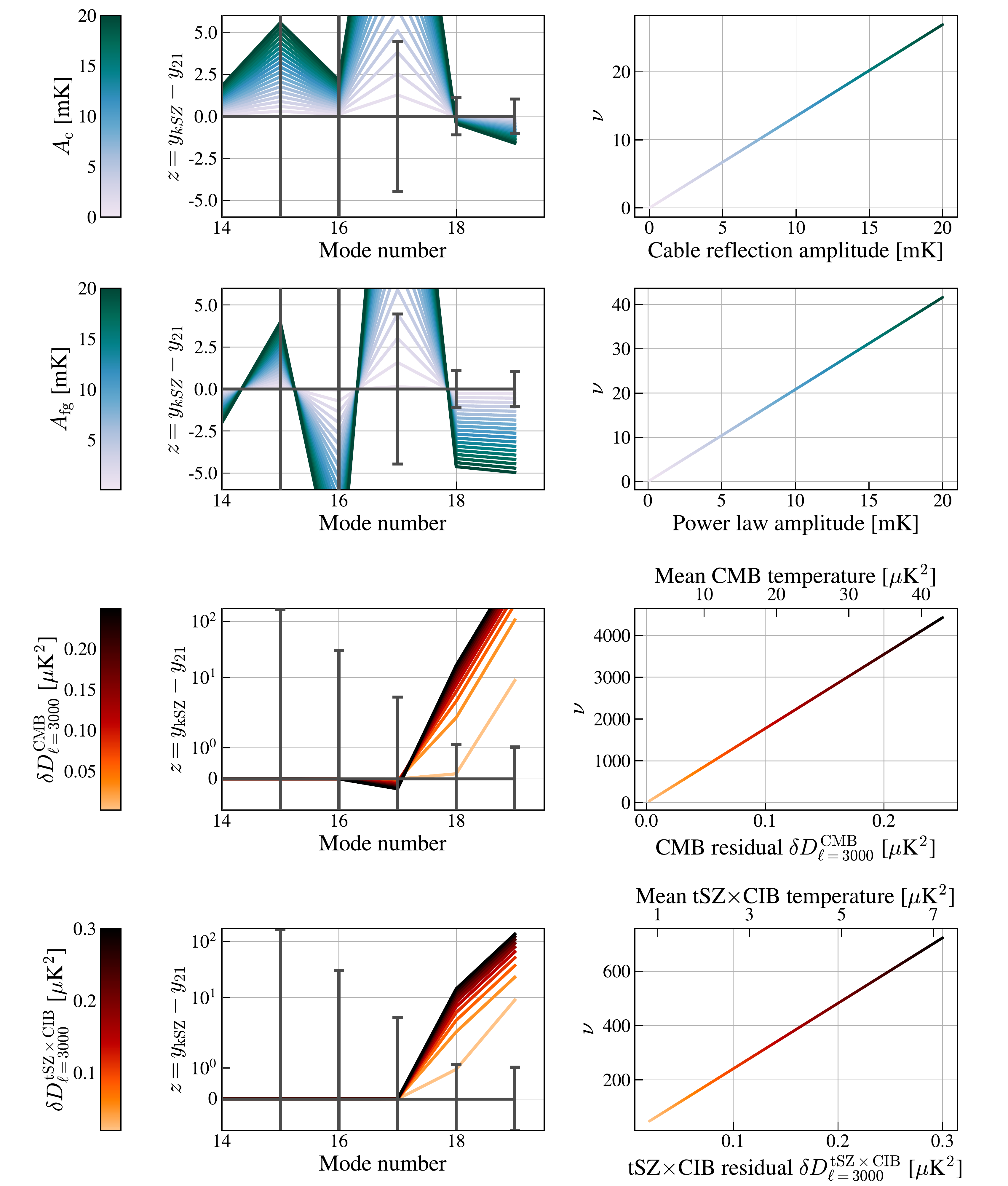}
    \caption{KL modes as a diagnostic for different systematics. From top to bottom: Amplitude of cable reflections and foreground residuals in the global 21\,cm signal, level of primary CMB and tSZxCIB cross-spectrum leakage in the measured patchy kSZ signal. \textit{Left:} Difference between the KL amplitudes obtained with a patchy kSZ and a 21\,cm global measurement. When both probes are free of systematics, there is no difference (in black). \textit{Right:} Statistical significance of a systematic detection using KL modes, given in terms of the number of sigmas $\nu$.}
    \label{fig:null_buster}
\end{figure*}

Casting our data into the KL basis confirms our intuition: there are modes of the ionization history that are best measured by the $21\,\textrm{cm}$ global signal and others by the kSZ. In the previous section, we saw that this complementarity enables improved constraints on the ionization history---provided there are no residual systematics in either dataset.

In this section, we explore techniques for detecting systematics by leveraging modes that are well measured by both probes. Consider, for example, the second-to-last mode in the bottom-right panel of Figure~\ref{fig:errorbars}. This mode has error bars of a similar order of magnitude for both the kSZ and the $21\,\textrm{cm}$ global signal, which means it can be measured to a similar precision using either probe. Such overlap modes can be useful for performing consistency checks between data sets. In a scenario where we have a measurement of the ionization history from both probes and we think that there may be some residual systematics in one of the two, we can decompose our data into the KL basis and compare the overlap modes. If the same KL expansion coefficients $\{ y_\alpha \}$ are not consistent between the kSZ and $21\,\textrm{cm}$ measurements at the level of expected uncertainties, then we are forced to conclude that there exist residual systematics in at least one of them.

In order to do this, we must first understand how the presence of systematics change the KL coefficients of each probe and whether we can distinguish residual systematics in the data from perturbations due to noise. We first do this qualitatively with a ``chi-by-eye'' test in Section~\ref{sec:chieye}, and then formalize the procedure with a linear matched filter in Section~\ref{sec:linearfilter}.

\subsection{Chi-by-eye \label{sec:chieye}}

Let us first introduce some notation. We define $\mathbf{x}_0$ to be the true ionization history, with a corresponding set of KL amplitudes $\mathbf{y}_0$ given by Equation~\eqref{eq:transformation}, i.e., 
\begin{equation}
    \mathbf{y}_0 = \mathbf{R}^{-1} \mathbf{x}_0.  
\end{equation}
Throughout this section, we take $\mathbf{x}_0$ to be the ``asymmetric" ionization history shown in the top-left panel of Figure~\ref{fig:errorbars} in orange, although the analysis framework that we develop does not rely on a specific model. A measurement of the KL modes using our two probes can be written as
\begin{align}
    &\mathbf{y}_{21} = \mathbf{y}_0 + \mathbf{n}_{21} + \mathbf{y}_{21,\text{sys}} \\
    &\mathbf{y}_{\rm kSZ} = \mathbf{y}_0 + \mathbf{n}_{\rm kSZ} + \mathbf{y}_{\rm kSZ,\text{sys}},
\end{align}
where $\mathbf{n}_{21}$ and $\mathbf{n}_{\rm kSZ}$ are instrumental noise contributions for the global $21\,\textrm{cm}$ signal and kSZ measurements, respectively, while $\mathbf{y}_{21,\text{sys}}$ and $\mathbf{y}_{\rm kSZ,\text{sys}}$ are corresponding systematic contributions. To hunt for systematics, one can imagine differencing $\mathbf{y}_{21}$ and $\mathbf{y}_{\rm kSZ}$ to form
\begin{eqnarray}
    \mathbf{z} &\equiv& \mathbf{y}_{\text{kSZ}} - \mathbf{y}_{21} \label{eq:z}\\
    &=& \mathbf{y}_{\rm kSZ,\text{sys}} - \mathbf{y}_{21,\text{sys}} + \mathbf{n}_z,
\end{eqnarray}
where $\mathbf{n}_z$ is a noise contribution to $\mathbf{z}$ with covariance given by
\begin{equation}
    \overline{\boldsymbol{\Sigma}} \equiv \overline{\mathbf{C}}_{\rm kSZ} + \overline{\mathbf{C}}_{21},
\end{equation}
where $\overline{\mathbf{C}}_{\rm kSZ} = \overline{\mathbf{F}}_{\rm kSZ}^{-1}$ and $\overline{\mathbf{C}}_{21} = \overline{\mathbf{F}}_{\rm 21}^{-1}$ are the noise covariances for the kSZ and $21\,\textrm{cm}$ measurements in the KL basis. Looking for the presence of systematics is then equivalent to asking whether $\mathbf{z}$ is consistent with noise that has a covariance matrix given by $\overline{\boldsymbol  \Sigma}$, since a measurement free of systematics would have $\mathbf{z} = \mathbf{n}_z$.

In principle, one could perform the same analysis with recovered ionization histories rather than KL amplitudes. However, working in KL space has several advantages. First, it highlights which modes of the ionization history are suitable for this type of consistency analysis. From the bottom right panel of Figure~\ref{fig:errorbars} we see that the vast majority modes are much better measured by the global $21\,\textrm{cm}$ signal. We thus effectively only have one measurement of these, disallowing the possibility of consistency checks. Only the overlap modes are useful in this regard. The second advantage of working in KL space is that $\overline{\boldsymbol{\Sigma}}$ is diagonal in this space, since Equations~\eqref{eq:cov_21_kl} and \eqref{eq:cov_ksz_kl} demonstrate that both $\overline{\mathbf{F}}_{\rm kSZ}$ and $\overline{\mathbf{F}}_{\rm 21}$ are diagonal. Each KL mode thus allows an independent consistency check that can be easily visualized.

As an illustration, we first consider cable reflections due to impedance mismatches in transmission lines, which are a common systematic present in 21cm datasets. A cable reflection might imprint a copy of an original signal at some time delay $\tau$, which in turn results in a sinusoidal perturbation in the measured spectrum of the form
\begin{equation}
T (\nu) = A_{\rm c} \sin (\nu \tau + \phi).
\end{equation}
Dividing by the appropriate prefactors from Equation~\eqref{eq:globalsignal_reduced} then converts this into a perturbation on the ionization history, which can in turn be cast in the KL basis using Equation~\eqref{eq:transformation}. 

In the top left panel of Figure~\ref{fig:null_buster} we plot in color $\mathbf{z}$ as defined in Equation~\eqref{eq:z}, where $\mathbf{y}_{21,\rm sys}$ are the KL amplitudes for a cable reflection with different values of $A_c$. In black, we plot $\mathbf{z}$ when both measurements are free of systematics, identically equal to zero, with the error bars given by $\overline{\boldsymbol{\Sigma}}$. We plot only the last few modes, since these correspond to modes that are well measured by both probes. If perturbations to $\mathbf{z}$ due to the presence of systematics fall within the error bars for a given mode, then this devation is consistent with a noise fluctuation. This is the case for example for the second to last mode plotted. In this scenario, we could not use this mode to perform a consistency check between the two datasets. In contrast, the last and third-to-last modes do exhibit deviations that fall outside of the error bars for large enough cable reflection amplitudes, and we would conclude that one measurement contains residual systematics. 

Another example of a common systematic in a $21\,\textrm{cm}$ measurement is foreground contamination. Although measures can be taken to model and remove contaminants \cite{Liu2013globalsig, Bernardi2015, LiuShaw2020, 2018ApJ...853..187T, 2019ApJ...883..126N, 2020ApJ...897..132T, 2020ApJ...897..174R, 2020ApJ...897..175T, 2020ApJ...905..113H, 2021ApJ...908..189B, 2021ApJ...915...66T, 2021MNRAS.502.4405B, 2021MNRAS.tmp.2933A, 2021MNRAS.506.2041A, 2016MNRAS.461.2847B, 2015MNRAS.449L..21H, 2014ApJ...793..102S, 2012MNRAS.419.1070H}, it is likely that some residuals will remain. Consider a scenario where the residual foreground spectrum $T_{\rm fg}$ takes the form of a power law 
\begin{equation}
T_{\rm fg} (\nu) = A_{\rm fg} \left( \frac{\nu}{\nu_*} \right)^\alpha,
\end{equation}
where $A_{\rm fg}$ is an amplitude parameter for the residuals, $\alpha$ is a power law index, and $\nu_{\star}$ is a frequency pivot scale. As we did for the cable reflections, we divide out the relevant conversion factors to cast this into an ionization history, which we then transform into KL space for adding to our fiducial measurement. The results are in the second panel from the top in Figure~\ref{fig:null_buster}, where we keep $\alpha$ and $\nu_{\star}$ fixed at $150\,\mathrm{MHz}$, but vary $A_{\rm fg}$ to explore the sensitivity of our tests. For both the $21\, \textrm{cm}$ systematics we consider residual amplitudes up to $20\,\mathrm{mK}$ since this is comparable in magnitude to the expected differential brightness temperature \citep{Pritchard2010}. We find that the inclusion of either of these systematics perturbs the KL amplitudes for some of the last few modes beyond what might be due to noise fluctuations.

Since our tests rely on self-consistency, they can equally well be used to detect systematics in a kSZ measurement. Consider, for example, the possibility that, because of an imperfect cosmological model, the kSZ measurement is contaminated by power from the primary CMB anisotropies. The precise way in which such a systematic would affect the ionization history is less straightforward than in the $21\,\textrm{cm}$ case because the mapping from the ionization history to the observable (the kSZ contribution to the angular power spectrum) is more complicated. Following Refs. \cite{Bernardi2015,Presley2015}, we can obtain an expression for how some residuals $\delta \mathbf{D}$ in the power spectrum affect the ionization history by computing
\begin{equation}
    (\delta \mathbf{x}_{\rm kSZ})_{\alpha} = \sum_{\beta} (\mathbf{F}^{-1}_{\rm kSZ})_{\alpha \beta} \frac{\partial \mathbf{D}^T}{\partial x_{i}(z_{\beta})} \boldsymbol{\Pi}^{-1}_{\rm kSZ} \delta  \mathbf{D},\label{eq:sys_ksz}
\end{equation}
where $\delta \mathbf{x}_{\rm kSZ}$ is the systematic-induced perturbation to the kSZ-derived ionization history, $\mathbf{D}$ is the fiducial patchy kSZ angular power spectrum, $\boldsymbol{\Pi}_{\rm kSZ}$ is the kSZ measurement covariance as defined in Section~\ref{sec:fisher}, and $\mathbf{F}_{\rm kSZ}$ is the kSZ Fisher matrix from Equation~\eqref{eq:ksz_fisher}. Since the Fisher matrices are diagonal, it can be computationally convenient to compute the perturbation directly in KL space. This has the added bonus that it avoids having to invert $\mathbf{F}_{\rm kSZ}$, which can be singular depending on how finely the redshift axis is binned\footnote{Notice from Equation~\eqref{eq:cov_ksz_kl} that $\overline{\mathbf{F}}_{\rm kSZ}$ can be formed without inverting $\mathbf{F}_{\rm kSZ}$. Because $\overline{\mathbf{F}}_{\rm kSZ}$ is diagonal, singular modes can simply be discarded without affecting the numerical stability of other modes.}. The equivalent expression to Equation~\eqref{eq:sys_ksz} in KL space is
\begin{equation}
    (\delta \mathbf{y}_{\rm kSZ})_{\alpha} = \sum_{\gamma} (\overline{\mathbf{F}}^{-1}_{\rm kSZ}\mathbf{R}^T)_{\alpha \gamma} \frac{\partial \mathbf{D}^T}{\partial x_i(z_{\beta})} \boldsymbol{\Pi}^{-1}_{\rm kSZ} \delta  \mathbf{D}.
\end{equation}

To simulate the CMB primary contaminating our kSZ measurement, we take $\delta \mathbf{D}$ to be a scaled primary CMB power spectrum. We allow the residual primary CMB temperature at $\ell = 3000$, $\delta D_{\ell = 3000}^{\rm CMB}$, to range up to $0.3\,\mu$K$^2$ and find that even for small primary CMB residuals, $\mathbf{z}$ is perturbed well outside the error bars for the overlap modes. This is unsurprising due to the large dynamic range of the CMB power spectrum over the range of $\ell$ that we are considering. Although a CMB temperature of $1\,\mu$K$^2$ at $\ell = 3000$ is of the same order as the kSZ signal at this $\ell$, the CMB can be up to two orders of magnitude brighter on the lower end of our $\ell$ range. 

Another potential systematic corresponds to the tSZ$\times$CIB cross-spectrum contaminating our measured kSZ spectrum. There is a long list of other high-multipole foregrounds to the primary CMB which could be considered, including the second, late-time component of the kSZ power itself. However, we limit our analysis to the tSZ$\times$CIB power, since it has similar amplitude and shape as the kSZ power and other potential contaminants, such as the thermal SZ spectrum. Using the same method as for the contamination from the CMB primary power, we see the same behavior for this systematic. Small residual amplitudes cause large changes in the last few KL modes, implying that even small residual tSZ$\times$CIB systematics will lead to discrepancies between the two datasets that can be immediately spotted.

\subsection{The Linear Matched Filter \label{sec:linearfilter}}

With the chi-by-eye test we can qualitatively predict whether it is appropriate to use the overlap modes to perform consistency checks. One limitation of this approach is that although we can assert the presence of residual systematics in one of our datasets, we cannot say what type of systematic might be causing the discrepancy. As a next step, we can use a linear matched filter (LMF) to search for a specific systematic within our dataset, as well as formalize the chi-by-eye results and quantify the statistical significance with which we can detect systematics. 

To employ the LMF, one supplies the filter with a template shape that is expected to be present in the data. Matched filtering has been used for example in gravitational wave astronomy to determine whether a characteristic gravitational wave signature is present in a signal \cite{Ligo_2016}. In our case, we define our LMF template $\mathbf{s}$ to be a vector containing the KL coefficients of some systematic that we suspect is contaminating our data. Then, the ``number of sigmas" $\nu$ by which the LMF is able to detect the template is given by
\begin{equation}
    \nu \equiv \frac{\vert \mathbf{s}^T \boldsymbol{\Sigma}^{-1}\mathbf{z} \vert}{\sqrt{\mathbf{s}^T \boldsymbol{\Sigma}^{-1}\mathbf{s}}}.
\end{equation}
Before proceeding with the results, we note a key property of this statistic: rescaling the amplitude of $\mathbf{s}$ leaves $\nu$ unchanged. This is particularly convenient as one often has reasonable priors on the shapes of various systematics without knowing the precise amplitude of the residuals\footnote{Indeed, if we knew both the shape and amplitude of a systematic precisely, we would simply subtract it out!}.

In the right column of Figure~\ref{fig:null_buster} we plot $\nu$ for the same four representative systematics that we studied in Section~\ref{sec:chieye}. The LMF can detect the presence of CMB and tSZ$\times$CIB residuals remarkably well, which agrees with the chi-by-eye result in the left column. We see that for the residual CMB and tSZxCIB amplitudes considered, $\mathbf{z}$ is perturbed well outside of the error bars. For residual spectra temperatures of $\sim 0.3 \mu\textrm{K}^2$ at $\ell = 3000$, the LMF can detect CMB residuals with a significance of $4000\sigma$, and $600\sigma$ for the tSZ$\times$CIB. These high significance detections are, as we discussed previously, in part due to the large variations in amplitude of the CMB and tSZxCIB spectra over the range of $\ell$ we consider. To illustrate this, in the bottom two LMF panels we include a secondary axis displaying the average of the spectra over the whole range of $\ell$. Indeed, although we have $\delta {D}^{\rm CMB}_{\ell = 3000} = 0.3 \mu\textrm{K}^2$, the average of the CMB spectrum is 40 $\mu\textrm{K}^2$.

\begin{figure}
    \centering
    \includegraphics[width=\linewidth,trim=0.0cm 0cm 0.cm 5cm,clip]{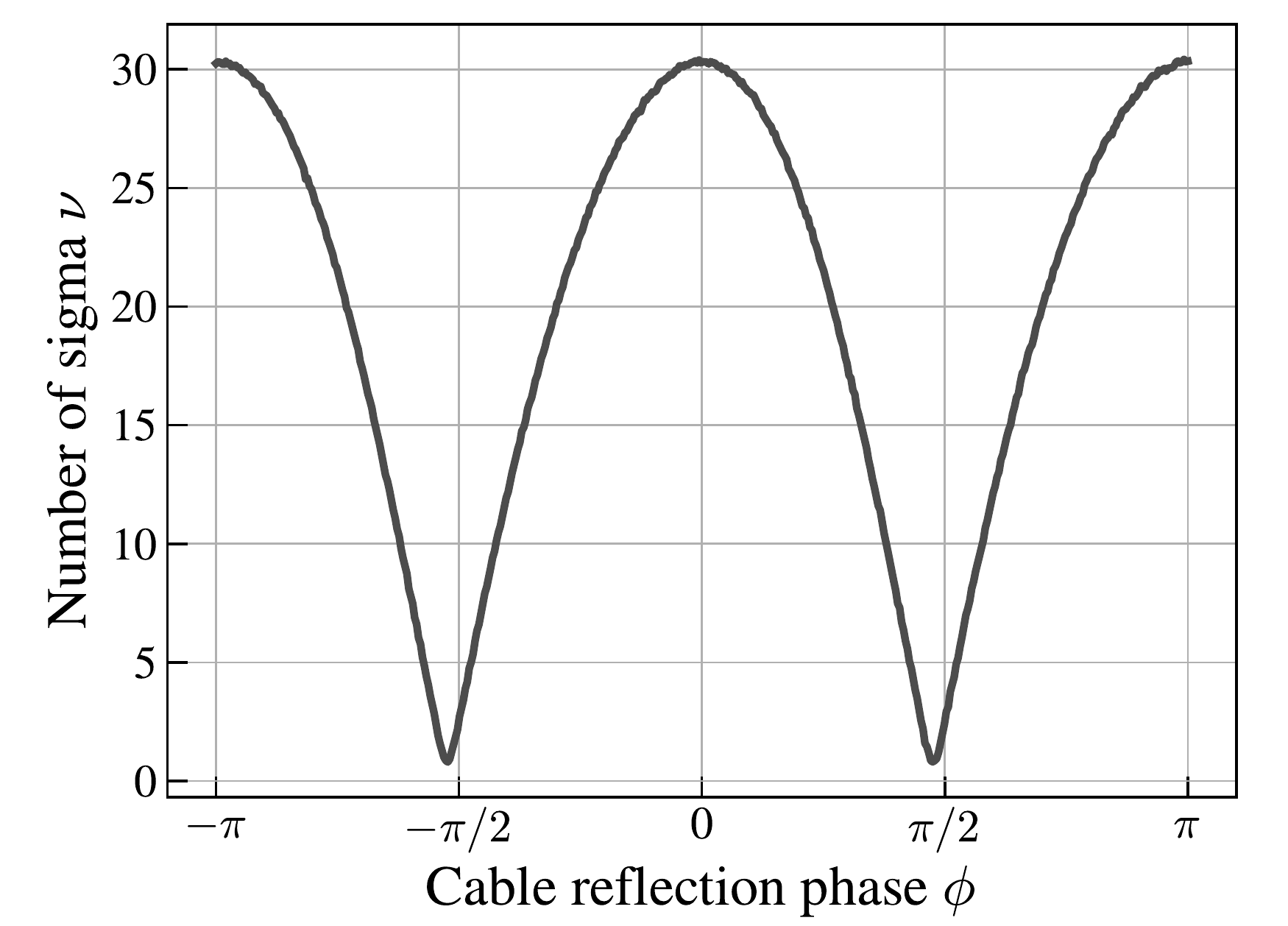}
    \caption{The number of sigma with which the LMF statistic can detect cable reflection systematics, where our systematic template differs from the true cable reflections by a phase. As expected, $\nu$ peaks when the template matches the true phase of the simulated systematic ($\phi = 0$) and falls off as the phases differ. For very large phase differences, the statistical significance rises again because $\nu$ is insensitive to the overall sign of our templates.}
    \label{fig:phase}
\end{figure}

In general, we find that $21\, \textrm{cm}$ systematics are more challenging to detect. For residual amplitudes of $20\,\textrm{mK}$, the LMF can detect cable reflections and foregrounds to $30\sigma$ and $40\sigma$, respectively. We emphasize that these detections are not the generic detection of outliers, but that of a specific template. As an illustration of this, consider the scenario where our LMF template for cable reflections differs from the true systematic by a phase. This is in fact a realistic situation, since in general one does not know the phase of one's reflections \emph{a priori}. Figure~\ref{fig:phase} shows the number of sigma with which a simulated cable reflection with $\phi = 0$ is detected for various phase choices in one's LMF template. As expected, $\nu$ peaks when the template's phase matches that of the true systematic, and drops on either side until one goes beyond $\pm \pi / 2$. At that point $\nu$ increases again because our statistic is insensitive to the overall sign of our template.

\section{What to do if there are signs of systematics in a probe\label{sec:systematics_remain}}

Ideally, upon identification of a systematic effect (for instance, using the techniques described in Section~\ref{sec:systematics}), one ought to track down their physical origin and simply eliminate them from future measurements. Failing that, we might choose to combine our datasets in a way that minimizes the influence of systematics. In this section, we introduce two methods for doing so: mode projection and automatic outlier detection.

\subsection{Mode projection}
\label{sec:modeprojection}
Suppose we suspect that our concatenated data vector $\mathbf{y}_{\rm conc}$ contains some residual systematic effect of the form $\textbf{s}$. A straightforward way to eliminate this mode from our data would be to assign infinite error to anything with the shape of $\textbf{s}$ in the noise covariance matrix. To do this, one can make the substitution
\begin{equation}
    \mathbf{N} \rightarrow \mathbf{N} + \epsilon \mathbf{s} \mathbf{s}^T .
\end{equation}
where $\epsilon$ is a free parameter that we will eventually send to infinity to signify that $\textbf{s}$ is not a mode to be trusted. If we wish to do so in the context of our least-squares estimator of Section~\ref{sec:leastsq}, the key quantity is $\mathbf{N}^{-1}$, since this is what enters the expression for $\widehat{\mathbf{x}}$. As $\epsilon \rightarrow \infty$, an application of the Woodbury identity reveals that $\mathbf{N}^{-1}$ is replaced by $ \mathbf{N}^{-1} \mathbf{P}$, where we have defined the projection matrix 
\begin{equation}
\mathbf{P} = \mathbf{I} - \frac{\mathbf{s}\mathbf{s}^T\mathbf{N}^{-1}}{\mathbf{s}^T\mathbf{N}^{-1}\mathbf{s}}.
\end{equation}
Replacing all copies of $\mathbf{N}^{-1}$ with $ \mathbf{N}^{-1} \mathbf{P}$ in Equation~\eqref{eq:comboest} then gives
\begin{equation}
    \widehat{\mathbf{x}} \rightarrow (\mathbf{A}^T\mathbf{N}^{-1}\mathbf{P}\mathbf{A})^{-1}\mathbf{A}^T \mathbf{N}^{-1} \mathbf{P}\mathbf{y}_{\rm conc} \label{eq:estimator_proj},
\end{equation}
with a correspondingly modified covariance for the estimator taking the form
\begin{equation}
  (\mathbf{A}^T\mathbf{N}^{-1}\mathbf{P}\mathbf{A})^{-1}\mathbf{A}^T\mathbf{N}^{-1}\mathbf{P}\mathbf{N}\mathbf{P}^{T}\mathbf{N}^{-1}\mathbf{A}(\mathbf{A}^T \mathbf{P}\mathbf{N}^{-1}\mathbf{A})^{-1}. \label{eq:cov_project}
\end{equation}

\begin{figure}
    \centering
    \includegraphics[width = \linewidth]{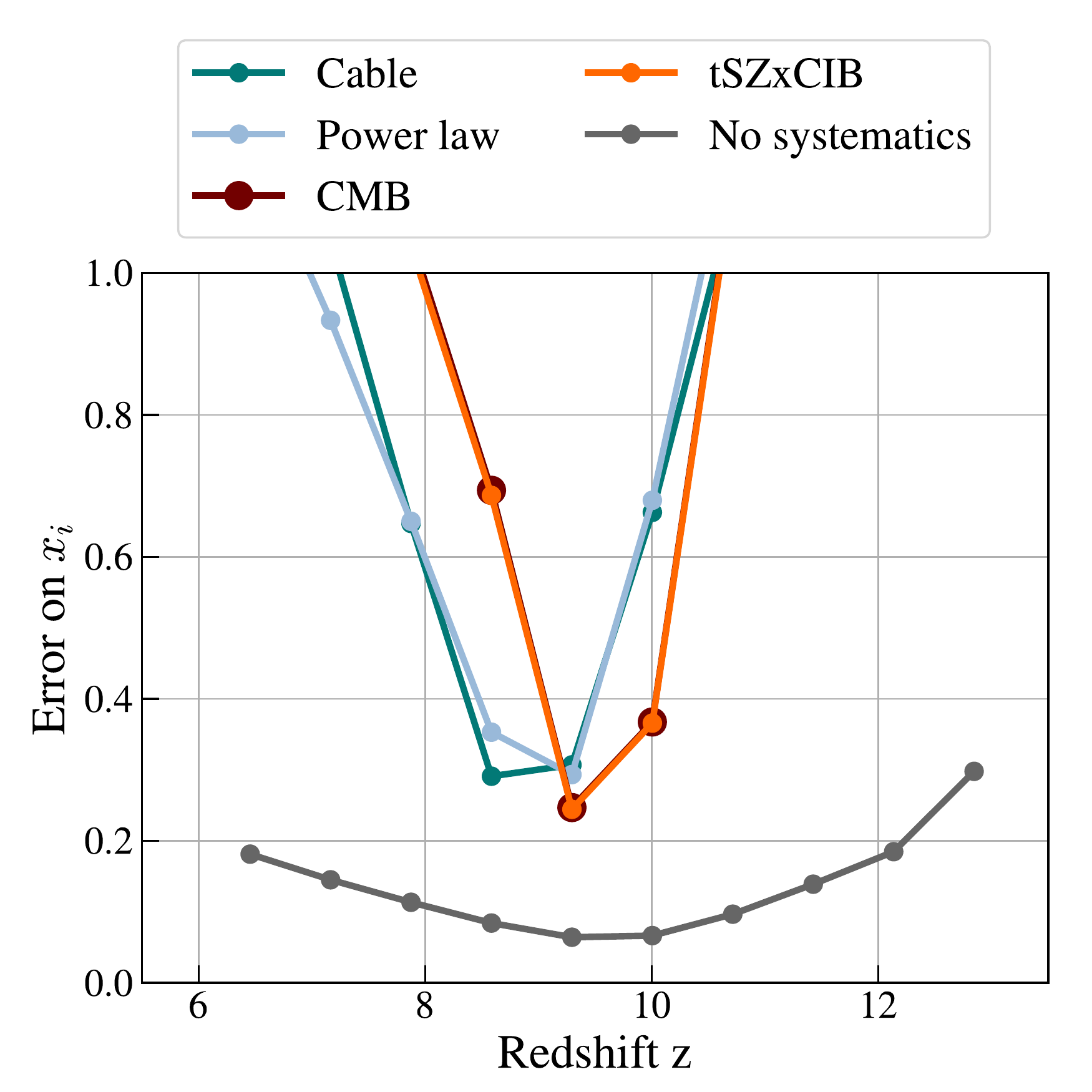}
    \caption{The errors on each redshift bin after projecting out systematics, computed with the covariance given in Equation~\ref{eq:cov_project}. When projecting out modes corresponding to any of our representative systematics, the errors on $x_i$ for the majority of the redshift bins we consider are increased significantly, so that most bins cannot be used to place constaints on the ionization history.}
    \label{fig:projecting}
\end{figure}

Figure~\ref{fig:projecting} illustrates the change in the error of our recovered ionization history after projecting out each of the systematics considered in Section~\ref{sec:systematics}. Regardless of which systematic is projected out, we see that the errors increase compared to the scenario where there are no systematic, in which case the errors are given by the inverse of the sum of the Fisher matrices as in Equation~\eqref{eq:cov_tot}. We limit the range of our plot to be between 0 and 1, since errors in a measurement of $x_i$ that are larger than order unity effectively give us no information about reionization. Projecting out any of the systematics we consider increases the errors on $x_i$ significantly, rendering most of the redshift bins we consider unmeasurable. This is to be expected, since the projecting out of a systematic mode removes information from the data. Ultimately, the complete projecting out of a mode is a rather drastic and heavy-handed way to remove a systematic, and one pays a steep price in the quality of one's ionization history reconstruction.

\subsection{Automatic outlier detection}

The projecting out of a suspicious mode is in some ways a rather arbitrary exercise that is appropriate under a fairly narrow set of circumstances. On one hand, one must not have precise knowledge of the exact form and amplitude of the systematic (otherwise, it would be simpler just to subtract out the systematic). On the other hand, one must have at least some vague knowledge of the shapes of the systematics in the data, and to have good reasons to suspect that the relevant systematics are present. Often this suspicion simply comes from seeing large amplitude signals in one's data, and thus the projecting out of systematics is a modified version of sigma clipping. This practice can be one of concern, as the data analyst is essentially placing artificial constraints on the probability distributions that the data are expected to follow, whether or not the real data actually obey these constraints\footnote{To be fair, the methods proposed in Section~\ref{sec:modeprojection} do have some safeguards against this. In particular, we note that we are discarding the relevant systematic modes in only one out of our two probes, and the projection only occurs for specific shapes. Our methods therefore do not constitute a literal sigma clipping process, but it is still incumbent on a data analyst to be cognizant of possible biases.}.

An alternative to mode projection is to use better, more expressive models for the likelihood that are able to account for outliers due to systematics. Different KL modes are statistically independent (i.e., $\mathbf{N}$ is diagonal), so our likelihood $\mathcal{L}$ factorizes into a product of constituent likelihood functions for the two different probes and the different KL modes, such that $\mathcal{L} = \prod_i \mathcal{L}^{\rm 21}_i \mathcal{L}^{\rm kSZ}_i$. If our proposed model has the value $y^{\rm model}_i$ for the $i$th mode, then our previous likelihood for this mode is given by
\begin{equation}
\mathcal{L}^{\rm in}_i = \frac{1}{\sqrt{2 \pi \sigma_i^2}} \exp \left[-\frac{(y_i - y_i^{\rm model})^2}{2 \sigma_i^2}\right],
\end{equation}
where $\sigma_i^2$ is the $i$th entry on the diagonal of $\mathbf{N}$. This likelihood applies equally well to a $21\,\textrm{cm}$ or a kSZ measurements, provided neither is an outlier. In contrast, saying that a particular measurement is an outlier is equivalent to saying that it is drawn from a different---much broader---distribution instead, such as
\begin{equation}
\mathcal{L}^{\rm out}_i = \frac{1}{\sqrt{2 \pi \gamma^2 \sigma_i^2}} \exp \left[-\frac{(y_i - y_i^{\rm model})^2}{2 \gamma^2 \sigma_i^2}\right],
\end{equation}
where $\gamma > 1$ is an error multiplier that makes this Gaussian broader than the previous one. Of course, in practice we will not know ahead of time whether a data point is an outlier or not. To get around this, one can introduce a nuisance parameter $g_i$ that enters into a more general likelihood $\mathcal{L}_i$ that is a mixture of the standard likelihood and the outlier likelihood, such that
\begin{equation}
\label{sec:mixturemod}
\mathcal{L}_i = g_i \mathcal{L}^{\rm in}_i + (1-g_i) \mathcal{L}^{\rm out}_i,
\end{equation}
which follows the suggestion in Ref.~\cite{Hogg2010fitting}. With such a likelihood, outliers can be identified in a data-driven way: the extra $g_i$ parameters (one for each mode and one for each probe) are parameters that can be fit from the data, under the restriction that they must lie between $0$ and $1$. If the posterior for a particular $g_i$ tends towards $1$, it is not an outlier; if it tends towards $0$, it is likely an outlier. Final constraints on the ionization history can then be obtained by marginalizing over $\{ g_i \}$ (and also $\gamma$, since we do not know \emph{a priori} how much of an outlier our outliers are). This will have the effect automatically discarding outliers in a statistically disciplined fashion.

With the large number of extra parameters, our inference problem is not well-defined unless we make further assumptions. To see this, suppose we were to parametrize the problem by specifying the ionization history redshift bin by redshift bin, as we did in Section~\ref{sec:leastsq}. Constraining the ionization history in $n$ redshift bins requires $n$ KL coefficients to be measured. Here, each KL mode independent and is measured by two probes. However, with just two measurements of each mode, outlier detection is impossible. Two measurements enables one to detect inconsistency, as we did in Section~\ref{sec:systematics}. However, with just two measurements per mode, there is simply not enough information to determine which measurement is the outlier that is causing the inconsistency, and which is the accurate measurement.

To make our inference of the ionization history a well-defined problem, it is necessary to make some assumptions. Here we recommend two. First, we can reduce the number of extra nuisance parameter that are introduced into the problem by only performing outlier detection for the overlap modes. Essentially, one is acknowledging that for a mode that is well-measured by only one probe, we have no choice but to accept a measurement even if it is an outlier, since we have no other way to obtain information about the mode. Second, rather than having the parameters in our inference problem be the bin-by-bin ionization history values, we might use a parametric form such as a $\tanh$ model. This effectively allows one to take advantage of a form of self-consistency \emph{between} different KL modes for automatic outlier detection---one cannot arbitrarily adjust a single KL mode without deviating from the assumed parametric form. In principle, tying together different KL modes in this way allows one to perform outlier detection over more modes than just the overlap modes, but in practice we recommend enacting both of the aforementioned measures for best results.

\begin{figure}
    \centering
    \includegraphics[width=0.5\textwidth,trim=0.0cm 0cm 1.cm 1cm,clip]{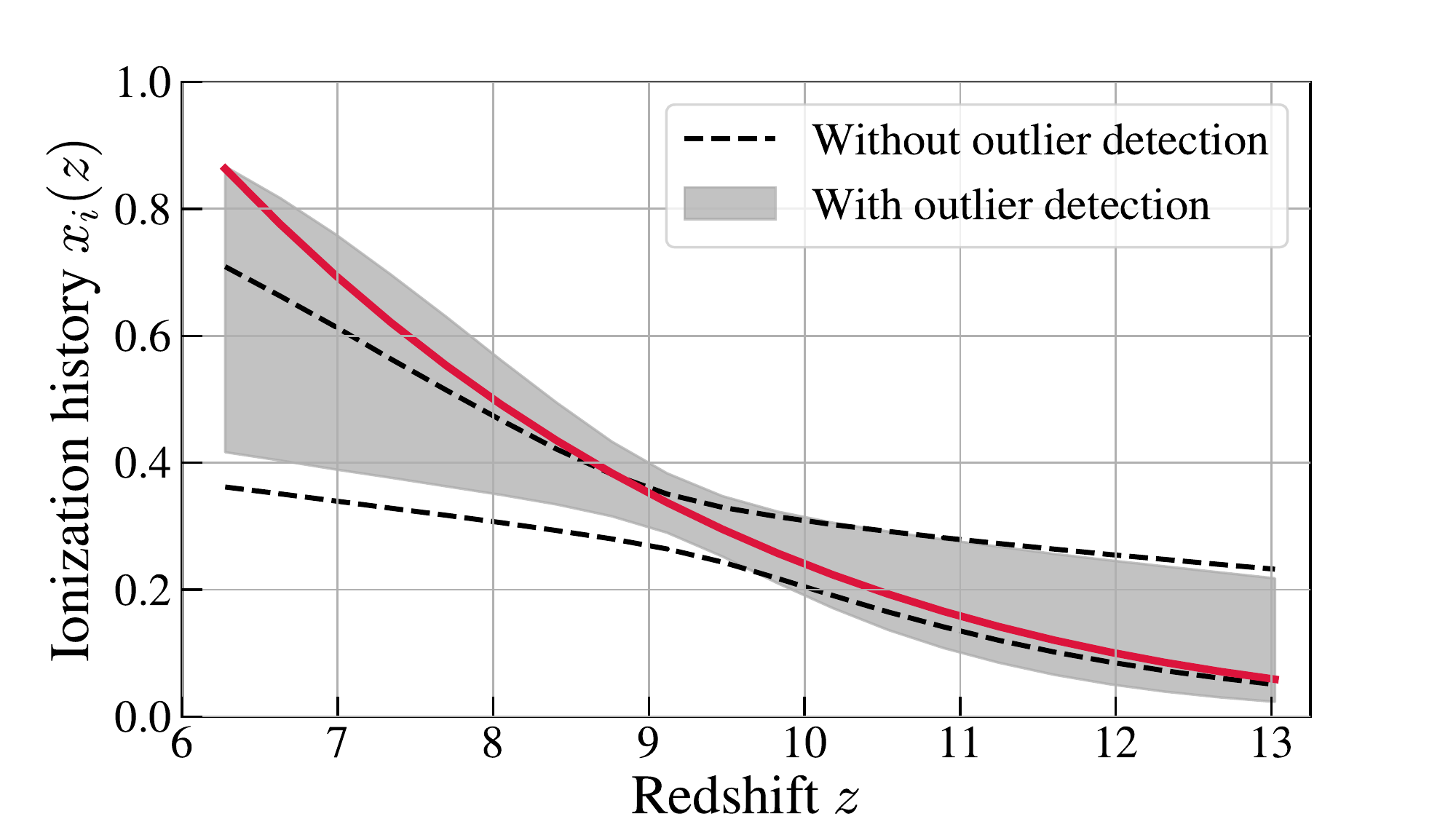}
    \caption{Inference of the fiducial asymmetric ionization history by fitting to a $\tanh$ parametrization assuming no outliers (dashed lines) or using automatic outlier detection (grey band), using mock data that has a $20\sigma$ outlier in the $19$th KL mode of the $21\,\textrm{cm}$ measurement. In each case the $95\%$ credibility interval is shown.}
    \label{fig:outlier}
\end{figure}

\begin{figure*}
    \centering
    \includegraphics[width=1.0\textwidth,trim=2.5cm 0.25cm 2.5cm 1cm,clip]{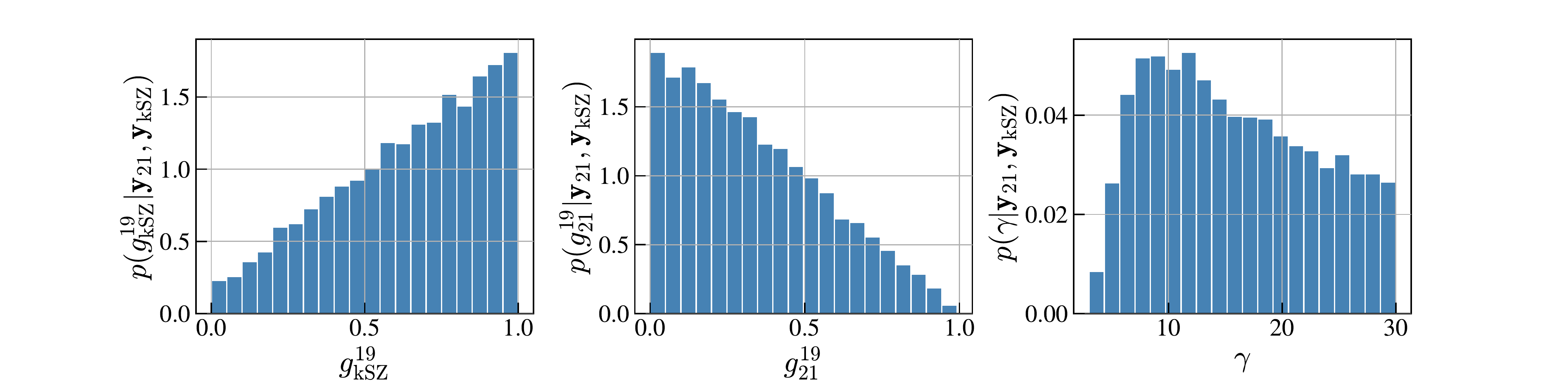}
    \caption{Marginalized posterior distributions for nuisance parameters in the fitting of a $\tanh$ ionization history model using a likelihood function that incorporates automatic outlier detection, i.e., Equation~\eqref{sec:mixturemod}. \emph{Left}: Distribution for $g_{\rm kSZ}^{19}$, the nuisance parameter that indicates whether the kSZ measurement of the $19$th KL mode is an outlier or not. \emph{Middle:} Distribution for $g_{\rm 21}^{19}$, the equivalent quantity on the $21\,\textrm{cm}$. \emph{Right:} Distribution for $\gamma$, the error multiplication factor controlling the width of a hypothetical distribution from which outliers are drawn. The automatic outlier detection successfully identifies the $21\,\textrm{cm}$ measurement as being the outlier in our simulation, with $\gamma > 1$ to indicate that it is in fact more aptly described by a wider distribution than one consistent with theoretically expected errors.}
    \label{fig:histogram}
\end{figure*}

In Figure~\ref{fig:outlier} we show the results of fitting a $\tanh$ model to data that has a $20\sigma$ outlier in the $19$th KL mode of the $21\,\textrm{cm}$ measurement. The true ionization history is shown in red, and two $95\%$ credibility regions are shown: the grey bands show the recovered constraint using the automatic outlier detection likelihood given by Equation~\eqref{sec:mixturemod} and the dashed lines show the results of analyzing data with the incorrect assumption that there are no outliers. Although the constraints of the $\tanh$ parametrization mean that neither gives appallingly inaccurate results, the latter clearly gives a biased result. In contrast, the former is able to identify---and effectively discard---the outlier in the $21\,\textrm{cm}$ measurement of the KL mode in question and to rely instead on the kSZ data point. This can be seen in Figure~\ref{fig:histogram}, where one sees that the marginalized posteriors in the nuisance parameters of our fit. The parameter  $g_{\rm 21}^{19}$ is skewed towards $0$, thus correctly identifying the $19$th KL coefficient derived from the $21\,\textrm{cm}$ data as coming from an outlier distribution with $\gamma > 1$. In contrast, $g_{\rm kSZ}^{19}$ is skewed towards $1$, suggesting that the corresponding data point from the kSZ side is not an outlier. Ultimately, the best course of action in any measurement process is of course to identify and eliminate experimental systematics. However, as another demonstration of the complementarity between $21\,\textrm{cm}$ and kSZ measurements, automatic outlier detection can be employed in the overlap modes as an additional safeguard.

\section{Conclusions \label{sec:conclusions}}

The global $21\,\textrm{cm}$ signal and the kinetic Sunyaev-Zel'dovich effect power spectrum are two promising probes in the study of cosmic reionization, but each is limited in constraining power. Indeed, the global $21\,\textrm{cm}$ signal is an average over the whole sky, whilst the kSZ effect is integrated over time. Additionally, because of the brightness of spectrally smooth foregrounds, the former is expected to be most sensitive to rapidly evolving ionization histories,  whereas the amplitude of the kSZ power increases with the duration of reionization, making this probe more sensitive to extended ionization histories. 

In this work, we have quantitatively confirmed this intuitive complementarity and demonstrated that it can be used to give tight constraints on the ionization history of our Universe. This was done by establishing a framework, based on an application of the Karhunen-Lo\`eve (KL) transformation, which decomposes the data into a basis whose eigenvalues and modes  describe the relative information content of each probe (Figure~\ref{fig:modes}). Moreover, the qualitiative aspects of this formalism are robust to different models of reionization, as illustrated in Figure~\ref{fig:errorbars}.
With this basis as a guide, we found that joint measurements of the global 21cm signal and the kSZ effect can considerably reduce errors and biases in recovered ionization histories, especially when physically motivated priors are imposed (Figure~\ref{fig:bayesianreconstruction}). 

The complementarity between the global $21\,\textrm{cm}$ and the kSZ can also be wielded to determine whether systematics are present in data, as well as to mitigate their effects. A wide range of potential systematics were considered, such as foreground leakage and cable reflections in the global $21\,\textrm{cm}$ signal, or leakage from various primary or secondary CMB anisotropies into the kSZ data. Exploiting the KL modes that are well measured by both probes enables consistency checks between datasets. Using a linear matched filter technique, we were able detect the presence of foreground residuals with an amplitude about four times smaller than the cosmological signal ($\sim 20\,\mathrm{mK}$) at $10 \sigma$ (Figure~\ref{fig:null_buster}). The detection significance is even better for kSZ systematics, with a $0.05\,\mu\mathrm{K}^2$ residual amplitude of the tSZxCIB cross-spectrum---about half of the error bars on current kSZ measurements \citep{Reichardt2021}---being picked up at $100\sigma$.

However, systematics detection alone is not sufficient, and one would ideally like to remove systematics from one's analysis in order to avoid biased constraints.
We found that the wholesale projecting out of systematic modes is not a viable option, as it results in a significant increase in errors and relies on good prior knowledge of the nature of the systematic projected out (Figure~\ref{fig:projecting}). Exploiting the fact that the KL basis diagonalizes measurement covariance matrices, we included outlier modelling in a fully Bayesian treatment of the data. In a proof-of-concept example, we were able to automatically identify and discard an injected outlier in the $21\,\textrm{cm}$ measurement of a KL mode, effectively relying on the information provided by the kSZ measurement of this mode alone to provide an unbiased constraint on the ionization history. Conversely, in the case where automatic outlier detection was not employed, the resulting ionization history was biased (Figure~\ref{fig:outlier}).

Having illustrated the complementarity between the global 21\,cm signal and the patchy kSZ angular power spectrum, future work should further examine the power of connecting the two probes in a realistic end-to-end simulation framework. With self-consistent sky realizations, one could capture correlated cosmic variance effects. Indeed, cosmic variance errors may be an obstacle to measurements of the global 21\,cm signal from reionization \cite{2021PhRvD.103b3512M}, and being a cosmological-signal sourced uncertainty, it would presumably correlate with a kSZ measurement, leading to off-diagonal covariance terms which we have not included in this analysis. End-to-end simulations also enable a more realistic quantification of systematic uncertainties, e.g., from foreground cleaning. On one hand, foreground component separation from sky models could lead to a degradation of our results, by increasing the error terms in the kSZ covariance \cite{Alvarez2020}. End-to-end foreground simulations also help to guard against any quantitative peculiarities of the specific foreground model used in this paper on the $21\,\textrm{cm}$ side. On the other hand, with realistic full sky maps there is the potential to leverage more advanced (and speculative) foreground mitigation techniques such as those based on detecting violations of statistical isotropy \cite{2006PhRvD..74b1301B,2016JCAP...09..034R,2018MNRAS.479.5577P}. Moreover, the techniques proposed in Section~\ref{sec:systematics} will allow for the detection of unexpected systematics and are therefore especially useful if the instrument is not perfectly known. Finally, one could also include other probes of reionization (e.g., from CMB polarization) in a more self-consistent way \cite{2022arXiv220208698G}, going beyond a simple $\tau$ prior.

Providing precision constraints on the ionization history is a challenging problem. As an essential step towards overcoming this crucial obstacle (and in order to prepare for upcoming precision measurements of the kSZ power spectrum \cite{CalabreseHlozek_2014,Alvarez2020}) in this paper we have established a framework for combining a particularly complementary pair of probes---the global $21\,\textrm{cm}$ signal and the kSZ effect---which together will serve to exemplify the power of a multi-pronged program to unlock the mysteries of the Epoch of Reionization.

\section*{Acknowledgements}

The authors are delighted to thank Marcelo Alvarez, Neil Bassett, Eamon Egan, Joshua Hibbard, Alexander Laroche, Jordan Mirocha, Morgan Presley, David Rapetti, Saurabh Singh, Keith Tauscher, and Oliver Zahn for useful discussions as well as the anonymous referee whose comments have improved this paper. AG's work was supported by the McGill Astrophysics Fellowship funded by the Trottier Chair in Astrophysics, as well as the Canadian Institute for Advanced Research (CIFAR) Azrieli Global Scholars program and the Canada 150 Programme. AL acknowledges support from the New Frontiers in Research Fund Exploration grant program, the Canadian Institute for Advanced Research (CIFAR) Azrieli Global Scholars program, a Natural Sciences and Engineering Research Council of Canada (NSERC) Discovery Grant and a Discovery Launch Supplement, the Sloan Research Fellowship, and the William Dawson Scholarship at McGill.

Results in this work were obtained with the help of \texttt{astropy} \citep{astropy,astropy2}; \texttt{matplotlib} \citep{matplotlib}; \texttt{scipy} \citep{scipy}---including \texttt{numpy} \citep{numpy}, and \texttt{emcee}, a Python implementation of the affine invariant MCMC ensemble sampler \citep{emcee}.


\bibliography{bib}

\begin{thebibliography}{98}%
\makeatletter
\providecommand \@ifxundefined [1]{%
 \@ifx{#1\undefined}
}%
\providecommand \@ifnum [1]{%
 \ifnum #1\expandafter \@firstoftwo
 \else \expandafter \@secondoftwo
 \fi
}%
\providecommand \@ifx [1]{%
 \ifx #1\expandafter \@firstoftwo
 \else \expandafter \@secondoftwo
 \fi
}%
\providecommand \natexlab [1]{#1}%
\providecommand \enquote  [1]{``#1''}%
\providecommand \bibnamefont  [1]{#1}%
\providecommand \bibfnamefont [1]{#1}%
\providecommand \citenamefont [1]{#1}%
\providecommand \href@noop [0]{\@secondoftwo}%
\providecommand \href [0]{\begingroup \@sanitize@url \@href}%
\providecommand \@href[1]{\@@startlink{#1}\@@href}%
\providecommand \@@href[1]{\endgroup#1\@@endlink}%
\providecommand \@sanitize@url [0]{\catcode `\\12\catcode `\$12\catcode
  `\&12\catcode `\#12\catcode `\^12\catcode `\_12\catcode `\%12\relax}%
\providecommand \@@startlink[1]{}%
\providecommand \@@endlink[0]{}%
\providecommand \url  [0]{\begingroup\@sanitize@url \@url }%
\providecommand \@url [1]{\endgroup\@href {#1}{\urlprefix }}%
\providecommand \urlprefix  [0]{URL }%
\providecommand \Eprint [0]{\href }%
\providecommand \doibase [0]{http://dx.doi.org/}%
\providecommand \selectlanguage [0]{\@gobble}%
\providecommand \bibinfo  [0]{\@secondoftwo}%
\providecommand \bibfield  [0]{\@secondoftwo}%
\providecommand \translation [1]{[#1]}%
\providecommand \BibitemOpen [0]{}%
\providecommand \bibitemStop [0]{}%
\providecommand \bibitemNoStop [0]{.\EOS\space}%
\providecommand \EOS [0]{\spacefactor3000\relax}%
\providecommand \BibitemShut  [1]{\csname bibitem#1\endcsname}%
\let\auto@bib@innerbib\@empty
\bibitem [{\citenamefont {{Planck Collaboration}}\ \emph
  {et~al.}(2020{\natexlab{a}})\citenamefont {{Planck Collaboration}},
  \citenamefont {{Aghanim}} \emph {et~al.}}]{Planck2018}%
  \BibitemOpen
  \bibfield  {author} {\bibinfo {author} {\bibnamefont {{Planck
  Collaboration}}}, \bibinfo {author} {\bibfnamefont {N.}~\bibnamefont
  {{Aghanim}}},  \emph {et~al.},\ }\href {\doibase 10.1051/0004-6361/201833880}
  {\bibfield  {journal} {\bibinfo  {journal} {\aap}\ }\textbf {\bibinfo
  {volume} {641}},\ \bibinfo {eid} {A1} (\bibinfo {year}
  {2020}{\natexlab{a}})},\ \Eprint {http://arxiv.org/abs/1807.06205}
  {arXiv:1807.06205 [astro-ph.CO]} \BibitemShut {NoStop}%
\bibitem [{\citenamefont {{Gorce}}\ \emph {et~al.}(2018)\citenamefont
  {{Gorce}}, \citenamefont {{Douspis}}, \citenamefont {{Aghanim}},\ and\
  \citenamefont {{Langer}}}]{Gorce2018}%
  \BibitemOpen
  \bibfield  {author} {\bibinfo {author} {\bibfnamefont {A.}~\bibnamefont
  {{Gorce}}}, \bibinfo {author} {\bibfnamefont {M.}~\bibnamefont {{Douspis}}},
  \bibinfo {author} {\bibfnamefont {N.}~\bibnamefont {{Aghanim}}}, \ and\
  \bibinfo {author} {\bibfnamefont {M.}~\bibnamefont {{Langer}}},\ }\href
  {\doibase 10.1051/0004-6361/201629661} {\bibfield  {journal} {\bibinfo
  {journal} {Astronomy \& Astrohysiccs}\ }\textbf {\bibinfo {volume} {616}},\
  \bibinfo {eid} {A113} (\bibinfo {year} {2018})},\ \Eprint
  {http://arxiv.org/abs/1710.04152} {arXiv:1710.04152 [astro-ph.CO]}
  \BibitemShut {NoStop}%
\bibitem [{\citenamefont {{Qin}}\ \emph {et~al.}(2020)\citenamefont {{Qin}},
  \citenamefont {{Poulin}}, \citenamefont {{Mesinger}}, \citenamefont
  {{Greig}}, \citenamefont {{Murray}},\ and\ \citenamefont {{Park}}}]{Qin2020}%
  \BibitemOpen
  \bibfield  {author} {\bibinfo {author} {\bibfnamefont {Y.}~\bibnamefont
  {{Qin}}}, \bibinfo {author} {\bibfnamefont {V.}~\bibnamefont {{Poulin}}},
  \bibinfo {author} {\bibfnamefont {A.}~\bibnamefont {{Mesinger}}}, \bibinfo
  {author} {\bibfnamefont {B.}~\bibnamefont {{Greig}}}, \bibinfo {author}
  {\bibfnamefont {S.}~\bibnamefont {{Murray}}}, \ and\ \bibinfo {author}
  {\bibfnamefont {J.}~\bibnamefont {{Park}}},\ }\href {\doibase
  10.1093/mnras/staa2797} {\bibfield  {journal} {\bibinfo  {journal} {Monthly
  Notices of the Royal Astronomical Society}\ }\textbf {\bibinfo {volume}
  {499}},\ \bibinfo {pages} {550} (\bibinfo {year} {2020})},\ \Eprint
  {http://arxiv.org/abs/2006.16828} {arXiv:2006.16828 [astro-ph.CO]}
  \BibitemShut {NoStop}%
\bibitem [{\citenamefont {Miranda}\ \emph {et~al.}(2017)\citenamefont
  {Miranda}, \citenamefont {Lidz}, \citenamefont {Heinrich},\ and\
  \citenamefont {Hu}}]{Miranda2017}%
  \BibitemOpen
  \bibfield  {author} {\bibinfo {author} {\bibfnamefont {V.}~\bibnamefont
  {Miranda}}, \bibinfo {author} {\bibfnamefont {A.}~\bibnamefont {Lidz}},
  \bibinfo {author} {\bibfnamefont {C.~H.}\ \bibnamefont {Heinrich}}, \ and\
  \bibinfo {author} {\bibfnamefont {W.}~\bibnamefont {Hu}},\ }\href {\doibase
  10.1093/mnras/stx306} {\bibfield  {journal} {\bibinfo  {journal} {Monthly
  Notices of the Royal Astronomical Society}\ }\textbf {\bibinfo {volume}
  {467}},\ \bibinfo {pages} {4050} (\bibinfo {year} {2017})}\BibitemShut
  {NoStop}%
\bibitem [{\citenamefont {Heinrich}\ and\ \citenamefont
  {Hu}(2018)}]{Heinrich2018}%
  \BibitemOpen
  \bibfield  {author} {\bibinfo {author} {\bibfnamefont {C.}~\bibnamefont
  {Heinrich}}\ and\ \bibinfo {author} {\bibfnamefont {W.}~\bibnamefont {Hu}},\
  }\href {\doibase 10.1103/PhysRevD.98.063514} {\bibfield  {journal} {\bibinfo
  {journal} {Phys. Rev. D}\ }\textbf {\bibinfo {volume} {98}},\ \bibinfo
  {pages} {063514} (\bibinfo {year} {2018})}\BibitemShut {NoStop}%
\bibitem [{\citenamefont {Gunn}\ and\ \citenamefont
  {Peterson}(1965)}]{Gunn_1965}%
  \BibitemOpen
  \bibfield  {author} {\bibinfo {author} {\bibfnamefont {J.~E.}\ \bibnamefont
  {Gunn}}\ and\ \bibinfo {author} {\bibfnamefont {B.~A.}\ \bibnamefont
  {Peterson}},\ }\href {\doibase 10.1086/148444} {\bibfield  {journal}
  {\bibinfo  {journal} {Astrophysical Journal}\ }\textbf {\bibinfo {volume}
  {142}},\ \bibinfo {pages} {1633} (\bibinfo {year} {1965})}\BibitemShut
  {NoStop}%
\bibitem [{\citenamefont {Fan}\ \emph {et~al.}(2006)\citenamefont {Fan} \emph
  {et~al.}}]{Fan_2006}%
  \BibitemOpen
  \bibfield  {author} {\bibinfo {author} {\bibfnamefont {X.}~\bibnamefont
  {Fan}} \emph {et~al.},\ }\href {\doibase 10.1086/504836} {\bibfield
  {journal} {\bibinfo  {journal} {The Astronomical Journal}\ }\textbf {\bibinfo
  {volume} {132}},\ \bibinfo {pages} {117} (\bibinfo {year}
  {2006})}\BibitemShut {NoStop}%
\bibitem [{\citenamefont {{Fan}}(2006)}]{Fan_2006_review}%
  \BibitemOpen
  \bibfield  {author} {\bibinfo {author} {\bibfnamefont {X.}~\bibnamefont
  {{Fan}}},\ }\href {\doibase 10.1016/j.newar.2006.06.077} {\bibfield
  {journal} {\bibinfo  {journal} {New Astronomy Reviews}\ }\textbf {\bibinfo
  {volume} {50}},\ \bibinfo {pages} {665} (\bibinfo {year} {2006})}\BibitemShut
  {NoStop}%
\bibitem [{\citenamefont {{Mesinger}}(2010)}]{Mesinger2010}%
  \BibitemOpen
  \bibfield  {author} {\bibinfo {author} {\bibfnamefont {A.}~\bibnamefont
  {{Mesinger}}},\ }\href {\doibase 10.1111/j.1365-2966.2010.16995.x} {\bibfield
   {journal} {\bibinfo  {journal} {Monthly Notices of the Royal Astronomical
  Society}\ }\textbf {\bibinfo {volume} {407}},\ \bibinfo {pages} {1328}
  (\bibinfo {year} {2010})},\ \Eprint {http://arxiv.org/abs/0910.4161}
  {arXiv:0910.4161 [astro-ph.CO]} \BibitemShut {NoStop}%
\bibitem [{\citenamefont {Wise}(2019)}]{Wise2019}%
  \BibitemOpen
  \bibfield  {author} {\bibinfo {author} {\bibfnamefont {J.~H.}\ \bibnamefont
  {Wise}},\ }\href {\doibase 10.1080/00107514.2019.1631548} {\bibfield
  {journal} {\bibinfo  {journal} {Contemporary Physics}\ }\textbf {\bibinfo
  {volume} {60}},\ \bibinfo {pages} {145} (\bibinfo {year} {2019})},\ \Eprint
  {http://arxiv.org/abs/https://doi.org/10.1080/00107514.2019.1631548}
  {https://doi.org/10.1080/00107514.2019.1631548} \BibitemShut {NoStop}%
\bibitem [{\citenamefont {{Planck Collaboration}}\ \emph
  {et~al.}(2016{\natexlab{a}})\citenamefont {{Planck Collaboration}},
  \citenamefont {{Ade, P. A. R.}} \emph {et~al.}}]{Planck15}%
  \BibitemOpen
  \bibfield  {author} {\bibinfo {author} {\bibnamefont {{Planck
  Collaboration}}}, \bibinfo {author} {\bibnamefont {{Ade, P. A. R.}}},  \emph
  {et~al.},\ }\href {\doibase 10.1051/0004-6361/201525830} {\bibfield
  {journal} {\bibinfo  {journal} {A\&A}\ }\textbf {\bibinfo {volume} {594}},\
  \bibinfo {pages} {A13} (\bibinfo {year} {2016}{\natexlab{a}})}\BibitemShut
  {NoStop}%
\bibitem [{\citenamefont {{Planck Collaboration}}\ \emph
  {et~al.}(2016{\natexlab{b}})\citenamefont {{Planck Collaboration}},
  \citenamefont {{Adam, R.}} \emph {et~al.}}]{Planck2016}%
  \BibitemOpen
  \bibfield  {author} {\bibinfo {author} {\bibnamefont {{Planck
  Collaboration}}}, \bibinfo {author} {\bibnamefont {{Adam, R.}}},  \emph
  {et~al.},\ }\href {\doibase 10.1051/0004-6361/201628897} {\bibfield
  {journal} {\bibinfo  {journal} {A\&A}\ }\textbf {\bibinfo {volume} {596}},\
  \bibinfo {pages} {A108} (\bibinfo {year} {2016}{\natexlab{b}})}\BibitemShut
  {NoStop}%
\bibitem [{\citenamefont {Zeldovich}\ and\ \citenamefont
  {Sunyaev}(1969)}]{Zeldovich_Sunyaev_1969}%
  \BibitemOpen
  \bibfield  {author} {\bibinfo {author} {\bibfnamefont {Y.~B.}\ \bibnamefont
  {Zeldovich}}\ and\ \bibinfo {author} {\bibfnamefont {R.~A.}\ \bibnamefont
  {Sunyaev}},\ }\href {\doibase 10.1007/BF00661821} {\bibfield  {journal}
  {\bibinfo  {journal} {Astrophysics and Space Science}\ }\textbf {\bibinfo
  {volume} {4}},\ \bibinfo {pages} {301} (\bibinfo {year} {1969})}\BibitemShut
  {NoStop}%
\bibitem [{\citenamefont {{Sunyaev}}\ and\ \citenamefont
  {{Zeldovich}}(1980)}]{Sunyaev_Zeldovich_1980}%
  \BibitemOpen
  \bibfield  {author} {\bibinfo {author} {\bibfnamefont {R.~A.}\ \bibnamefont
  {{Sunyaev}}}\ and\ \bibinfo {author} {\bibfnamefont {I.~B.}\ \bibnamefont
  {{Zeldovich}}},\ }\href {\doibase 10.1146/annurev.aa.18.090180.002541}
  {\bibfield  {journal} {\bibinfo  {journal} {Annual Review of Astronomy and
  Astrophysics}\ }\textbf {\bibinfo {volume} {18}},\ \bibinfo {pages} {537}
  (\bibinfo {year} {1980})}\BibitemShut {NoStop}%
\bibitem [{\citenamefont {Battaglia}\ \emph {et~al.}(2013)\citenamefont
  {Battaglia}, \citenamefont {Natarajan}, \citenamefont {Trac}, \citenamefont
  {Cen},\ and\ \citenamefont {Loeb}}]{Battaglia2013}%
  \BibitemOpen
  \bibfield  {author} {\bibinfo {author} {\bibfnamefont {N.}~\bibnamefont
  {Battaglia}}, \bibinfo {author} {\bibfnamefont {A.}~\bibnamefont
  {Natarajan}}, \bibinfo {author} {\bibfnamefont {H.}~\bibnamefont {Trac}},
  \bibinfo {author} {\bibfnamefont {R.}~\bibnamefont {Cen}}, \ and\ \bibinfo
  {author} {\bibfnamefont {A.}~\bibnamefont {Loeb}},\ }\href {\doibase
  10.1088/0004-637x/776/2/83} {\bibfield  {journal} {\bibinfo  {journal} {The
  Astrophysical Journal}\ }\textbf {\bibinfo {volume} {776}},\ \bibinfo {pages}
  {83} (\bibinfo {year} {2013})}\BibitemShut {NoStop}%
\bibitem [{\citenamefont {Mesinger}\ \emph {et~al.}(2012)\citenamefont
  {Mesinger}, \citenamefont {McQuinn},\ and\ \citenamefont
  {Spergel}}]{Mesinger2012}%
  \BibitemOpen
  \bibfield  {author} {\bibinfo {author} {\bibfnamefont {A.}~\bibnamefont
  {Mesinger}}, \bibinfo {author} {\bibfnamefont {M.}~\bibnamefont {McQuinn}}, \
  and\ \bibinfo {author} {\bibfnamefont {D.~N.}\ \bibnamefont {Spergel}},\
  }\href {\doibase 10.1111/j.1365-2966.2012.20713.x} {\bibfield  {journal}
  {\bibinfo  {journal} {Monthly Notices of the Royal Astronomical Society}\
  }\textbf {\bibinfo {volume} {422}},\ \bibinfo {pages} {1403} (\bibinfo {year}
  {2012})},\ \Eprint
  {http://arxiv.org/abs/https://academic.oup.com/mnras/article-pdf/422/2/1403/3487085/mnras0422-1403.pdf}
  {https://academic.oup.com/mnras/article-pdf/422/2/1403/3487085/mnras0422-1403.pdf}
  \BibitemShut {NoStop}%
\bibitem [{\citenamefont {Park}\ \emph {et~al.}(2013)\citenamefont {Park},
  \citenamefont {Shapiro}, \citenamefont {Komatsu}, \citenamefont {Iliev},
  \citenamefont {Ahn},\ and\ \citenamefont {Mellema}}]{Park2013}%
  \BibitemOpen
  \bibfield  {author} {\bibinfo {author} {\bibfnamefont {H.}~\bibnamefont
  {Park}}, \bibinfo {author} {\bibfnamefont {P.~R.}\ \bibnamefont {Shapiro}},
  \bibinfo {author} {\bibfnamefont {E.}~\bibnamefont {Komatsu}}, \bibinfo
  {author} {\bibfnamefont {I.~T.}\ \bibnamefont {Iliev}}, \bibinfo {author}
  {\bibfnamefont {K.}~\bibnamefont {Ahn}}, \ and\ \bibinfo {author}
  {\bibfnamefont {G.}~\bibnamefont {Mellema}},\ }\href {\doibase
  10.1088/0004-637x/769/2/93} {\bibfield  {journal} {\bibinfo  {journal} {The
  Astrophysical Journal}\ }\textbf {\bibinfo {volume} {769}},\ \bibinfo {pages}
  {93} (\bibinfo {year} {2013})}\BibitemShut {NoStop}%
\bibitem [{\citenamefont {{Choudhury}}\ \emph {et~al.}(2021)\citenamefont
  {{Choudhury}}, \citenamefont {{Mukherjee}},\ and\ \citenamefont
  {{Paul}}}]{ChoudhuryMukherjee_2021}%
  \BibitemOpen
  \bibfield  {author} {\bibinfo {author} {\bibfnamefont {T.~R.}\ \bibnamefont
  {{Choudhury}}}, \bibinfo {author} {\bibfnamefont {S.}~\bibnamefont
  {{Mukherjee}}}, \ and\ \bibinfo {author} {\bibfnamefont {S.}~\bibnamefont
  {{Paul}}},\ }\href {\doibase 10.1093/mnrasl/slaa185} {\bibfield  {journal}
  {\bibinfo  {journal} {\mnras}\ }\textbf {\bibinfo {volume} {501}},\ \bibinfo
  {pages} {L7} (\bibinfo {year} {2021})},\ \Eprint
  {http://arxiv.org/abs/2007.03705} {arXiv:2007.03705 [astro-ph.CO]}
  \BibitemShut {NoStop}%
\bibitem [{\citenamefont {Reichardt}\ \emph {et~al.}(2021)\citenamefont
  {Reichardt} \emph {et~al.}}]{Reichardt2021}%
  \BibitemOpen
  \bibfield  {author} {\bibinfo {author} {\bibfnamefont {C.~L.}\ \bibnamefont
  {Reichardt}} \emph {et~al.},\ }\href {\doibase 10.3847/1538-4357/abd407}
  {\bibfield  {journal} {\bibinfo  {journal} {The Astrophysical Journal}\
  }\textbf {\bibinfo {volume} {908}},\ \bibinfo {pages} {199} (\bibinfo {year}
  {2021})}\BibitemShut {NoStop}%
\bibitem [{\citenamefont {{Ade}}\ \emph {et~al.}(2019)\citenamefont {{Ade}},
  \citenamefont {{Aguirre}}, \citenamefont {{Ahmed}}, \citenamefont {{Aiola}},
  \citenamefont {{Ali}}, \citenamefont {{Alonso}}, \citenamefont {{Alvarez}},
  \citenamefont {{Arnold}}, \citenamefont {{Ashton}}, \citenamefont
  {{Austermann}},\ and\ \citenamefont {et~al.}}]{AdeAguirre_2019}%
  \BibitemOpen
  \bibfield  {author} {\bibinfo {author} {\bibfnamefont {P.}~\bibnamefont
  {{Ade}}}, \bibinfo {author} {\bibfnamefont {J.}~\bibnamefont {{Aguirre}}},
  \bibinfo {author} {\bibfnamefont {Z.}~\bibnamefont {{Ahmed}}}, \bibinfo
  {author} {\bibfnamefont {S.}~\bibnamefont {{Aiola}}}, \bibinfo {author}
  {\bibfnamefont {A.}~\bibnamefont {{Ali}}}, \bibinfo {author} {\bibfnamefont
  {D.}~\bibnamefont {{Alonso}}}, \bibinfo {author} {\bibfnamefont {M.~A.}\
  \bibnamefont {{Alvarez}}}, \bibinfo {author} {\bibfnamefont {K.}~\bibnamefont
  {{Arnold}}}, \bibinfo {author} {\bibfnamefont {P.}~\bibnamefont {{Ashton}}},
  \bibinfo {author} {\bibfnamefont {J.}~\bibnamefont {{Austermann}}}, \ and\
  \bibinfo {author} {\bibnamefont {et~al.}},\ }\href {\doibase
  10.1088/1475-7516/2019/02/056} {\bibfield  {journal} {\bibinfo  {journal}
  {JCAP}\ }\textbf {\bibinfo {volume} {2019}},\ \bibinfo {eid} {056} (\bibinfo
  {year} {2019})},\ \Eprint {http://arxiv.org/abs/1808.07445} {arXiv:1808.07445
  [astro-ph.CO]} \BibitemShut {NoStop}%
\bibitem [{\citenamefont {{Abazajian}}\ \emph {et~al.}(2016)\citenamefont
  {{Abazajian}}, \citenamefont {{Adshead}}, \citenamefont {{Ahmed}} \emph
  {et~al.}}]{CMB-S4}%
  \BibitemOpen
  \bibfield  {author} {\bibinfo {author} {\bibfnamefont {K.~N.}\ \bibnamefont
  {{Abazajian}}}, \bibinfo {author} {\bibfnamefont {P.}~\bibnamefont
  {{Adshead}}}, \bibinfo {author} {\bibfnamefont {Z.}~\bibnamefont {{Ahmed}}},
  \emph {et~al.},\ }\href@noop {} {\bibfield  {journal} {\bibinfo  {journal}
  {arXiv e-prints}\ ,\ \bibinfo {eid} {arXiv:1610.02743}} (\bibinfo {year}
  {2016})},\ \Eprint {http://arxiv.org/abs/1610.02743} {arXiv:1610.02743
  [astro-ph.CO]} \BibitemShut {NoStop}%
\bibitem [{\citenamefont {Alvarez}\ \emph {et~al.}(2021)\citenamefont
  {Alvarez}, \citenamefont {Ferraro}, \citenamefont {Hill}, \citenamefont
  {Hlo\ifmmode~\check{z}\else \v{z}\fi{}ek},\ and\ \citenamefont
  {Ikape}}]{Alvarez2020}%
  \BibitemOpen
  \bibfield  {author} {\bibinfo {author} {\bibfnamefont {M.~A.}\ \bibnamefont
  {Alvarez}}, \bibinfo {author} {\bibfnamefont {S.}~\bibnamefont {Ferraro}},
  \bibinfo {author} {\bibfnamefont {J.~C.}\ \bibnamefont {Hill}}, \bibinfo
  {author} {\bibfnamefont {R.}~\bibnamefont {Hlo\ifmmode~\check{z}\else
  \v{z}\fi{}ek}}, \ and\ \bibinfo {author} {\bibfnamefont {M.}~\bibnamefont
  {Ikape}},\ }\href {\doibase 10.1103/PhysRevD.103.063518} {\bibfield
  {journal} {\bibinfo  {journal} {Phys. Rev. D}\ }\textbf {\bibinfo {volume}
  {103}},\ \bibinfo {pages} {063518} (\bibinfo {year} {2021})}\BibitemShut
  {NoStop}%
\bibitem [{\citenamefont {Gorce}\ \emph {et~al.}(2020)\citenamefont {Gorce},
  \citenamefont {Ili{\'{c}}}, \citenamefont {Douspis}, \citenamefont {Aubert},\
  and\ \citenamefont {Langer}}]{Gorce2020}%
  \BibitemOpen
  \bibfield  {author} {\bibinfo {author} {\bibfnamefont {A.}~\bibnamefont
  {Gorce}}, \bibinfo {author} {\bibfnamefont {S.}~\bibnamefont {Ili{\'{c}}}},
  \bibinfo {author} {\bibfnamefont {M.}~\bibnamefont {Douspis}}, \bibinfo
  {author} {\bibfnamefont {D.}~\bibnamefont {Aubert}}, \ and\ \bibinfo {author}
  {\bibfnamefont {M.}~\bibnamefont {Langer}},\ }\href {\doibase
  10.1051/0004-6361/202038170} {\bibfield  {journal} {\bibinfo  {journal}
  {Astronomy {\&} Astrophysics}\ }\textbf {\bibinfo {volume} {640}},\ \bibinfo
  {pages} {A90} (\bibinfo {year} {2020})}\BibitemShut {NoStop}%
\bibitem [{\citenamefont {{Shaver}}\ \emph {et~al.}(1999)\citenamefont
  {{Shaver}}, \citenamefont {{Windhorst}}, \citenamefont {{Madau}},\ and\
  \citenamefont {{de Bruyn}}}]{Shaver1999}%
  \BibitemOpen
  \bibfield  {author} {\bibinfo {author} {\bibfnamefont {P.~A.}\ \bibnamefont
  {{Shaver}}}, \bibinfo {author} {\bibfnamefont {R.~A.}\ \bibnamefont
  {{Windhorst}}}, \bibinfo {author} {\bibfnamefont {P.}~\bibnamefont
  {{Madau}}}, \ and\ \bibinfo {author} {\bibfnamefont {A.~G.}\ \bibnamefont
  {{de Bruyn}}},\ }\href@noop {} {\bibfield  {journal} {\bibinfo  {journal}
  {\aap}\ }\textbf {\bibinfo {volume} {345}},\ \bibinfo {pages} {380} (\bibinfo
  {year} {1999})},\ \Eprint {http://arxiv.org/abs/astro-ph/9901320}
  {arXiv:astro-ph/9901320 [astro-ph]} \BibitemShut {NoStop}%
\bibitem [{\citenamefont {{Furlanetto}}(2006)}]{Furlanetto2006globalsig}%
  \BibitemOpen
  \bibfield  {author} {\bibinfo {author} {\bibfnamefont {S.~R.}\ \bibnamefont
  {{Furlanetto}}},\ }\href {\doibase 10.1111/j.1365-2966.2006.10725.x}
  {\bibfield  {journal} {\bibinfo  {journal} {\mnras}\ }\textbf {\bibinfo
  {volume} {371}},\ \bibinfo {pages} {867} (\bibinfo {year} {2006})},\ \Eprint
  {http://arxiv.org/abs/astro-ph/0604040} {arXiv:astro-ph/0604040 [astro-ph]}
  \BibitemShut {NoStop}%
\bibitem [{\citenamefont {Pritchard}\ and\ \citenamefont
  {Loeb}(2010)}]{Pritchard2010}%
  \BibitemOpen
  \bibfield  {author} {\bibinfo {author} {\bibfnamefont {J.~R.}\ \bibnamefont
  {Pritchard}}\ and\ \bibinfo {author} {\bibfnamefont {A.}~\bibnamefont
  {Loeb}},\ }\href {\doibase 10.1103/physrevd.82.023006} {\bibfield  {journal}
  {\bibinfo  {journal} {Physical Review D}\ }\textbf {\bibinfo {volume} {82}}
  (\bibinfo {year} {2010}),\ 10.1103/physrevd.82.023006}\BibitemShut {NoStop}%
\bibitem [{\citenamefont {{Morandi}}\ and\ \citenamefont
  {{Barkana}}(2012)}]{2012MNRAS.424.2551M}%
  \BibitemOpen
  \bibfield  {author} {\bibinfo {author} {\bibfnamefont {A.}~\bibnamefont
  {{Morandi}}}\ and\ \bibinfo {author} {\bibfnamefont {R.}~\bibnamefont
  {{Barkana}}},\ }\href {\doibase 10.1111/j.1365-2966.2012.21240.x} {\bibfield
  {journal} {\bibinfo  {journal} {\mnras}\ }\textbf {\bibinfo {volume} {424}},\
  \bibinfo {pages} {2551} (\bibinfo {year} {2012})},\ \Eprint
  {http://arxiv.org/abs/1102.2378} {arXiv:1102.2378 [astro-ph.CO]} \BibitemShut
  {NoStop}%
\bibitem [{\citenamefont {Bowman}\ \emph {et~al.}(2018)\citenamefont {Bowman},
  \citenamefont {Rogers}, \citenamefont {Monsalve}, \citenamefont {Mozdzen},\
  and\ \citenamefont {Mahesh}}]{Bowman2018}%
  \BibitemOpen
  \bibfield  {author} {\bibinfo {author} {\bibfnamefont {J.~D.}\ \bibnamefont
  {Bowman}}, \bibinfo {author} {\bibfnamefont {A.~E.~E.}\ \bibnamefont
  {Rogers}}, \bibinfo {author} {\bibfnamefont {R.~A.}\ \bibnamefont
  {Monsalve}}, \bibinfo {author} {\bibfnamefont {T.~J.}\ \bibnamefont
  {Mozdzen}}, \ and\ \bibinfo {author} {\bibfnamefont {N.}~\bibnamefont
  {Mahesh}},\ }\href {\doibase 10.1038/nature25792} {\bibfield  {journal}
  {\bibinfo  {journal} {Nature}\ }\textbf {\bibinfo {volume} {555}},\ \bibinfo
  {pages} {67} (\bibinfo {year} {2018})}\BibitemShut {NoStop}%
\bibitem [{\citenamefont {Philip}\ \emph {et~al.}(2019)\citenamefont {Philip}
  \emph {et~al.}}]{Philip2019}%
  \BibitemOpen
  \bibfield  {author} {\bibinfo {author} {\bibfnamefont {L.}~\bibnamefont
  {Philip}} \emph {et~al.},\ }\href {\doibase 10.1142/s2251171719500041}
  {\bibfield  {journal} {\bibinfo  {journal} {Journal of Astronomical
  Instrumentation}\ }\textbf {\bibinfo {volume} {08}},\ \bibinfo {pages}
  {1950004} (\bibinfo {year} {2019})}\BibitemShut {NoStop}%
\bibitem [{\citenamefont {{Price}}\ \emph {et~al.}(2018)\citenamefont
  {{Price}}, \citenamefont {{Greenhill}}, \citenamefont {{Fialkov}},
  \citenamefont {{Bernardi}}, \citenamefont {{Garsden}}, \citenamefont
  {{Barsdell}}, \citenamefont {{Kocz}}, \citenamefont {{Anderson}},
  \citenamefont {{Bourke}}, \citenamefont {{Craig}}, \citenamefont {{Dexter}},
  \citenamefont {{Dowell}}, \citenamefont {{Eastwood}}, \citenamefont
  {{Eftekhari}}, \citenamefont {{Ellingson}}, \citenamefont {{Hallinan}},
  \citenamefont {{Hartman}}, \citenamefont {{Kimberk}}, \citenamefont
  {{Lazio}}, \citenamefont {{Leiker}}, \citenamefont {{MacMahon}},
  \citenamefont {{Monroe}}, \citenamefont {{Schinzel}}, \citenamefont
  {{Taylor}}, \citenamefont {{Tong}}, \citenamefont {{Werthimer}},\ and\
  \citenamefont {{Woody}}}]{Price2018LEDA}%
  \BibitemOpen
  \bibfield  {author} {\bibinfo {author} {\bibfnamefont {D.~C.}\ \bibnamefont
  {{Price}}}, \bibinfo {author} {\bibfnamefont {L.~J.}\ \bibnamefont
  {{Greenhill}}}, \bibinfo {author} {\bibfnamefont {A.}~\bibnamefont
  {{Fialkov}}}, \bibinfo {author} {\bibfnamefont {G.}~\bibnamefont
  {{Bernardi}}}, \bibinfo {author} {\bibfnamefont {H.}~\bibnamefont
  {{Garsden}}}, \bibinfo {author} {\bibfnamefont {B.~R.}\ \bibnamefont
  {{Barsdell}}}, \bibinfo {author} {\bibfnamefont {J.}~\bibnamefont {{Kocz}}},
  \bibinfo {author} {\bibfnamefont {M.~M.}\ \bibnamefont {{Anderson}}},
  \bibinfo {author} {\bibfnamefont {S.~A.}\ \bibnamefont {{Bourke}}}, \bibinfo
  {author} {\bibfnamefont {J.}~\bibnamefont {{Craig}}}, \bibinfo {author}
  {\bibfnamefont {M.~R.}\ \bibnamefont {{Dexter}}}, \bibinfo {author}
  {\bibfnamefont {J.}~\bibnamefont {{Dowell}}}, \bibinfo {author}
  {\bibfnamefont {M.~W.}\ \bibnamefont {{Eastwood}}}, \bibinfo {author}
  {\bibfnamefont {T.}~\bibnamefont {{Eftekhari}}}, \bibinfo {author}
  {\bibfnamefont {S.~W.}\ \bibnamefont {{Ellingson}}}, \bibinfo {author}
  {\bibfnamefont {G.}~\bibnamefont {{Hallinan}}}, \bibinfo {author}
  {\bibfnamefont {J.~M.}\ \bibnamefont {{Hartman}}}, \bibinfo {author}
  {\bibfnamefont {R.}~\bibnamefont {{Kimberk}}}, \bibinfo {author}
  {\bibfnamefont {T.~J.~W.}\ \bibnamefont {{Lazio}}}, \bibinfo {author}
  {\bibfnamefont {S.}~\bibnamefont {{Leiker}}}, \bibinfo {author}
  {\bibfnamefont {D.}~\bibnamefont {{MacMahon}}}, \bibinfo {author}
  {\bibfnamefont {R.}~\bibnamefont {{Monroe}}}, \bibinfo {author}
  {\bibfnamefont {F.}~\bibnamefont {{Schinzel}}}, \bibinfo {author}
  {\bibfnamefont {G.~B.}\ \bibnamefont {{Taylor}}}, \bibinfo {author}
  {\bibfnamefont {E.}~\bibnamefont {{Tong}}}, \bibinfo {author} {\bibfnamefont
  {D.}~\bibnamefont {{Werthimer}}}, \ and\ \bibinfo {author} {\bibfnamefont
  {D.~P.}\ \bibnamefont {{Woody}}},\ }\href {\doibase 10.1093/mnras/sty1244}
  {\bibfield  {journal} {\bibinfo  {journal} {\mnras}\ }\textbf {\bibinfo
  {volume} {478}},\ \bibinfo {pages} {4193} (\bibinfo {year} {2018})},\ \Eprint
  {http://arxiv.org/abs/1709.09313} {arXiv:1709.09313 [astro-ph.IM]}
  \BibitemShut {NoStop}%
\bibitem [{\citenamefont {{Cumner}}\ \emph {et~al.}(2021)\citenamefont
  {{Cumner}}, \citenamefont {{De Lera Acedo}}, \citenamefont {{de Villiers}},
  \citenamefont {{Anstey}}, \citenamefont {{Kolitsidas}}, \citenamefont
  {{Gurdon}}, \citenamefont {{Fagnoni}}, \citenamefont {{Alexander}},
  \citenamefont {{Bernardi}}, \citenamefont {{Bevins}}, \citenamefont
  {{Carey}}, \citenamefont {{Cavillot}}, \citenamefont {{Chiello}},
  \citenamefont {{Craeye}}, \citenamefont {{Croukamp}}, \citenamefont {{Ely}},
  \citenamefont {{Fialkov}}, \citenamefont {{Gessey-Jones}}, \citenamefont
  {{Gueuning}}, \citenamefont {{Handley}}, \citenamefont {{Hills}},
  \citenamefont {{Josaitis}}, \citenamefont {{Kulkarni}}, \citenamefont
  {{Magro}}, \citenamefont {{Maiolino}}, \citenamefont {{Meerburg}},
  \citenamefont {{Mittal}}, \citenamefont {{Pritchard}}, \citenamefont
  {{Puchwein}}, \citenamefont {{Razavi-Ghods}}, \citenamefont {{Roque}},
  \citenamefont {{Saxena}}, \citenamefont {{Scheutwinkel}}, \citenamefont
  {{Shen}}, \citenamefont {{Sims}}, \citenamefont {{Smirnov}}, \citenamefont
  {{Spinelli}},\ and\ \citenamefont {{Zarb-Adami}}}]{2021arXiv210910098C}%
  \BibitemOpen
  \bibfield  {author} {\bibinfo {author} {\bibfnamefont {J.}~\bibnamefont
  {{Cumner}}}, \bibinfo {author} {\bibfnamefont {E.}~\bibnamefont {{De Lera
  Acedo}}}, \bibinfo {author} {\bibfnamefont {D.~I.~L.}\ \bibnamefont {{de
  Villiers}}}, \bibinfo {author} {\bibfnamefont {D.}~\bibnamefont {{Anstey}}},
  \bibinfo {author} {\bibfnamefont {C.~I.}\ \bibnamefont {{Kolitsidas}}},
  \bibinfo {author} {\bibfnamefont {B.}~\bibnamefont {{Gurdon}}}, \bibinfo
  {author} {\bibfnamefont {N.}~\bibnamefont {{Fagnoni}}}, \bibinfo {author}
  {\bibfnamefont {P.}~\bibnamefont {{Alexander}}}, \bibinfo {author}
  {\bibfnamefont {G.}~\bibnamefont {{Bernardi}}}, \bibinfo {author}
  {\bibfnamefont {H.~T.~J.}\ \bibnamefont {{Bevins}}}, \bibinfo {author}
  {\bibfnamefont {S.}~\bibnamefont {{Carey}}}, \bibinfo {author} {\bibfnamefont
  {J.}~\bibnamefont {{Cavillot}}}, \bibinfo {author} {\bibfnamefont
  {R.}~\bibnamefont {{Chiello}}}, \bibinfo {author} {\bibfnamefont
  {C.}~\bibnamefont {{Craeye}}}, \bibinfo {author} {\bibfnamefont
  {W.}~\bibnamefont {{Croukamp}}}, \bibinfo {author} {\bibfnamefont {J.~A.}\
  \bibnamefont {{Ely}}}, \bibinfo {author} {\bibfnamefont {A.}~\bibnamefont
  {{Fialkov}}}, \bibinfo {author} {\bibfnamefont {T.}~\bibnamefont
  {{Gessey-Jones}}}, \bibinfo {author} {\bibfnamefont {Q.}~\bibnamefont
  {{Gueuning}}}, \bibinfo {author} {\bibfnamefont {W.}~\bibnamefont
  {{Handley}}}, \bibinfo {author} {\bibfnamefont {R.}~\bibnamefont {{Hills}}},
  \bibinfo {author} {\bibfnamefont {A.~T.}\ \bibnamefont {{Josaitis}}},
  \bibinfo {author} {\bibfnamefont {G.}~\bibnamefont {{Kulkarni}}}, \bibinfo
  {author} {\bibfnamefont {A.}~\bibnamefont {{Magro}}}, \bibinfo {author}
  {\bibfnamefont {R.}~\bibnamefont {{Maiolino}}}, \bibinfo {author}
  {\bibfnamefont {P.~D.}\ \bibnamefont {{Meerburg}}}, \bibinfo {author}
  {\bibfnamefont {S.}~\bibnamefont {{Mittal}}}, \bibinfo {author}
  {\bibfnamefont {J.~R.}\ \bibnamefont {{Pritchard}}}, \bibinfo {author}
  {\bibfnamefont {E.}~\bibnamefont {{Puchwein}}}, \bibinfo {author}
  {\bibfnamefont {N.}~\bibnamefont {{Razavi-Ghods}}}, \bibinfo {author}
  {\bibfnamefont {I.~L.~V.}\ \bibnamefont {{Roque}}}, \bibinfo {author}
  {\bibfnamefont {A.}~\bibnamefont {{Saxena}}}, \bibinfo {author}
  {\bibfnamefont {K.~H.}\ \bibnamefont {{Scheutwinkel}}}, \bibinfo {author}
  {\bibfnamefont {E.}~\bibnamefont {{Shen}}}, \bibinfo {author} {\bibfnamefont
  {P.~H.}\ \bibnamefont {{Sims}}}, \bibinfo {author} {\bibfnamefont
  {O.}~\bibnamefont {{Smirnov}}}, \bibinfo {author} {\bibfnamefont
  {M.}~\bibnamefont {{Spinelli}}}, \ and\ \bibinfo {author} {\bibfnamefont
  {K.}~\bibnamefont {{Zarb-Adami}}},\ }\href@noop {} {\bibfield  {journal}
  {\bibinfo  {journal} {arXiv e-prints}\ ,\ \bibinfo {eid} {arXiv:2109.10098}}
  (\bibinfo {year} {2021})},\ \Eprint {http://arxiv.org/abs/2109.10098}
  {arXiv:2109.10098 [astro-ph.IM]} \BibitemShut {NoStop}%
\bibitem [{\citenamefont {Singh}\ \emph {et~al.}(2018)\citenamefont {Singh}
  \emph {et~al.}}]{Singh2018}%
  \BibitemOpen
  \bibfield  {author} {\bibinfo {author} {\bibfnamefont {S.}~\bibnamefont
  {Singh}} \emph {et~al.},\ }\href {\doibase 10.3847/1538-4357/aabae1}
  {\bibfield  {journal} {\bibinfo  {journal} {The Astrophysical Journal}\
  }\textbf {\bibinfo {volume} {858}},\ \bibinfo {pages} {54} (\bibinfo {year}
  {2018})}\BibitemShut {NoStop}%
\bibitem [{\citenamefont {Chapman}\ and\ \citenamefont
  {Jelić}(2019)}]{ChapmanJelic_2019}%
  \BibitemOpen
  \bibfield  {author} {\bibinfo {author} {\bibfnamefont {E.}~\bibnamefont
  {Chapman}}\ and\ \bibinfo {author} {\bibfnamefont {V.}~\bibnamefont
  {Jelić}},\ }in\ \href {\doibase 10.1088/2514-3433/ab4a73ch6} {\emph
  {\bibinfo {booktitle} {The Cosmic 21-cm Revolution}}},\ \bibinfo {series and
  number} {2514-3433}\ (\bibinfo  {publisher} {IOP Publishing},\ \bibinfo
  {year} {2019})\ pp.\ \bibinfo {pages} {6--1 to 6--29}\BibitemShut {NoStop}%
\bibitem [{\citenamefont {{Jeli{\'c}}}\ \emph {et~al.}(2010)\citenamefont
  {{Jeli{\'c}}}, \citenamefont {{Zaroubi}}, \citenamefont {{Aghanim}},
  \citenamefont {{Douspis}}, \citenamefont {{Koopmans}}, \citenamefont
  {{Langer}}, \citenamefont {{Mellema}}, \citenamefont {{Tashiro}},\ and\
  \citenamefont {{Thomas}}}]{JelicZaroubi_2010}%
  \BibitemOpen
  \bibfield  {author} {\bibinfo {author} {\bibfnamefont {V.}~\bibnamefont
  {{Jeli{\'c}}}}, \bibinfo {author} {\bibfnamefont {S.}~\bibnamefont
  {{Zaroubi}}}, \bibinfo {author} {\bibfnamefont {N.}~\bibnamefont
  {{Aghanim}}}, \bibinfo {author} {\bibfnamefont {M.}~\bibnamefont
  {{Douspis}}}, \bibinfo {author} {\bibfnamefont {L.~V.~E.}\ \bibnamefont
  {{Koopmans}}}, \bibinfo {author} {\bibfnamefont {M.}~\bibnamefont
  {{Langer}}}, \bibinfo {author} {\bibfnamefont {G.}~\bibnamefont {{Mellema}}},
  \bibinfo {author} {\bibfnamefont {H.}~\bibnamefont {{Tashiro}}}, \ and\
  \bibinfo {author} {\bibfnamefont {R.~M.}\ \bibnamefont {{Thomas}}},\ }\href
  {\doibase 10.1111/j.1365-2966.2009.16086.x} {\bibfield  {journal} {\bibinfo
  {journal} {\mnras}\ }\textbf {\bibinfo {volume} {402}},\ \bibinfo {pages}
  {2279} (\bibinfo {year} {2010})},\ \Eprint {http://arxiv.org/abs/0907.5179}
  {arXiv:0907.5179 [astro-ph.CO]} \BibitemShut {NoStop}%
\bibitem [{\citenamefont {{Tashiro}}\ \emph {et~al.}(2011)\citenamefont
  {{Tashiro}}, \citenamefont {{Aghanim}}, \citenamefont {{Langer}},
  \citenamefont {{Douspis}}, \citenamefont {{Zaroubi}},\ and\ \citenamefont
  {{Jeli{\'c}}}}]{TashiroAghanim_2011}%
  \BibitemOpen
  \bibfield  {author} {\bibinfo {author} {\bibfnamefont {H.}~\bibnamefont
  {{Tashiro}}}, \bibinfo {author} {\bibfnamefont {N.}~\bibnamefont
  {{Aghanim}}}, \bibinfo {author} {\bibfnamefont {M.}~\bibnamefont {{Langer}}},
  \bibinfo {author} {\bibfnamefont {M.}~\bibnamefont {{Douspis}}}, \bibinfo
  {author} {\bibfnamefont {S.}~\bibnamefont {{Zaroubi}}}, \ and\ \bibinfo
  {author} {\bibfnamefont {V.}~\bibnamefont {{Jeli{\'c}}}},\ }\href {\doibase
  10.1111/j.1365-2966.2011.18644.x} {\bibfield  {journal} {\bibinfo  {journal}
  {\mnras}\ }\textbf {\bibinfo {volume} {414}},\ \bibinfo {pages} {3424}
  (\bibinfo {year} {2011})}\BibitemShut {NoStop}%
\bibitem [{\citenamefont {{Paul}}\ \emph {et~al.}(2021)\citenamefont {{Paul}},
  \citenamefont {{Mukherjee}},\ and\ \citenamefont
  {{Choudhury}}}]{PaulMukherjee_2021}%
  \BibitemOpen
  \bibfield  {author} {\bibinfo {author} {\bibfnamefont {S.}~\bibnamefont
  {{Paul}}}, \bibinfo {author} {\bibfnamefont {S.}~\bibnamefont {{Mukherjee}}},
  \ and\ \bibinfo {author} {\bibfnamefont {T.~R.}\ \bibnamefont
  {{Choudhury}}},\ }\href {\doibase 10.1093/mnras/staa3221} {\bibfield
  {journal} {\bibinfo  {journal} {\mnras}\ }\textbf {\bibinfo {volume} {500}},\
  \bibinfo {pages} {232} (\bibinfo {year} {2021})},\ \Eprint
  {http://arxiv.org/abs/2005.05327} {arXiv:2005.05327 [astro-ph.CO]}
  \BibitemShut {NoStop}%
\bibitem [{\citenamefont {Plante}\ \emph {et~al.}(2020)\citenamefont {Plante},
  \citenamefont {Lidz}, \citenamefont {Aguirre},\ and\ \citenamefont
  {Kohn}}]{LaPlante_2020}%
  \BibitemOpen
  \bibfield  {author} {\bibinfo {author} {\bibfnamefont {P.~L.}\ \bibnamefont
  {Plante}}, \bibinfo {author} {\bibfnamefont {A.}~\bibnamefont {Lidz}},
  \bibinfo {author} {\bibfnamefont {J.}~\bibnamefont {Aguirre}}, \ and\
  \bibinfo {author} {\bibfnamefont {S.}~\bibnamefont {Kohn}},\ }\href {\doibase
  10.3847/1538-4357/aba2ed} {\bibfield  {journal} {\bibinfo  {journal} {The
  Astrophysical Journal}\ }\textbf {\bibinfo {volume} {899}},\ \bibinfo {pages}
  {40} (\bibinfo {year} {2020})}\BibitemShut {NoStop}%
\bibitem [{\citenamefont {Ma}\ \emph {et~al.}(2018)\citenamefont {Ma},
  \citenamefont {Helgason}, \citenamefont {Komatsu}, \citenamefont {Ciardi},\
  and\ \citenamefont {Ferrara}}]{Ma2018}%
  \BibitemOpen
  \bibfield  {author} {\bibinfo {author} {\bibfnamefont {Q.}~\bibnamefont
  {Ma}}, \bibinfo {author} {\bibfnamefont {K.}~\bibnamefont {Helgason}},
  \bibinfo {author} {\bibfnamefont {E.}~\bibnamefont {Komatsu}}, \bibinfo
  {author} {\bibfnamefont {B.}~\bibnamefont {Ciardi}}, \ and\ \bibinfo {author}
  {\bibfnamefont {A.}~\bibnamefont {Ferrara}},\ }\href {\doibase
  10.1093/mnras/sty543} {\bibfield  {journal} {\bibinfo  {journal} {Monthly
  Notices of the Royal Astronomical Society}\ }\textbf {\bibinfo {volume}
  {476}},\ \bibinfo {pages} {4025} (\bibinfo {year} {2018})},\ \Eprint
  {http://arxiv.org/abs/https://academic.oup.com/mnras/article-pdf/476/3/4025/24525331/sty543.pdf}
  {https://academic.oup.com/mnras/article-pdf/476/3/4025/24525331/sty543.pdf}
  \BibitemShut {NoStop}%
\bibitem [{\citenamefont {Alvarez}\ \emph {et~al.}(2006)\citenamefont
  {Alvarez}, \citenamefont {Komatsu}, \citenamefont {Dore},\ and\ \citenamefont
  {Shapiro}}]{Alvarez_2006}%
  \BibitemOpen
  \bibfield  {author} {\bibinfo {author} {\bibfnamefont {M.~A.}\ \bibnamefont
  {Alvarez}}, \bibinfo {author} {\bibfnamefont {E.}~\bibnamefont {Komatsu}},
  \bibinfo {author} {\bibfnamefont {O.}~\bibnamefont {Dore}}, \ and\ \bibinfo
  {author} {\bibfnamefont {P.~R.}\ \bibnamefont {Shapiro}},\ }\href {\doibase
  10.1086/504888} {\bibfield  {journal} {\bibinfo  {journal} {The Astrophysical
  Journal}\ }\textbf {\bibinfo {volume} {647}},\ \bibinfo {pages} {840}
  (\bibinfo {year} {2006})}\BibitemShut {NoStop}%
\bibitem [{\citenamefont {Furlanetto}\ \emph {et~al.}(2006)\citenamefont
  {Furlanetto}, \citenamefont {Oh},\ and\ \citenamefont
  {Briggs}}]{Furlanetto2006}%
  \BibitemOpen
  \bibfield  {author} {\bibinfo {author} {\bibfnamefont {S.~R.}\ \bibnamefont
  {Furlanetto}}, \bibinfo {author} {\bibfnamefont {S.~P.}\ \bibnamefont {Oh}},
  \ and\ \bibinfo {author} {\bibfnamefont {F.~H.}\ \bibnamefont {Briggs}},\
  }\href {\doibase 10.1016/j.physrep.2006.08.002} {\bibfield  {journal}
  {\bibinfo  {journal} {Science Reports}\ }\textbf {\bibinfo {volume} {433}},\
  \bibinfo {pages} {181} (\bibinfo {year} {2006})}\BibitemShut {NoStop}%
\bibitem [{\citenamefont {Morales}\ and\ \citenamefont
  {Wyithe}(2010)}]{Morales2010}%
  \BibitemOpen
  \bibfield  {author} {\bibinfo {author} {\bibfnamefont {M.~F.}\ \bibnamefont
  {Morales}}\ and\ \bibinfo {author} {\bibfnamefont {J.~S.~B.}\ \bibnamefont
  {Wyithe}},\ }\href {\doibase 10.1146/annurev-astro-081309-130936} {\bibfield
  {journal} {\bibinfo  {journal} {Annual Reviews of Astronomy and
  Astrophysics}\ }\textbf {\bibinfo {volume} {48}},\ \bibinfo {pages} {127}
  (\bibinfo {year} {2010})}\BibitemShut {NoStop}%
\bibitem [{\citenamefont {Pritchard}\ and\ \citenamefont
  {Loeb}(2012)}]{Pritchard2012}%
  \BibitemOpen
  \bibfield  {author} {\bibinfo {author} {\bibfnamefont {J.~R.}\ \bibnamefont
  {Pritchard}}\ and\ \bibinfo {author} {\bibfnamefont {A.}~\bibnamefont
  {Loeb}},\ }\href@noop {} {\bibfield  {journal} {\bibinfo  {journal} {Reports
  on Progress in Physics}\ }\textbf {\bibinfo {volume} {75}} (\bibinfo {year}
  {2012})}\BibitemShut {NoStop}%
\bibitem [{\citenamefont {{Mirocha}}\ \emph {et~al.}(2017)\citenamefont
  {{Mirocha}}, \citenamefont {{Furlanetto}},\ and\ \citenamefont
  {{Sun}}}]{MirochaFurlanetto_2017}%
  \BibitemOpen
  \bibfield  {author} {\bibinfo {author} {\bibfnamefont {J.}~\bibnamefont
  {{Mirocha}}}, \bibinfo {author} {\bibfnamefont {S.~R.}\ \bibnamefont
  {{Furlanetto}}}, \ and\ \bibinfo {author} {\bibfnamefont {G.}~\bibnamefont
  {{Sun}}},\ }\href {\doibase 10.1093/mnras/stw2412} {\bibfield  {journal}
  {\bibinfo  {journal} {\mnras}\ }\textbf {\bibinfo {volume} {464}},\ \bibinfo
  {pages} {1365} (\bibinfo {year} {2017})},\ \Eprint
  {http://arxiv.org/abs/1607.00386} {arXiv:1607.00386 [astro-ph.GA]}
  \BibitemShut {NoStop}%
\bibitem [{\citenamefont {{Heneka}}\ and\ \citenamefont
  {{Mesinger}}(2020)}]{HenekaMesinger_2020}%
  \BibitemOpen
  \bibfield  {author} {\bibinfo {author} {\bibfnamefont {C.}~\bibnamefont
  {{Heneka}}}\ and\ \bibinfo {author} {\bibfnamefont {A.}~\bibnamefont
  {{Mesinger}}},\ }\href {\doibase 10.1093/mnras/staa1517} {\bibfield
  {journal} {\bibinfo  {journal} {\mnras}\ }\textbf {\bibinfo {volume} {496}},\
  \bibinfo {pages} {581} (\bibinfo {year} {2020})},\ \Eprint
  {http://arxiv.org/abs/2004.10097} {arXiv:2004.10097 [astro-ph.CO]}
  \BibitemShut {NoStop}%
\bibitem [{\citenamefont {{Rogers}}\ and\ \citenamefont
  {{Bowman}}(2008)}]{2008AJ....136..641R}%
  \BibitemOpen
  \bibfield  {author} {\bibinfo {author} {\bibfnamefont {A.~E.~E.}\
  \bibnamefont {{Rogers}}}\ and\ \bibinfo {author} {\bibfnamefont {J.~D.}\
  \bibnamefont {{Bowman}}},\ }\href {\doibase 10.1088/0004-6256/136/2/641}
  {\bibfield  {journal} {\bibinfo  {journal} {\aj}\ }\textbf {\bibinfo {volume}
  {136}},\ \bibinfo {pages} {641} (\bibinfo {year} {2008})},\ \Eprint
  {http://arxiv.org/abs/0806.2868} {arXiv:0806.2868 [astro-ph]} \BibitemShut
  {NoStop}%
\bibitem [{\citenamefont {{Liu}}\ \emph {et~al.}(2013)\citenamefont {{Liu}},
  \citenamefont {{Pritchard}}, \citenamefont {{Tegmark}},\ and\ \citenamefont
  {{Loeb}}}]{Liu2013globalsig}%
  \BibitemOpen
  \bibfield  {author} {\bibinfo {author} {\bibfnamefont {A.}~\bibnamefont
  {{Liu}}}, \bibinfo {author} {\bibfnamefont {J.~R.}\ \bibnamefont
  {{Pritchard}}}, \bibinfo {author} {\bibfnamefont {M.}~\bibnamefont
  {{Tegmark}}}, \ and\ \bibinfo {author} {\bibfnamefont {A.}~\bibnamefont
  {{Loeb}}},\ }\href {\doibase 10.1103/PhysRevD.87.043002} {\bibfield
  {journal} {\bibinfo  {journal} {\prd}\ }\textbf {\bibinfo {volume} {87}},\
  \bibinfo {eid} {043002} (\bibinfo {year} {2013})},\ \Eprint
  {http://arxiv.org/abs/1211.3743} {arXiv:1211.3743 [astro-ph.CO]} \BibitemShut
  {NoStop}%
\bibitem [{\citenamefont {Liu}\ and\ \citenamefont {Tegmark}(2012)}]{Liu2012}%
  \BibitemOpen
  \bibfield  {author} {\bibinfo {author} {\bibfnamefont {A.}~\bibnamefont
  {Liu}}\ and\ \bibinfo {author} {\bibfnamefont {M.}~\bibnamefont {Tegmark}},\
  }\href {\doibase 10.1111/j.1365-2966.2011.19989.x} {\bibfield  {journal}
  {\bibinfo  {journal} {Monthly Notices of the Royal Astronomical Society}\
  }\textbf {\bibinfo {volume} {419}},\ \bibinfo {pages} {3491} (\bibinfo {year}
  {2012})},\ \Eprint
  {http://arxiv.org/abs/https://academic.oup.com/mnras/article-pdf/419/4/3491/9508051/mnras0419-3491.pdf}
  {https://academic.oup.com/mnras/article-pdf/419/4/3491/9508051/mnras0419-3491.pdf}
  \BibitemShut {NoStop}%
\bibitem [{\citenamefont {{Calabrese}}\ \emph {et~al.}(2014)\citenamefont
  {{Calabrese}}, \citenamefont {{Hlo{\v{z}}ek}}, \citenamefont {{Battaglia}},
  \citenamefont {{Bond}}, \citenamefont {{de Bernardis}}, \citenamefont
  {{Devlin}}, \citenamefont {{Hajian}}, \citenamefont {{Henderson}},
  \citenamefont {{Hil}}, \citenamefont {{Kosowsky}}, \citenamefont {{Louis}},
  \citenamefont {{McMahon}}, \citenamefont {{Moodley}}, \citenamefont
  {{Newburgh}}, \citenamefont {{Niemack}}, \citenamefont {{Page}},
  \citenamefont {{Partridge}}, \citenamefont {{Sehgal}}, \citenamefont
  {{Sievers}}, \citenamefont {{Spergel}}, \citenamefont {{Staggs}},
  \citenamefont {{Switzer}}, \citenamefont {{Trac}},\ and\ \citenamefont
  {{Wollack}}}]{CalabreseHlozek_2014}%
  \BibitemOpen
  \bibfield  {author} {\bibinfo {author} {\bibfnamefont {E.}~\bibnamefont
  {{Calabrese}}}, \bibinfo {author} {\bibfnamefont {R.}~\bibnamefont
  {{Hlo{\v{z}}ek}}}, \bibinfo {author} {\bibfnamefont {N.}~\bibnamefont
  {{Battaglia}}}, \bibinfo {author} {\bibfnamefont {J.~R.}\ \bibnamefont
  {{Bond}}}, \bibinfo {author} {\bibfnamefont {F.}~\bibnamefont {{de
  Bernardis}}}, \bibinfo {author} {\bibfnamefont {M.~J.}\ \bibnamefont
  {{Devlin}}}, \bibinfo {author} {\bibfnamefont {A.}~\bibnamefont {{Hajian}}},
  \bibinfo {author} {\bibfnamefont {S.}~\bibnamefont {{Henderson}}}, \bibinfo
  {author} {\bibfnamefont {J.~C.}\ \bibnamefont {{Hil}}}, \bibinfo {author}
  {\bibfnamefont {A.}~\bibnamefont {{Kosowsky}}}, \bibinfo {author}
  {\bibfnamefont {T.}~\bibnamefont {{Louis}}}, \bibinfo {author} {\bibfnamefont
  {J.}~\bibnamefont {{McMahon}}}, \bibinfo {author} {\bibfnamefont
  {K.}~\bibnamefont {{Moodley}}}, \bibinfo {author} {\bibfnamefont
  {L.}~\bibnamefont {{Newburgh}}}, \bibinfo {author} {\bibfnamefont {M.~D.}\
  \bibnamefont {{Niemack}}}, \bibinfo {author} {\bibfnamefont {L.~A.}\
  \bibnamefont {{Page}}}, \bibinfo {author} {\bibfnamefont {B.}~\bibnamefont
  {{Partridge}}}, \bibinfo {author} {\bibfnamefont {N.}~\bibnamefont
  {{Sehgal}}}, \bibinfo {author} {\bibfnamefont {J.~L.}\ \bibnamefont
  {{Sievers}}}, \bibinfo {author} {\bibfnamefont {D.~N.}\ \bibnamefont
  {{Spergel}}}, \bibinfo {author} {\bibfnamefont {S.~T.}\ \bibnamefont
  {{Staggs}}}, \bibinfo {author} {\bibfnamefont {E.~R.}\ \bibnamefont
  {{Switzer}}}, \bibinfo {author} {\bibfnamefont {H.}~\bibnamefont {{Trac}}}, \
  and\ \bibinfo {author} {\bibfnamefont {E.~J.}\ \bibnamefont {{Wollack}}},\
  }\href {\doibase 10.1088/1475-7516/2014/08/010} {\bibfield  {journal}
  {\bibinfo  {journal} {JCAP}\ }\textbf {\bibinfo {volume} {2014}},\ \bibinfo
  {eid} {010} (\bibinfo {year} {2014})},\ \Eprint
  {http://arxiv.org/abs/1406.4794} {arXiv:1406.4794 [astro-ph.CO]} \BibitemShut
  {NoStop}%
\bibitem [{\citenamefont {{Addison}}\ \emph {et~al.}(2012)\citenamefont
  {{Addison}}, \citenamefont {{Dunkley}},\ and\ \citenamefont
  {{Spergel}}}]{AddisonDunkley_2012}%
  \BibitemOpen
  \bibfield  {author} {\bibinfo {author} {\bibfnamefont {G.~E.}\ \bibnamefont
  {{Addison}}}, \bibinfo {author} {\bibfnamefont {J.}~\bibnamefont
  {{Dunkley}}}, \ and\ \bibinfo {author} {\bibfnamefont {D.~N.}\ \bibnamefont
  {{Spergel}}},\ }\href {\doibase 10.1111/j.1365-2966.2012.21664.x} {\bibfield
  {journal} {\bibinfo  {journal} {\mnras}\ }\textbf {\bibinfo {volume} {427}},\
  \bibinfo {pages} {1741} (\bibinfo {year} {2012})},\ \Eprint
  {http://arxiv.org/abs/1204.5927} {arXiv:1204.5927 [astro-ph.CO]} \BibitemShut
  {NoStop}%
\bibitem [{\citenamefont {{Reichardt}}\ \emph {et~al.}(2012)\citenamefont
  {{Reichardt}}, \citenamefont {{Shaw}}, \citenamefont {{Zahn}}, \citenamefont
  {{Aird}}, \citenamefont {{Benson}}, \citenamefont {{Bleem}}, \citenamefont
  {{Carlstrom}}, \citenamefont {{Chang}}, \citenamefont {{Cho}}, \citenamefont
  {{Crawford}}, \citenamefont {{Crites}}, \citenamefont {{de Haan}},
  \citenamefont {{Dobbs}}, \citenamefont {{Dudley}}, \citenamefont {{George}},
  \citenamefont {{Halverson}}, \citenamefont {{Holder}}, \citenamefont
  {{Holzapfel}}, \citenamefont {{Hoover}}, \citenamefont {{Hou}}, \citenamefont
  {{Hrubes}}, \citenamefont {{Joy}}, \citenamefont {{Keisler}}, \citenamefont
  {{Knox}}, \citenamefont {{Lee}}, \citenamefont {{Leitch}}, \citenamefont
  {{Lueker}}, \citenamefont {{Luong-Van}}, \citenamefont {{McMahon}},
  \citenamefont {{Mehl}}, \citenamefont {{Meyer}}, \citenamefont {{Millea}},
  \citenamefont {{Mohr}}, \citenamefont {{Montroy}}, \citenamefont {{Natoli}},
  \citenamefont {{Padin}}, \citenamefont {{Plagge}}, \citenamefont {{Pryke}},
  \citenamefont {{Ruhl}}, \citenamefont {{Schaffer}}, \citenamefont
  {{Shirokoff}}, \citenamefont {{Spieler}}, \citenamefont {{Staniszewski}},
  \citenamefont {{Stark}}, \citenamefont {{Story}}, \citenamefont {{van
  Engelen}}, \citenamefont {{Vanderlinde}}, \citenamefont {{Vieira}},\ and\
  \citenamefont {{Williamson}}}]{ReichardtShaw_2012}%
  \BibitemOpen
  \bibfield  {author} {\bibinfo {author} {\bibfnamefont {C.~L.}\ \bibnamefont
  {{Reichardt}}}, \bibinfo {author} {\bibfnamefont {L.}~\bibnamefont {{Shaw}}},
  \bibinfo {author} {\bibfnamefont {O.}~\bibnamefont {{Zahn}}}, \bibinfo
  {author} {\bibfnamefont {K.~A.}\ \bibnamefont {{Aird}}}, \bibinfo {author}
  {\bibfnamefont {B.~A.}\ \bibnamefont {{Benson}}}, \bibinfo {author}
  {\bibfnamefont {L.~E.}\ \bibnamefont {{Bleem}}}, \bibinfo {author}
  {\bibfnamefont {J.~E.}\ \bibnamefont {{Carlstrom}}}, \bibinfo {author}
  {\bibfnamefont {C.~L.}\ \bibnamefont {{Chang}}}, \bibinfo {author}
  {\bibfnamefont {H.~M.}\ \bibnamefont {{Cho}}}, \bibinfo {author}
  {\bibfnamefont {T.~M.}\ \bibnamefont {{Crawford}}}, \bibinfo {author}
  {\bibfnamefont {A.~T.}\ \bibnamefont {{Crites}}}, \bibinfo {author}
  {\bibfnamefont {T.}~\bibnamefont {{de Haan}}}, \bibinfo {author}
  {\bibfnamefont {M.~A.}\ \bibnamefont {{Dobbs}}}, \bibinfo {author}
  {\bibfnamefont {J.}~\bibnamefont {{Dudley}}}, \bibinfo {author}
  {\bibfnamefont {E.~M.}\ \bibnamefont {{George}}}, \bibinfo {author}
  {\bibfnamefont {N.~W.}\ \bibnamefont {{Halverson}}}, \bibinfo {author}
  {\bibfnamefont {G.~P.}\ \bibnamefont {{Holder}}}, \bibinfo {author}
  {\bibfnamefont {W.~L.}\ \bibnamefont {{Holzapfel}}}, \bibinfo {author}
  {\bibfnamefont {S.}~\bibnamefont {{Hoover}}}, \bibinfo {author}
  {\bibfnamefont {Z.}~\bibnamefont {{Hou}}}, \bibinfo {author} {\bibfnamefont
  {J.~D.}\ \bibnamefont {{Hrubes}}}, \bibinfo {author} {\bibfnamefont
  {M.}~\bibnamefont {{Joy}}}, \bibinfo {author} {\bibfnamefont
  {R.}~\bibnamefont {{Keisler}}}, \bibinfo {author} {\bibfnamefont
  {L.}~\bibnamefont {{Knox}}}, \bibinfo {author} {\bibfnamefont {A.~T.}\
  \bibnamefont {{Lee}}}, \bibinfo {author} {\bibfnamefont {E.~M.}\ \bibnamefont
  {{Leitch}}}, \bibinfo {author} {\bibfnamefont {M.}~\bibnamefont {{Lueker}}},
  \bibinfo {author} {\bibfnamefont {D.}~\bibnamefont {{Luong-Van}}}, \bibinfo
  {author} {\bibfnamefont {J.~J.}\ \bibnamefont {{McMahon}}}, \bibinfo {author}
  {\bibfnamefont {J.}~\bibnamefont {{Mehl}}}, \bibinfo {author} {\bibfnamefont
  {S.~S.}\ \bibnamefont {{Meyer}}}, \bibinfo {author} {\bibfnamefont
  {M.}~\bibnamefont {{Millea}}}, \bibinfo {author} {\bibfnamefont {J.~J.}\
  \bibnamefont {{Mohr}}}, \bibinfo {author} {\bibfnamefont {T.~E.}\
  \bibnamefont {{Montroy}}}, \bibinfo {author} {\bibfnamefont {T.}~\bibnamefont
  {{Natoli}}}, \bibinfo {author} {\bibfnamefont {S.}~\bibnamefont {{Padin}}},
  \bibinfo {author} {\bibfnamefont {T.}~\bibnamefont {{Plagge}}}, \bibinfo
  {author} {\bibfnamefont {C.}~\bibnamefont {{Pryke}}}, \bibinfo {author}
  {\bibfnamefont {J.~E.}\ \bibnamefont {{Ruhl}}}, \bibinfo {author}
  {\bibfnamefont {K.~K.}\ \bibnamefont {{Schaffer}}}, \bibinfo {author}
  {\bibfnamefont {E.}~\bibnamefont {{Shirokoff}}}, \bibinfo {author}
  {\bibfnamefont {H.~G.}\ \bibnamefont {{Spieler}}}, \bibinfo {author}
  {\bibfnamefont {Z.}~\bibnamefont {{Staniszewski}}}, \bibinfo {author}
  {\bibfnamefont {A.~A.}\ \bibnamefont {{Stark}}}, \bibinfo {author}
  {\bibfnamefont {K.}~\bibnamefont {{Story}}}, \bibinfo {author} {\bibfnamefont
  {A.}~\bibnamefont {{van Engelen}}}, \bibinfo {author} {\bibfnamefont
  {K.}~\bibnamefont {{Vanderlinde}}}, \bibinfo {author} {\bibfnamefont {J.~D.}\
  \bibnamefont {{Vieira}}}, \ and\ \bibinfo {author} {\bibfnamefont
  {R.}~\bibnamefont {{Williamson}}},\ }\href {\doibase
  10.1088/0004-637X/755/1/70} {\bibfield  {journal} {\bibinfo  {journal}
  {\apj}\ }\textbf {\bibinfo {volume} {755}},\ \bibinfo {eid} {70} (\bibinfo
  {year} {2012})},\ \Eprint {http://arxiv.org/abs/1111.0932} {arXiv:1111.0932
  [astro-ph.CO]} \BibitemShut {NoStop}%
\bibitem [{\citenamefont {{Sievers}}\ \emph {et~al.}(2013)\citenamefont
  {{Sievers}}, \citenamefont {{Hlozek}}, \citenamefont {{Nolta}}, \citenamefont
  {{Acquaviva}}, \citenamefont {{Addison}}, \citenamefont {{Ade}},
  \citenamefont {{Aguirre}}, \citenamefont {{Amiri}}, \citenamefont {{Appel}},
  \citenamefont {{Barrientos}}, \citenamefont {{Battistelli}}, \citenamefont
  {{Battaglia}}, \citenamefont {{Bond}}, \citenamefont {{Brown}}, \citenamefont
  {{Burger}}, \citenamefont {{Calabrese}}, \citenamefont {{Chervenak}},
  \citenamefont {{Crichton}}, \citenamefont {{Das}}, \citenamefont {{Devlin}},
  \citenamefont {{Dicker}}, \citenamefont {{Bertrand Doriese}}, \citenamefont
  {{Dunkley}}, \citenamefont {{D{\"u}nner}}, \citenamefont
  {{Essinger-Hileman}}, \citenamefont {{Faber}}, \citenamefont {{Fisher}},
  \citenamefont {{Fowler}}, \citenamefont {{Gallardo}}, \citenamefont
  {{Gordon}}, \citenamefont {{Gralla}}, \citenamefont {{Hajian}}, \citenamefont
  {{Halpern}}, \citenamefont {{Hasselfield}}, \citenamefont
  {{Hern{\'a}ndez-Monteagudo}}, \citenamefont {{Hill}}, \citenamefont
  {{Hilton}}, \citenamefont {{Hilton}}, \citenamefont {{Hincks}}, \citenamefont
  {{Holtz}}, \citenamefont {{Huffenberger}}, \citenamefont {{Hughes}},
  \citenamefont {{Hughes}}, \citenamefont {{Infante}}, \citenamefont {{Irwin}},
  \citenamefont {{Jacobson}}, \citenamefont {{Johnstone}}, \citenamefont
  {{Baptiste Juin}}, \citenamefont {{Kaul}}, \citenamefont {{Klein}},
  \citenamefont {{Kosowsky}}, \citenamefont {{Lau}}, \citenamefont {{Limon}},
  \citenamefont {{Lin}}, \citenamefont {{Louis}}, \citenamefont {{Lupton}},
  \citenamefont {{Marriage}}, \citenamefont {{Marsden}}, \citenamefont
  {{Martocci}}, \citenamefont {{Mauskopf}}, \citenamefont {{McLaren}},
  \citenamefont {{Menanteau}}, \citenamefont {{Moodley}}, \citenamefont
  {{Moseley}}, \citenamefont {{Netterfield}}, \citenamefont {{Niemack}},
  \citenamefont {{Page}}, \citenamefont {{Page}}, \citenamefont {{Parker}},
  \citenamefont {{Partridge}}, \citenamefont {{Plimpton}}, \citenamefont
  {{Quintana}}, \citenamefont {{Reese}}, \citenamefont {{Reid}}, \citenamefont
  {{Rojas}}, \citenamefont {{Sehgal}}, \citenamefont {{Sherwin}}, \citenamefont
  {{Schmitt}}, \citenamefont {{Spergel}}, \citenamefont {{Staggs}},
  \citenamefont {{Stryzak}}, \citenamefont {{Swetz}}, \citenamefont
  {{Switzer}}, \citenamefont {{Thornton}}, \citenamefont {{Trac}},
  \citenamefont {{Tucker}}, \citenamefont {{Uehara}}, \citenamefont
  {{Visnjic}}, \citenamefont {{Warne}}, \citenamefont {{Wilson}}, \citenamefont
  {{Wollack}}, \citenamefont {{Zhao}},\ and\ \citenamefont
  {{Zunckel}}}]{SieversHlozek_2013}%
  \BibitemOpen
  \bibfield  {author} {\bibinfo {author} {\bibfnamefont {J.~L.}\ \bibnamefont
  {{Sievers}}}, \bibinfo {author} {\bibfnamefont {R.~A.}\ \bibnamefont
  {{Hlozek}}}, \bibinfo {author} {\bibfnamefont {M.~R.}\ \bibnamefont
  {{Nolta}}}, \bibinfo {author} {\bibfnamefont {V.}~\bibnamefont
  {{Acquaviva}}}, \bibinfo {author} {\bibfnamefont {G.~E.}\ \bibnamefont
  {{Addison}}}, \bibinfo {author} {\bibfnamefont {P.~A.~R.}\ \bibnamefont
  {{Ade}}}, \bibinfo {author} {\bibfnamefont {P.}~\bibnamefont {{Aguirre}}},
  \bibinfo {author} {\bibfnamefont {M.}~\bibnamefont {{Amiri}}}, \bibinfo
  {author} {\bibfnamefont {J.~W.}\ \bibnamefont {{Appel}}}, \bibinfo {author}
  {\bibfnamefont {L.~F.}\ \bibnamefont {{Barrientos}}}, \bibinfo {author}
  {\bibfnamefont {E.~S.}\ \bibnamefont {{Battistelli}}}, \bibinfo {author}
  {\bibfnamefont {N.}~\bibnamefont {{Battaglia}}}, \bibinfo {author}
  {\bibfnamefont {J.~R.}\ \bibnamefont {{Bond}}}, \bibinfo {author}
  {\bibfnamefont {B.}~\bibnamefont {{Brown}}}, \bibinfo {author} {\bibfnamefont
  {B.}~\bibnamefont {{Burger}}}, \bibinfo {author} {\bibfnamefont
  {E.}~\bibnamefont {{Calabrese}}}, \bibinfo {author} {\bibfnamefont
  {J.}~\bibnamefont {{Chervenak}}}, \bibinfo {author} {\bibfnamefont
  {D.}~\bibnamefont {{Crichton}}}, \bibinfo {author} {\bibfnamefont
  {S.}~\bibnamefont {{Das}}}, \bibinfo {author} {\bibfnamefont {M.~J.}\
  \bibnamefont {{Devlin}}}, \bibinfo {author} {\bibfnamefont {S.~R.}\
  \bibnamefont {{Dicker}}}, \bibinfo {author} {\bibfnamefont {W.}~\bibnamefont
  {{Bertrand Doriese}}}, \bibinfo {author} {\bibfnamefont {J.}~\bibnamefont
  {{Dunkley}}}, \bibinfo {author} {\bibfnamefont {R.}~\bibnamefont
  {{D{\"u}nner}}}, \bibinfo {author} {\bibfnamefont {T.}~\bibnamefont
  {{Essinger-Hileman}}}, \bibinfo {author} {\bibfnamefont {D.}~\bibnamefont
  {{Faber}}}, \bibinfo {author} {\bibfnamefont {R.~P.}\ \bibnamefont
  {{Fisher}}}, \bibinfo {author} {\bibfnamefont {J.~W.}\ \bibnamefont
  {{Fowler}}}, \bibinfo {author} {\bibfnamefont {P.}~\bibnamefont
  {{Gallardo}}}, \bibinfo {author} {\bibfnamefont {M.~S.}\ \bibnamefont
  {{Gordon}}}, \bibinfo {author} {\bibfnamefont {M.~B.}\ \bibnamefont
  {{Gralla}}}, \bibinfo {author} {\bibfnamefont {A.}~\bibnamefont {{Hajian}}},
  \bibinfo {author} {\bibfnamefont {M.}~\bibnamefont {{Halpern}}}, \bibinfo
  {author} {\bibfnamefont {M.}~\bibnamefont {{Hasselfield}}}, \bibinfo {author}
  {\bibfnamefont {C.}~\bibnamefont {{Hern{\'a}ndez-Monteagudo}}}, \bibinfo
  {author} {\bibfnamefont {J.~C.}\ \bibnamefont {{Hill}}}, \bibinfo {author}
  {\bibfnamefont {G.~C.}\ \bibnamefont {{Hilton}}}, \bibinfo {author}
  {\bibfnamefont {M.}~\bibnamefont {{Hilton}}}, \bibinfo {author}
  {\bibfnamefont {A.~D.}\ \bibnamefont {{Hincks}}}, \bibinfo {author}
  {\bibfnamefont {D.}~\bibnamefont {{Holtz}}}, \bibinfo {author} {\bibfnamefont
  {K.~M.}\ \bibnamefont {{Huffenberger}}}, \bibinfo {author} {\bibfnamefont
  {D.~H.}\ \bibnamefont {{Hughes}}}, \bibinfo {author} {\bibfnamefont {J.~P.}\
  \bibnamefont {{Hughes}}}, \bibinfo {author} {\bibfnamefont {L.}~\bibnamefont
  {{Infante}}}, \bibinfo {author} {\bibfnamefont {K.~D.}\ \bibnamefont
  {{Irwin}}}, \bibinfo {author} {\bibfnamefont {D.~R.}\ \bibnamefont
  {{Jacobson}}}, \bibinfo {author} {\bibfnamefont {B.}~\bibnamefont
  {{Johnstone}}}, \bibinfo {author} {\bibfnamefont {J.}~\bibnamefont {{Baptiste
  Juin}}}, \bibinfo {author} {\bibfnamefont {M.}~\bibnamefont {{Kaul}}},
  \bibinfo {author} {\bibfnamefont {J.}~\bibnamefont {{Klein}}}, \bibinfo
  {author} {\bibfnamefont {A.}~\bibnamefont {{Kosowsky}}}, \bibinfo {author}
  {\bibfnamefont {J.~M.}\ \bibnamefont {{Lau}}}, \bibinfo {author}
  {\bibfnamefont {M.}~\bibnamefont {{Limon}}}, \bibinfo {author} {\bibfnamefont
  {Y.-T.}\ \bibnamefont {{Lin}}}, \bibinfo {author} {\bibfnamefont
  {T.}~\bibnamefont {{Louis}}}, \bibinfo {author} {\bibfnamefont {R.~H.}\
  \bibnamefont {{Lupton}}}, \bibinfo {author} {\bibfnamefont {T.~A.}\
  \bibnamefont {{Marriage}}}, \bibinfo {author} {\bibfnamefont
  {D.}~\bibnamefont {{Marsden}}}, \bibinfo {author} {\bibfnamefont
  {K.}~\bibnamefont {{Martocci}}}, \bibinfo {author} {\bibfnamefont
  {P.}~\bibnamefont {{Mauskopf}}}, \bibinfo {author} {\bibfnamefont
  {M.}~\bibnamefont {{McLaren}}}, \bibinfo {author} {\bibfnamefont
  {F.}~\bibnamefont {{Menanteau}}}, \bibinfo {author} {\bibfnamefont
  {K.}~\bibnamefont {{Moodley}}}, \bibinfo {author} {\bibfnamefont
  {H.}~\bibnamefont {{Moseley}}}, \bibinfo {author} {\bibfnamefont {C.~B.}\
  \bibnamefont {{Netterfield}}}, \bibinfo {author} {\bibfnamefont {M.~D.}\
  \bibnamefont {{Niemack}}}, \bibinfo {author} {\bibfnamefont {L.~A.}\
  \bibnamefont {{Page}}}, \bibinfo {author} {\bibfnamefont {W.~A.}\
  \bibnamefont {{Page}}}, \bibinfo {author} {\bibfnamefont {L.}~\bibnamefont
  {{Parker}}}, \bibinfo {author} {\bibfnamefont {B.}~\bibnamefont
  {{Partridge}}}, \bibinfo {author} {\bibfnamefont {R.}~\bibnamefont
  {{Plimpton}}}, \bibinfo {author} {\bibfnamefont {H.}~\bibnamefont
  {{Quintana}}}, \bibinfo {author} {\bibfnamefont {E.~D.}\ \bibnamefont
  {{Reese}}}, \bibinfo {author} {\bibfnamefont {B.}~\bibnamefont {{Reid}}},
  \bibinfo {author} {\bibfnamefont {F.}~\bibnamefont {{Rojas}}}, \bibinfo
  {author} {\bibfnamefont {N.}~\bibnamefont {{Sehgal}}}, \bibinfo {author}
  {\bibfnamefont {B.~D.}\ \bibnamefont {{Sherwin}}}, \bibinfo {author}
  {\bibfnamefont {B.~L.}\ \bibnamefont {{Schmitt}}}, \bibinfo {author}
  {\bibfnamefont {D.~N.}\ \bibnamefont {{Spergel}}}, \bibinfo {author}
  {\bibfnamefont {S.~T.}\ \bibnamefont {{Staggs}}}, \bibinfo {author}
  {\bibfnamefont {O.}~\bibnamefont {{Stryzak}}}, \bibinfo {author}
  {\bibfnamefont {D.~S.}\ \bibnamefont {{Swetz}}}, \bibinfo {author}
  {\bibfnamefont {E.~R.}\ \bibnamefont {{Switzer}}}, \bibinfo {author}
  {\bibfnamefont {R.}~\bibnamefont {{Thornton}}}, \bibinfo {author}
  {\bibfnamefont {H.}~\bibnamefont {{Trac}}}, \bibinfo {author} {\bibfnamefont
  {C.}~\bibnamefont {{Tucker}}}, \bibinfo {author} {\bibfnamefont
  {M.}~\bibnamefont {{Uehara}}}, \bibinfo {author} {\bibfnamefont
  {K.}~\bibnamefont {{Visnjic}}}, \bibinfo {author} {\bibfnamefont
  {R.}~\bibnamefont {{Warne}}}, \bibinfo {author} {\bibfnamefont
  {G.}~\bibnamefont {{Wilson}}}, \bibinfo {author} {\bibfnamefont
  {E.}~\bibnamefont {{Wollack}}}, \bibinfo {author} {\bibfnamefont
  {Y.}~\bibnamefont {{Zhao}}}, \ and\ \bibinfo {author} {\bibfnamefont
  {C.}~\bibnamefont {{Zunckel}}},\ }\href {\doibase
  10.1088/1475-7516/2013/10/060} {\bibfield  {journal} {\bibinfo  {journal}
  {\jcap}\ }\textbf {\bibinfo {volume} {2013}},\ \bibinfo {eid} {060} (\bibinfo
  {year} {2013})},\ \Eprint {http://arxiv.org/abs/1301.0824} {arXiv:1301.0824
  [astro-ph.CO]} \BibitemShut {NoStop}%
\bibitem [{\citenamefont {{George}}\ \emph {et~al.}(2015)\citenamefont
  {{George}}, \citenamefont {{Reichardt}}, \citenamefont {{Aird}},
  \citenamefont {{Benson}}, \citenamefont {{Bleem}}, \citenamefont
  {{Carlstrom}}, \citenamefont {{Chang}}, \citenamefont {{Cho}}, \citenamefont
  {{Crawford}}, \citenamefont {{Crites}}, \citenamefont {{de Haan}},
  \citenamefont {{Dobbs}}, \citenamefont {{Dudley}}, \citenamefont
  {{Halverson}}, \citenamefont {{Harrington}}, \citenamefont {{Holder}},
  \citenamefont {{Holzapfel}}, \citenamefont {{Hou}}, \citenamefont {{Hrubes}},
  \citenamefont {{Keisler}}, \citenamefont {{Knox}}, \citenamefont {{Lee}},
  \citenamefont {{Leitch}}, \citenamefont {{Lueker}}, \citenamefont
  {{Luong-Van}}, \citenamefont {{McMahon}}, \citenamefont {{Mehl}},
  \citenamefont {{Meyer}}, \citenamefont {{Millea}}, \citenamefont {{Mocanu}},
  \citenamefont {{Mohr}}, \citenamefont {{Montroy}}, \citenamefont {{Padin}},
  \citenamefont {{Plagge}}, \citenamefont {{Pryke}}, \citenamefont {{Ruhl}},
  \citenamefont {{Schaffer}}, \citenamefont {{Shaw}}, \citenamefont
  {{Shirokoff}}, \citenamefont {{Spieler}}, \citenamefont {{Staniszewski}},
  \citenamefont {{Stark}}, \citenamefont {{Story}}, \citenamefont {{van
  Engelen}}, \citenamefont {{Vanderlinde}}, \citenamefont {{Vieira}},
  \citenamefont {{Williamson}},\ and\ \citenamefont
  {{Zahn}}}]{GeorgeReichardt_2015}%
  \BibitemOpen
  \bibfield  {author} {\bibinfo {author} {\bibfnamefont {E.~M.}\ \bibnamefont
  {{George}}}, \bibinfo {author} {\bibfnamefont {C.~L.}\ \bibnamefont
  {{Reichardt}}}, \bibinfo {author} {\bibfnamefont {K.~A.}\ \bibnamefont
  {{Aird}}}, \bibinfo {author} {\bibfnamefont {B.~A.}\ \bibnamefont
  {{Benson}}}, \bibinfo {author} {\bibfnamefont {L.~E.}\ \bibnamefont
  {{Bleem}}}, \bibinfo {author} {\bibfnamefont {J.~E.}\ \bibnamefont
  {{Carlstrom}}}, \bibinfo {author} {\bibfnamefont {C.~L.}\ \bibnamefont
  {{Chang}}}, \bibinfo {author} {\bibfnamefont {H.~M.}\ \bibnamefont {{Cho}}},
  \bibinfo {author} {\bibfnamefont {T.~M.}\ \bibnamefont {{Crawford}}},
  \bibinfo {author} {\bibfnamefont {A.~T.}\ \bibnamefont {{Crites}}}, \bibinfo
  {author} {\bibfnamefont {T.}~\bibnamefont {{de Haan}}}, \bibinfo {author}
  {\bibfnamefont {M.~A.}\ \bibnamefont {{Dobbs}}}, \bibinfo {author}
  {\bibfnamefont {J.}~\bibnamefont {{Dudley}}}, \bibinfo {author}
  {\bibfnamefont {N.~W.}\ \bibnamefont {{Halverson}}}, \bibinfo {author}
  {\bibfnamefont {N.~L.}\ \bibnamefont {{Harrington}}}, \bibinfo {author}
  {\bibfnamefont {G.~P.}\ \bibnamefont {{Holder}}}, \bibinfo {author}
  {\bibfnamefont {W.~L.}\ \bibnamefont {{Holzapfel}}}, \bibinfo {author}
  {\bibfnamefont {Z.}~\bibnamefont {{Hou}}}, \bibinfo {author} {\bibfnamefont
  {J.~D.}\ \bibnamefont {{Hrubes}}}, \bibinfo {author} {\bibfnamefont
  {R.}~\bibnamefont {{Keisler}}}, \bibinfo {author} {\bibfnamefont
  {L.}~\bibnamefont {{Knox}}}, \bibinfo {author} {\bibfnamefont {A.~T.}\
  \bibnamefont {{Lee}}}, \bibinfo {author} {\bibfnamefont {E.~M.}\ \bibnamefont
  {{Leitch}}}, \bibinfo {author} {\bibfnamefont {M.}~\bibnamefont {{Lueker}}},
  \bibinfo {author} {\bibfnamefont {D.}~\bibnamefont {{Luong-Van}}}, \bibinfo
  {author} {\bibfnamefont {J.~J.}\ \bibnamefont {{McMahon}}}, \bibinfo {author}
  {\bibfnamefont {J.}~\bibnamefont {{Mehl}}}, \bibinfo {author} {\bibfnamefont
  {S.~S.}\ \bibnamefont {{Meyer}}}, \bibinfo {author} {\bibfnamefont
  {M.}~\bibnamefont {{Millea}}}, \bibinfo {author} {\bibfnamefont {L.~M.}\
  \bibnamefont {{Mocanu}}}, \bibinfo {author} {\bibfnamefont {J.~J.}\
  \bibnamefont {{Mohr}}}, \bibinfo {author} {\bibfnamefont {T.~E.}\
  \bibnamefont {{Montroy}}}, \bibinfo {author} {\bibfnamefont {S.}~\bibnamefont
  {{Padin}}}, \bibinfo {author} {\bibfnamefont {T.}~\bibnamefont {{Plagge}}},
  \bibinfo {author} {\bibfnamefont {C.}~\bibnamefont {{Pryke}}}, \bibinfo
  {author} {\bibfnamefont {J.~E.}\ \bibnamefont {{Ruhl}}}, \bibinfo {author}
  {\bibfnamefont {K.~K.}\ \bibnamefont {{Schaffer}}}, \bibinfo {author}
  {\bibfnamefont {L.}~\bibnamefont {{Shaw}}}, \bibinfo {author} {\bibfnamefont
  {E.}~\bibnamefont {{Shirokoff}}}, \bibinfo {author} {\bibfnamefont {H.~G.}\
  \bibnamefont {{Spieler}}}, \bibinfo {author} {\bibfnamefont {Z.}~\bibnamefont
  {{Staniszewski}}}, \bibinfo {author} {\bibfnamefont {A.~A.}\ \bibnamefont
  {{Stark}}}, \bibinfo {author} {\bibfnamefont {K.~T.}\ \bibnamefont
  {{Story}}}, \bibinfo {author} {\bibfnamefont {A.}~\bibnamefont {{van
  Engelen}}}, \bibinfo {author} {\bibfnamefont {K.}~\bibnamefont
  {{Vanderlinde}}}, \bibinfo {author} {\bibfnamefont {J.~D.}\ \bibnamefont
  {{Vieira}}}, \bibinfo {author} {\bibfnamefont {R.}~\bibnamefont
  {{Williamson}}}, \ and\ \bibinfo {author} {\bibfnamefont {O.}~\bibnamefont
  {{Zahn}}},\ }\href {\doibase 10.1088/0004-637X/799/2/177} {\bibfield
  {journal} {\bibinfo  {journal} {ApJ}\ }\textbf {\bibinfo {volume} {799}},\
  \bibinfo {eid} {177} (\bibinfo {year} {2015})},\ \Eprint
  {http://arxiv.org/abs/1408.3161} {arXiv:1408.3161 [astro-ph.CO]} \BibitemShut
  {NoStop}%
\bibitem [{\citenamefont {{Keisler}}\ \emph {et~al.}(2011)\citenamefont
  {{Keisler}}, \citenamefont {{Reichardt}}, \citenamefont {{Aird}},
  \citenamefont {{Benson}}, \citenamefont {{Bleem}}, \citenamefont
  {{Carlstrom}}, \citenamefont {{Chang}}, \citenamefont {{Cho}}, \citenamefont
  {{Crawford}}, \citenamefont {{Crites}}, \citenamefont {{de Haan}},
  \citenamefont {{Dobbs}}, \citenamefont {{Dudley}}, \citenamefont {{George}},
  \citenamefont {{Halverson}}, \citenamefont {{Holder}}, \citenamefont
  {{Holzapfel}}, \citenamefont {{Hoover}}, \citenamefont {{Hou}}, \citenamefont
  {{Hrubes}}, \citenamefont {{Joy}}, \citenamefont {{Knox}}, \citenamefont
  {{Lee}}, \citenamefont {{Leitch}}, \citenamefont {{Lueker}}, \citenamefont
  {{Luong-Van}}, \citenamefont {{McMahon}}, \citenamefont {{Mehl}},
  \citenamefont {{Meyer}}, \citenamefont {{Millea}}, \citenamefont {{Mohr}},
  \citenamefont {{Montroy}}, \citenamefont {{Natoli}}, \citenamefont {{Padin}},
  \citenamefont {{Plagge}}, \citenamefont {{Pryke}}, \citenamefont {{Ruhl}},
  \citenamefont {{Schaffer}}, \citenamefont {{Shaw}}, \citenamefont
  {{Shirokoff}}, \citenamefont {{Spieler}}, \citenamefont {{Staniszewski}},
  \citenamefont {{Stark}}, \citenamefont {{Story}}, \citenamefont {{van
  Engelen}}, \citenamefont {{Vanderlinde}}, \citenamefont {{Vieira}},
  \citenamefont {{Williamson}},\ and\ \citenamefont
  {{Zahn}}}]{KeislerReichardt_2011}%
  \BibitemOpen
  \bibfield  {author} {\bibinfo {author} {\bibfnamefont {R.}~\bibnamefont
  {{Keisler}}}, \bibinfo {author} {\bibfnamefont {C.~L.}\ \bibnamefont
  {{Reichardt}}}, \bibinfo {author} {\bibfnamefont {K.~A.}\ \bibnamefont
  {{Aird}}}, \bibinfo {author} {\bibfnamefont {B.~A.}\ \bibnamefont
  {{Benson}}}, \bibinfo {author} {\bibfnamefont {L.~E.}\ \bibnamefont
  {{Bleem}}}, \bibinfo {author} {\bibfnamefont {J.~E.}\ \bibnamefont
  {{Carlstrom}}}, \bibinfo {author} {\bibfnamefont {C.~L.}\ \bibnamefont
  {{Chang}}}, \bibinfo {author} {\bibfnamefont {H.~M.}\ \bibnamefont {{Cho}}},
  \bibinfo {author} {\bibfnamefont {T.~M.}\ \bibnamefont {{Crawford}}},
  \bibinfo {author} {\bibfnamefont {A.~T.}\ \bibnamefont {{Crites}}}, \bibinfo
  {author} {\bibfnamefont {T.}~\bibnamefont {{de Haan}}}, \bibinfo {author}
  {\bibfnamefont {M.~A.}\ \bibnamefont {{Dobbs}}}, \bibinfo {author}
  {\bibfnamefont {J.}~\bibnamefont {{Dudley}}}, \bibinfo {author}
  {\bibfnamefont {E.~M.}\ \bibnamefont {{George}}}, \bibinfo {author}
  {\bibfnamefont {N.~W.}\ \bibnamefont {{Halverson}}}, \bibinfo {author}
  {\bibfnamefont {G.~P.}\ \bibnamefont {{Holder}}}, \bibinfo {author}
  {\bibfnamefont {W.~L.}\ \bibnamefont {{Holzapfel}}}, \bibinfo {author}
  {\bibfnamefont {S.}~\bibnamefont {{Hoover}}}, \bibinfo {author}
  {\bibfnamefont {Z.}~\bibnamefont {{Hou}}}, \bibinfo {author} {\bibfnamefont
  {J.~D.}\ \bibnamefont {{Hrubes}}}, \bibinfo {author} {\bibfnamefont
  {M.}~\bibnamefont {{Joy}}}, \bibinfo {author} {\bibfnamefont
  {L.}~\bibnamefont {{Knox}}}, \bibinfo {author} {\bibfnamefont {A.~T.}\
  \bibnamefont {{Lee}}}, \bibinfo {author} {\bibfnamefont {E.~M.}\ \bibnamefont
  {{Leitch}}}, \bibinfo {author} {\bibfnamefont {M.}~\bibnamefont {{Lueker}}},
  \bibinfo {author} {\bibfnamefont {D.}~\bibnamefont {{Luong-Van}}}, \bibinfo
  {author} {\bibfnamefont {J.~J.}\ \bibnamefont {{McMahon}}}, \bibinfo {author}
  {\bibfnamefont {J.}~\bibnamefont {{Mehl}}}, \bibinfo {author} {\bibfnamefont
  {S.~S.}\ \bibnamefont {{Meyer}}}, \bibinfo {author} {\bibfnamefont
  {M.}~\bibnamefont {{Millea}}}, \bibinfo {author} {\bibfnamefont {J.~J.}\
  \bibnamefont {{Mohr}}}, \bibinfo {author} {\bibfnamefont {T.~E.}\
  \bibnamefont {{Montroy}}}, \bibinfo {author} {\bibfnamefont {T.}~\bibnamefont
  {{Natoli}}}, \bibinfo {author} {\bibfnamefont {S.}~\bibnamefont {{Padin}}},
  \bibinfo {author} {\bibfnamefont {T.}~\bibnamefont {{Plagge}}}, \bibinfo
  {author} {\bibfnamefont {C.}~\bibnamefont {{Pryke}}}, \bibinfo {author}
  {\bibfnamefont {J.~E.}\ \bibnamefont {{Ruhl}}}, \bibinfo {author}
  {\bibfnamefont {K.~K.}\ \bibnamefont {{Schaffer}}}, \bibinfo {author}
  {\bibfnamefont {L.}~\bibnamefont {{Shaw}}}, \bibinfo {author} {\bibfnamefont
  {E.}~\bibnamefont {{Shirokoff}}}, \bibinfo {author} {\bibfnamefont {H.~G.}\
  \bibnamefont {{Spieler}}}, \bibinfo {author} {\bibfnamefont {Z.}~\bibnamefont
  {{Staniszewski}}}, \bibinfo {author} {\bibfnamefont {A.~A.}\ \bibnamefont
  {{Stark}}}, \bibinfo {author} {\bibfnamefont {K.}~\bibnamefont {{Story}}},
  \bibinfo {author} {\bibfnamefont {A.}~\bibnamefont {{van Engelen}}}, \bibinfo
  {author} {\bibfnamefont {K.}~\bibnamefont {{Vanderlinde}}}, \bibinfo {author}
  {\bibfnamefont {J.~D.}\ \bibnamefont {{Vieira}}}, \bibinfo {author}
  {\bibfnamefont {R.}~\bibnamefont {{Williamson}}}, \ and\ \bibinfo {author}
  {\bibfnamefont {O.}~\bibnamefont {{Zahn}}},\ }\href {\doibase
  10.1088/0004-637X/743/1/28} {\bibfield  {journal} {\bibinfo  {journal}
  {\apj}\ }\textbf {\bibinfo {volume} {743}},\ \bibinfo {eid} {28} (\bibinfo
  {year} {2011})},\ \Eprint {http://arxiv.org/abs/1105.3182} {arXiv:1105.3182
  [astro-ph.CO]} \BibitemShut {NoStop}%
\bibitem [{\citenamefont {{Tegmark}}\ \emph {et~al.}(2006)\citenamefont
  {{Tegmark}}, \citenamefont {{Eisenstein}}, \citenamefont {{Strauss}},
  \citenamefont {{Weinberg}}, \citenamefont {{Blanton}}, \citenamefont
  {{Frieman}}, \citenamefont {{Fukugita}}, \citenamefont {{Gunn}},
  \citenamefont {{Hamilton}}, \citenamefont {{Knapp}}, \citenamefont
  {{Nichol}}, \citenamefont {{Ostriker}}, \citenamefont {{Padmanabhan}},
  \citenamefont {{Percival}}, \citenamefont {{Schlegel}}, \citenamefont
  {{Schneider}}, \citenamefont {{Scoccimarro}}, \citenamefont {{Seljak}},
  \citenamefont {{Seo}}, \citenamefont {{Swanson}}, \citenamefont {{Szalay}},
  \citenamefont {{Vogeley}}, \citenamefont {{Yoo}}, \citenamefont {{Zehavi}},
  \citenamefont {{Abazajian}}, \citenamefont {{Anderson}}, \citenamefont
  {{Annis}}, \citenamefont {{Bahcall}}, \citenamefont {{Bassett}},
  \citenamefont {{Berlind}}, \citenamefont {{Brinkmann}}, \citenamefont
  {{Budavari}}, \citenamefont {{Castander}}, \citenamefont {{Connolly}},
  \citenamefont {{Csabai}}, \citenamefont {{Doi}}, \citenamefont
  {{Finkbeiner}}, \citenamefont {{Gillespie}}, \citenamefont {{Glazebrook}},
  \citenamefont {{Hennessy}}, \citenamefont {{Hogg}}, \citenamefont
  {{Ivezi{\'c}}}, \citenamefont {{Jain}}, \citenamefont {{Johnston}},
  \citenamefont {{Kent}}, \citenamefont {{Lamb}}, \citenamefont {{Lee}},
  \citenamefont {{Lin}}, \citenamefont {{Loveday}}, \citenamefont {{Lupton}},
  \citenamefont {{Munn}}, \citenamefont {{Pan}}, \citenamefont {{Park}},
  \citenamefont {{Peoples}}, \citenamefont {{Pier}}, \citenamefont {{Pope}},
  \citenamefont {{Richmond}}, \citenamefont {{Rockosi}}, \citenamefont
  {{Scranton}}, \citenamefont {{Sheth}}, \citenamefont {{Stebbins}},
  \citenamefont {{Stoughton}}, \citenamefont {{Szapudi}}, \citenamefont
  {{Tucker}}, \citenamefont {{vanden Berk}}, \citenamefont {{Yanny}},\ and\
  \citenamefont {{York}}}]{Tegmark2006}%
  \BibitemOpen
  \bibfield  {author} {\bibinfo {author} {\bibfnamefont {M.}~\bibnamefont
  {{Tegmark}}}, \bibinfo {author} {\bibfnamefont {D.~J.}\ \bibnamefont
  {{Eisenstein}}}, \bibinfo {author} {\bibfnamefont {M.~A.}\ \bibnamefont
  {{Strauss}}}, \bibinfo {author} {\bibfnamefont {D.~H.}\ \bibnamefont
  {{Weinberg}}}, \bibinfo {author} {\bibfnamefont {M.~R.}\ \bibnamefont
  {{Blanton}}}, \bibinfo {author} {\bibfnamefont {J.~A.}\ \bibnamefont
  {{Frieman}}}, \bibinfo {author} {\bibfnamefont {M.}~\bibnamefont
  {{Fukugita}}}, \bibinfo {author} {\bibfnamefont {J.~E.}\ \bibnamefont
  {{Gunn}}}, \bibinfo {author} {\bibfnamefont {A.~J.~S.}\ \bibnamefont
  {{Hamilton}}}, \bibinfo {author} {\bibfnamefont {G.~R.}\ \bibnamefont
  {{Knapp}}}, \bibinfo {author} {\bibfnamefont {R.~C.}\ \bibnamefont
  {{Nichol}}}, \bibinfo {author} {\bibfnamefont {J.~P.}\ \bibnamefont
  {{Ostriker}}}, \bibinfo {author} {\bibfnamefont {N.}~\bibnamefont
  {{Padmanabhan}}}, \bibinfo {author} {\bibfnamefont {W.~J.}\ \bibnamefont
  {{Percival}}}, \bibinfo {author} {\bibfnamefont {D.~J.}\ \bibnamefont
  {{Schlegel}}}, \bibinfo {author} {\bibfnamefont {D.~P.}\ \bibnamefont
  {{Schneider}}}, \bibinfo {author} {\bibfnamefont {R.}~\bibnamefont
  {{Scoccimarro}}}, \bibinfo {author} {\bibfnamefont {U.}~\bibnamefont
  {{Seljak}}}, \bibinfo {author} {\bibfnamefont {H.-J.}\ \bibnamefont {{Seo}}},
  \bibinfo {author} {\bibfnamefont {M.}~\bibnamefont {{Swanson}}}, \bibinfo
  {author} {\bibfnamefont {A.~S.}\ \bibnamefont {{Szalay}}}, \bibinfo {author}
  {\bibfnamefont {M.~S.}\ \bibnamefont {{Vogeley}}}, \bibinfo {author}
  {\bibfnamefont {J.}~\bibnamefont {{Yoo}}}, \bibinfo {author} {\bibfnamefont
  {I.}~\bibnamefont {{Zehavi}}}, \bibinfo {author} {\bibfnamefont
  {K.}~\bibnamefont {{Abazajian}}}, \bibinfo {author} {\bibfnamefont {S.~F.}\
  \bibnamefont {{Anderson}}}, \bibinfo {author} {\bibfnamefont
  {J.}~\bibnamefont {{Annis}}}, \bibinfo {author} {\bibfnamefont {N.~A.}\
  \bibnamefont {{Bahcall}}}, \bibinfo {author} {\bibfnamefont {B.}~\bibnamefont
  {{Bassett}}}, \bibinfo {author} {\bibfnamefont {A.}~\bibnamefont
  {{Berlind}}}, \bibinfo {author} {\bibfnamefont {J.}~\bibnamefont
  {{Brinkmann}}}, \bibinfo {author} {\bibfnamefont {T.}~\bibnamefont
  {{Budavari}}}, \bibinfo {author} {\bibfnamefont {F.}~\bibnamefont
  {{Castander}}}, \bibinfo {author} {\bibfnamefont {A.}~\bibnamefont
  {{Connolly}}}, \bibinfo {author} {\bibfnamefont {I.}~\bibnamefont
  {{Csabai}}}, \bibinfo {author} {\bibfnamefont {M.}~\bibnamefont {{Doi}}},
  \bibinfo {author} {\bibfnamefont {D.~P.}\ \bibnamefont {{Finkbeiner}}},
  \bibinfo {author} {\bibfnamefont {B.}~\bibnamefont {{Gillespie}}}, \bibinfo
  {author} {\bibfnamefont {K.}~\bibnamefont {{Glazebrook}}}, \bibinfo {author}
  {\bibfnamefont {G.~S.}\ \bibnamefont {{Hennessy}}}, \bibinfo {author}
  {\bibfnamefont {D.~W.}\ \bibnamefont {{Hogg}}}, \bibinfo {author}
  {\bibfnamefont {{\v{Z}}.}~\bibnamefont {{Ivezi{\'c}}}}, \bibinfo {author}
  {\bibfnamefont {B.}~\bibnamefont {{Jain}}}, \bibinfo {author} {\bibfnamefont
  {D.}~\bibnamefont {{Johnston}}}, \bibinfo {author} {\bibfnamefont
  {S.}~\bibnamefont {{Kent}}}, \bibinfo {author} {\bibfnamefont {D.~Q.}\
  \bibnamefont {{Lamb}}}, \bibinfo {author} {\bibfnamefont {B.~C.}\
  \bibnamefont {{Lee}}}, \bibinfo {author} {\bibfnamefont {H.}~\bibnamefont
  {{Lin}}}, \bibinfo {author} {\bibfnamefont {J.}~\bibnamefont {{Loveday}}},
  \bibinfo {author} {\bibfnamefont {R.~H.}\ \bibnamefont {{Lupton}}}, \bibinfo
  {author} {\bibfnamefont {J.~A.}\ \bibnamefont {{Munn}}}, \bibinfo {author}
  {\bibfnamefont {K.}~\bibnamefont {{Pan}}}, \bibinfo {author} {\bibfnamefont
  {C.}~\bibnamefont {{Park}}}, \bibinfo {author} {\bibfnamefont
  {J.}~\bibnamefont {{Peoples}}}, \bibinfo {author} {\bibfnamefont {J.~R.}\
  \bibnamefont {{Pier}}}, \bibinfo {author} {\bibfnamefont {A.}~\bibnamefont
  {{Pope}}}, \bibinfo {author} {\bibfnamefont {M.}~\bibnamefont {{Richmond}}},
  \bibinfo {author} {\bibfnamefont {C.}~\bibnamefont {{Rockosi}}}, \bibinfo
  {author} {\bibfnamefont {R.}~\bibnamefont {{Scranton}}}, \bibinfo {author}
  {\bibfnamefont {R.~K.}\ \bibnamefont {{Sheth}}}, \bibinfo {author}
  {\bibfnamefont {A.}~\bibnamefont {{Stebbins}}}, \bibinfo {author}
  {\bibfnamefont {C.}~\bibnamefont {{Stoughton}}}, \bibinfo {author}
  {\bibfnamefont {I.}~\bibnamefont {{Szapudi}}}, \bibinfo {author}
  {\bibfnamefont {D.~L.}\ \bibnamefont {{Tucker}}}, \bibinfo {author}
  {\bibfnamefont {D.~E.}\ \bibnamefont {{vanden Berk}}}, \bibinfo {author}
  {\bibfnamefont {B.}~\bibnamefont {{Yanny}}}, \ and\ \bibinfo {author}
  {\bibfnamefont {D.~G.}\ \bibnamefont {{York}}},\ }\href {\doibase
  10.1103/PhysRevD.74.123507} {\bibfield  {journal} {\bibinfo  {journal}
  {\prd}\ }\textbf {\bibinfo {volume} {74}},\ \bibinfo {eid} {123507} (\bibinfo
  {year} {2006})},\ \Eprint {http://arxiv.org/abs/astro-ph/0608632}
  {arXiv:astro-ph/0608632 [astro-ph]} \BibitemShut {NoStop}%
\bibitem [{\citenamefont {{Zablocki}}\ and\ \citenamefont
  {{Dodelson}}(2016)}]{Zablocki2016}%
  \BibitemOpen
  \bibfield  {author} {\bibinfo {author} {\bibfnamefont {A.}~\bibnamefont
  {{Zablocki}}}\ and\ \bibinfo {author} {\bibfnamefont {S.}~\bibnamefont
  {{Dodelson}}},\ }\href {\doibase 10.1103/PhysRevD.93.083525} {\bibfield
  {journal} {\bibinfo  {journal} {PRD}\ }\textbf {\bibinfo {volume} {93}},\
  \bibinfo {eid} {083525} (\bibinfo {year} {2016})},\ \Eprint
  {http://arxiv.org/abs/1512.00072} {arXiv:1512.00072 [astro-ph.CO]}
  \BibitemShut {NoStop}%
\bibitem [{\citenamefont {{Szalay}}\ \emph {et~al.}(2003)\citenamefont
  {{Szalay}}, \citenamefont {{Jain}}, \citenamefont {{Matsubara}},
  \citenamefont {{Scranton}}, \citenamefont {{Vogeley}}, \citenamefont
  {{Connolly}}, \citenamefont {{Dodelson}}, \citenamefont {{Eisenstein}},
  \citenamefont {{Frieman}}, \citenamefont {{Gunn}}, \citenamefont {{Hui}},
  \citenamefont {{Johnston}}, \citenamefont {{Kent}}, \citenamefont
  {{Kerscher}}, \citenamefont {{Loveday}}, \citenamefont {{Meiksin}},
  \citenamefont {{Narayanan}}, \citenamefont {{Nichol}}, \citenamefont
  {{O'Connell}}, \citenamefont {{Pope}}, \citenamefont {{Scoccimarro}},
  \citenamefont {{Sheth}}, \citenamefont {{Stebbins}}, \citenamefont
  {{Strauss}}, \citenamefont {{Szapudi}}, \citenamefont {{Tegmark}},
  \citenamefont {{Zehavi}}, \citenamefont {{Annis}}, \citenamefont {{Bahcall}},
  \citenamefont {{Brinkmann}}, \citenamefont {{Csabai}}, \citenamefont
  {{Fukugita}}, \citenamefont {{Hennessy}}, \citenamefont {{Ivezic}},
  \citenamefont {{Knapp}}, \citenamefont {{Kunszt}}, \citenamefont {{Lamb}},
  \citenamefont {{Lee}}, \citenamefont {{Lupton}}, \citenamefont {{Munn}},
  \citenamefont {{Peoples}}, \citenamefont {{Pier}}, \citenamefont {{Rockosi}},
  \citenamefont {{Schlegel}}, \citenamefont {{Stoughton}}, \citenamefont
  {{Tucker}}, \citenamefont {{Yanny}}, \citenamefont {{York}},\ and\
  \citenamefont {{SDSS Collaboration}}}]{Szalay2003}%
  \BibitemOpen
  \bibfield  {author} {\bibinfo {author} {\bibfnamefont {A.~S.}\ \bibnamefont
  {{Szalay}}}, \bibinfo {author} {\bibfnamefont {B.}~\bibnamefont {{Jain}}},
  \bibinfo {author} {\bibfnamefont {T.}~\bibnamefont {{Matsubara}}}, \bibinfo
  {author} {\bibfnamefont {R.}~\bibnamefont {{Scranton}}}, \bibinfo {author}
  {\bibfnamefont {M.~S.}\ \bibnamefont {{Vogeley}}}, \bibinfo {author}
  {\bibfnamefont {A.}~\bibnamefont {{Connolly}}}, \bibinfo {author}
  {\bibfnamefont {S.}~\bibnamefont {{Dodelson}}}, \bibinfo {author}
  {\bibfnamefont {D.}~\bibnamefont {{Eisenstein}}}, \bibinfo {author}
  {\bibfnamefont {J.~A.}\ \bibnamefont {{Frieman}}}, \bibinfo {author}
  {\bibfnamefont {J.~E.}\ \bibnamefont {{Gunn}}}, \bibinfo {author}
  {\bibfnamefont {L.}~\bibnamefont {{Hui}}}, \bibinfo {author} {\bibfnamefont
  {D.}~\bibnamefont {{Johnston}}}, \bibinfo {author} {\bibfnamefont
  {S.}~\bibnamefont {{Kent}}}, \bibinfo {author} {\bibfnamefont
  {M.}~\bibnamefont {{Kerscher}}}, \bibinfo {author} {\bibfnamefont
  {J.}~\bibnamefont {{Loveday}}}, \bibinfo {author} {\bibfnamefont
  {A.}~\bibnamefont {{Meiksin}}}, \bibinfo {author} {\bibfnamefont
  {V.}~\bibnamefont {{Narayanan}}}, \bibinfo {author} {\bibfnamefont {R.~C.}\
  \bibnamefont {{Nichol}}}, \bibinfo {author} {\bibfnamefont {L.}~\bibnamefont
  {{O'Connell}}}, \bibinfo {author} {\bibfnamefont {A.}~\bibnamefont {{Pope}}},
  \bibinfo {author} {\bibfnamefont {R.}~\bibnamefont {{Scoccimarro}}}, \bibinfo
  {author} {\bibfnamefont {R.~K.}\ \bibnamefont {{Sheth}}}, \bibinfo {author}
  {\bibfnamefont {A.}~\bibnamefont {{Stebbins}}}, \bibinfo {author}
  {\bibfnamefont {M.~A.}\ \bibnamefont {{Strauss}}}, \bibinfo {author}
  {\bibfnamefont {I.}~\bibnamefont {{Szapudi}}}, \bibinfo {author}
  {\bibfnamefont {M.}~\bibnamefont {{Tegmark}}}, \bibinfo {author}
  {\bibfnamefont {I.}~\bibnamefont {{Zehavi}}}, \bibinfo {author}
  {\bibfnamefont {J.}~\bibnamefont {{Annis}}}, \bibinfo {author} {\bibfnamefont
  {N.}~\bibnamefont {{Bahcall}}}, \bibinfo {author} {\bibfnamefont
  {J.}~\bibnamefont {{Brinkmann}}}, \bibinfo {author} {\bibfnamefont
  {I.}~\bibnamefont {{Csabai}}}, \bibinfo {author} {\bibfnamefont
  {M.}~\bibnamefont {{Fukugita}}}, \bibinfo {author} {\bibfnamefont
  {G.}~\bibnamefont {{Hennessy}}}, \bibinfo {author} {\bibfnamefont
  {Z.}~\bibnamefont {{Ivezic}}}, \bibinfo {author} {\bibfnamefont {G.~R.}\
  \bibnamefont {{Knapp}}}, \bibinfo {author} {\bibfnamefont {P.~Z.}\
  \bibnamefont {{Kunszt}}}, \bibinfo {author} {\bibfnamefont {D.~Q.}\
  \bibnamefont {{Lamb}}}, \bibinfo {author} {\bibfnamefont {B.~C.}\
  \bibnamefont {{Lee}}}, \bibinfo {author} {\bibfnamefont {R.~H.}\ \bibnamefont
  {{Lupton}}}, \bibinfo {author} {\bibfnamefont {J.~R.}\ \bibnamefont
  {{Munn}}}, \bibinfo {author} {\bibfnamefont {J.}~\bibnamefont {{Peoples}}},
  \bibinfo {author} {\bibfnamefont {J.~R.}\ \bibnamefont {{Pier}}}, \bibinfo
  {author} {\bibfnamefont {C.}~\bibnamefont {{Rockosi}}}, \bibinfo {author}
  {\bibfnamefont {D.}~\bibnamefont {{Schlegel}}}, \bibinfo {author}
  {\bibfnamefont {C.}~\bibnamefont {{Stoughton}}}, \bibinfo {author}
  {\bibfnamefont {D.~L.}\ \bibnamefont {{Tucker}}}, \bibinfo {author}
  {\bibfnamefont {B.}~\bibnamefont {{Yanny}}}, \bibinfo {author} {\bibfnamefont
  {D.~G.}\ \bibnamefont {{York}}}, \ and\ \bibinfo {author} {\bibnamefont
  {{SDSS Collaboration}}},\ }\href {\doibase 10.1086/375264} {\bibfield
  {journal} {\bibinfo  {journal} {ApJ}\ }\textbf {\bibinfo {volume} {591}},\
  \bibinfo {pages} {1} (\bibinfo {year} {2003})},\ \Eprint
  {http://arxiv.org/abs/astro-ph/0107419} {arXiv:astro-ph/0107419 [astro-ph]}
  \BibitemShut {NoStop}%
\bibitem [{\citenamefont {{Douspis}}\ \emph {et~al.}(2015)\citenamefont
  {{Douspis}}, \citenamefont {{Aghanim}}, \citenamefont {{Ili{\'c}}},\ and\
  \citenamefont {{Langer}}}]{Douspis2015}%
  \BibitemOpen
  \bibfield  {author} {\bibinfo {author} {\bibfnamefont {M.}~\bibnamefont
  {{Douspis}}}, \bibinfo {author} {\bibfnamefont {N.}~\bibnamefont
  {{Aghanim}}}, \bibinfo {author} {\bibfnamefont {S.}~\bibnamefont
  {{Ili{\'c}}}}, \ and\ \bibinfo {author} {\bibfnamefont {M.}~\bibnamefont
  {{Langer}}},\ }\href {\doibase 10.1051/0004-6361/201526543} {\bibfield
  {journal} {\bibinfo  {journal} {\aap}\ }\textbf {\bibinfo {volume} {580}},\
  \bibinfo {eid} {L4} (\bibinfo {year} {2015})},\ \Eprint
  {http://arxiv.org/abs/1509.02785} {arXiv:1509.02785 [astro-ph.CO]}
  \BibitemShut {NoStop}%
\bibitem [{\citenamefont {Lewis}(2008)}]{Lewis2008}%
  \BibitemOpen
  \bibfield  {author} {\bibinfo {author} {\bibfnamefont {A.}~\bibnamefont
  {Lewis}},\ }\href {\doibase 10.1103/PhysRevD.78.023002} {\bibfield  {journal}
  {\bibinfo  {journal} {Phys. Rev. D}\ }\textbf {\bibinfo {volume} {78}},\
  \bibinfo {pages} {023002} (\bibinfo {year} {2008})}\BibitemShut {NoStop}%
\bibitem [{\citenamefont {{Tegmark}}(1997)}]{T97maps}%
  \BibitemOpen
  \bibfield  {author} {\bibinfo {author} {\bibfnamefont {M.}~\bibnamefont
  {{Tegmark}}},\ }\href {\doibase 10.1086/310631} {\bibfield  {journal}
  {\bibinfo  {journal} {\apjl}\ }\textbf {\bibinfo {volume} {480}},\ \bibinfo
  {pages} {L87} (\bibinfo {year} {1997})},\ \Eprint
  {http://arxiv.org/abs/astro-ph/9611130} {arXiv:astro-ph/9611130 [astro-ph]}
  \BibitemShut {NoStop}%
\bibitem [{\citenamefont {{Monsalve}}\ \emph {et~al.}(2017)\citenamefont
  {{Monsalve}}, \citenamefont {{Rogers}}, \citenamefont {{Bowman}},\ and\
  \citenamefont {{Mozdzen}}}]{Monsalve2017reionconstraints}%
  \BibitemOpen
  \bibfield  {author} {\bibinfo {author} {\bibfnamefont {R.~A.}\ \bibnamefont
  {{Monsalve}}}, \bibinfo {author} {\bibfnamefont {A.~E.~E.}\ \bibnamefont
  {{Rogers}}}, \bibinfo {author} {\bibfnamefont {J.~D.}\ \bibnamefont
  {{Bowman}}}, \ and\ \bibinfo {author} {\bibfnamefont {T.~J.}\ \bibnamefont
  {{Mozdzen}}},\ }\href {\doibase 10.3847/1538-4357/aa88d1} {\bibfield
  {journal} {\bibinfo  {journal} {\apj}\ }\textbf {\bibinfo {volume} {847}},\
  \bibinfo {eid} {64} (\bibinfo {year} {2017})},\ \Eprint
  {http://arxiv.org/abs/1708.05817} {arXiv:1708.05817 [astro-ph.CO]}
  \BibitemShut {NoStop}%
\bibitem [{\citenamefont {Wyithe}\ and\ \citenamefont
  {Loeb}(2003)}]{Wyithe:2002qu}%
  \BibitemOpen
  \bibfield  {author} {\bibinfo {author} {\bibfnamefont {J.~S.~B.}\
  \bibnamefont {Wyithe}}\ and\ \bibinfo {author} {\bibfnamefont
  {A.}~\bibnamefont {Loeb}},\ }\href {\doibase 10.1086/367721} {\bibfield
  {journal} {\bibinfo  {journal} {\apj}\ }\textbf {\bibinfo {volume} {586}},\
  \bibinfo {pages} {693} (\bibinfo {year} {2003})},\ \Eprint
  {http://arxiv.org/abs/astro-ph/0209056} {arXiv:astro-ph/0209056} \BibitemShut
  {NoStop}%
\bibitem [{\citenamefont {{Cen}}(2003)}]{Cen2003}%
  \BibitemOpen
  \bibfield  {author} {\bibinfo {author} {\bibfnamefont {R.}~\bibnamefont
  {{Cen}}},\ }\href {\doibase 10.1086/375217} {\bibfield  {journal} {\bibinfo
  {journal} {\apj}\ }\textbf {\bibinfo {volume} {591}},\ \bibinfo {pages} {12}
  (\bibinfo {year} {2003})},\ \Eprint {http://arxiv.org/abs/astro-ph/0210473}
  {arXiv:astro-ph/0210473 [astro-ph]} \BibitemShut {NoStop}%
\bibitem [{\citenamefont {Furlanetto}\ and\ \citenamefont
  {Loeb}(2005)}]{Furlanetto:2004nt}%
  \BibitemOpen
  \bibfield  {author} {\bibinfo {author} {\bibfnamefont {S.}~\bibnamefont
  {Furlanetto}}\ and\ \bibinfo {author} {\bibfnamefont {A.}~\bibnamefont
  {Loeb}},\ }\href {\doibase 10.1086/429080} {\bibfield  {journal} {\bibinfo
  {journal} {\apj}\ }\textbf {\bibinfo {volume} {634}},\ \bibinfo {pages} {1}
  (\bibinfo {year} {2005})},\ \Eprint {http://arxiv.org/abs/astro-ph/0409656}
  {arXiv:astro-ph/0409656} \BibitemShut {NoStop}%
\bibitem [{\citenamefont {{Hu}}\ and\ \citenamefont
  {{Holder}}(2003)}]{HuHolder_2003}%
  \BibitemOpen
  \bibfield  {author} {\bibinfo {author} {\bibfnamefont {W.}~\bibnamefont
  {{Hu}}}\ and\ \bibinfo {author} {\bibfnamefont {G.~P.}\ \bibnamefont
  {{Holder}}},\ }\href {\doibase 10.1103/PhysRevD.68.023001} {\bibfield
  {journal} {\bibinfo  {journal} {\prd}\ }\textbf {\bibinfo {volume} {68}},\
  \bibinfo {eid} {023001} (\bibinfo {year} {2003})},\ \Eprint
  {http://arxiv.org/abs/astro-ph/0303400} {arXiv:astro-ph/0303400 [astro-ph]}
  \BibitemShut {NoStop}%
\bibitem [{\citenamefont {{Millea}}\ and\ \citenamefont
  {{Bouchet}}(2018)}]{MilleaBouchet_2018}%
  \BibitemOpen
  \bibfield  {author} {\bibinfo {author} {\bibfnamefont {M.}~\bibnamefont
  {{Millea}}}\ and\ \bibinfo {author} {\bibfnamefont {F.}~\bibnamefont
  {{Bouchet}}},\ }\href {\doibase 10.1051/0004-6361/201833288} {\bibfield
  {journal} {\bibinfo  {journal} {\aap}\ }\textbf {\bibinfo {volume} {617}},\
  \bibinfo {eid} {A96} (\bibinfo {year} {2018})},\ \Eprint
  {http://arxiv.org/abs/1804.08476} {arXiv:1804.08476 [astro-ph.CO]}
  \BibitemShut {NoStop}%
\bibitem [{\citenamefont {{Planck Collaboration}}\ \emph
  {et~al.}(2020{\natexlab{b}})\citenamefont {{Planck Collaboration}},
  \citenamefont {{Aghanim}} \emph {et~al.}}]{Planck2018_cosmo_params}%
  \BibitemOpen
  \bibfield  {author} {\bibinfo {author} {\bibnamefont {{Planck
  Collaboration}}}, \bibinfo {author} {\bibfnamefont {N.}~\bibnamefont
  {{Aghanim}}},  \emph {et~al.},\ }\href {\doibase 10.1051/0004-6361/201833910}
  {\bibfield  {journal} {\bibinfo  {journal} {\aap}\ }\textbf {\bibinfo
  {volume} {641}},\ \bibinfo {eid} {A6} (\bibinfo {year}
  {2020}{\natexlab{b}})},\ \Eprint {http://arxiv.org/abs/1807.06209}
  {arXiv:1807.06209 [astro-ph.CO]} \BibitemShut {NoStop}%
\bibitem [{\citenamefont {{Trac}}\ \emph {et~al.}(2021)\citenamefont {{Trac}},
  \citenamefont {{Chen}}, \citenamefont {{Holst}}, \citenamefont {{Alvarez}},\
  and\ \citenamefont {{Cen}}}]{Trac2021}%
  \BibitemOpen
  \bibfield  {author} {\bibinfo {author} {\bibfnamefont {H.}~\bibnamefont
  {{Trac}}}, \bibinfo {author} {\bibfnamefont {N.}~\bibnamefont {{Chen}}},
  \bibinfo {author} {\bibfnamefont {I.}~\bibnamefont {{Holst}}}, \bibinfo
  {author} {\bibfnamefont {M.~A.}\ \bibnamefont {{Alvarez}}}, \ and\ \bibinfo
  {author} {\bibfnamefont {R.}~\bibnamefont {{Cen}}},\ }\href@noop {}
  {\bibfield  {journal} {\bibinfo  {journal} {arXiv e-prints}\ ,\ \bibinfo
  {eid} {arXiv:2109.10375}} (\bibinfo {year} {2021})},\ \Eprint
  {http://arxiv.org/abs/2109.10375} {arXiv:2109.10375 [astro-ph.CO]}
  \BibitemShut {NoStop}%
\bibitem [{\citenamefont {{Bernardi}}\ \emph {et~al.}(2015)\citenamefont
  {{Bernardi}}, \citenamefont {{McQuinn}},\ and\ \citenamefont
  {{Greenhill}}}]{Bernardi2015}%
  \BibitemOpen
  \bibfield  {author} {\bibinfo {author} {\bibfnamefont {G.}~\bibnamefont
  {{Bernardi}}}, \bibinfo {author} {\bibfnamefont {M.}~\bibnamefont
  {{McQuinn}}}, \ and\ \bibinfo {author} {\bibfnamefont {L.~J.}\ \bibnamefont
  {{Greenhill}}},\ }\href {\doibase 10.1088/0004-637X/799/1/90} {\bibfield
  {journal} {\bibinfo  {journal} {\apj}\ }\textbf {\bibinfo {volume} {799}},\
  \bibinfo {eid} {90} (\bibinfo {year} {2015})},\ \Eprint
  {http://arxiv.org/abs/1404.0887} {arXiv:1404.0887 [astro-ph.CO]} \BibitemShut
  {NoStop}%
\bibitem [{\citenamefont {{Liu}}\ and\ \citenamefont
  {{Shaw}}(2020)}]{LiuShaw2020}%
  \BibitemOpen
  \bibfield  {author} {\bibinfo {author} {\bibfnamefont {A.}~\bibnamefont
  {{Liu}}}\ and\ \bibinfo {author} {\bibfnamefont {J.~R.}\ \bibnamefont
  {{Shaw}}},\ }\href {\doibase 10.1088/1538-3873/ab5bfd} {\bibfield  {journal}
  {\bibinfo  {journal} {\pasp}\ }\textbf {\bibinfo {volume} {132}},\ \bibinfo
  {eid} {062001} (\bibinfo {year} {2020})},\ \Eprint
  {http://arxiv.org/abs/1907.08211} {arXiv:1907.08211 [astro-ph.IM]}
  \BibitemShut {NoStop}%
\bibitem [{\citenamefont {{Tauscher}}\ \emph {et~al.}(2018)\citenamefont
  {{Tauscher}}, \citenamefont {{Rapetti}}, \citenamefont {{Burns}},\ and\
  \citenamefont {{Switzer}}}]{2018ApJ...853..187T}%
  \BibitemOpen
  \bibfield  {author} {\bibinfo {author} {\bibfnamefont {K.}~\bibnamefont
  {{Tauscher}}}, \bibinfo {author} {\bibfnamefont {D.}~\bibnamefont
  {{Rapetti}}}, \bibinfo {author} {\bibfnamefont {J.~O.}\ \bibnamefont
  {{Burns}}}, \ and\ \bibinfo {author} {\bibfnamefont {E.}~\bibnamefont
  {{Switzer}}},\ }\href {\doibase 10.3847/1538-4357/aaa41f} {\bibfield
  {journal} {\bibinfo  {journal} {\apj}\ }\textbf {\bibinfo {volume} {853}},\
  \bibinfo {eid} {187} (\bibinfo {year} {2018})},\ \Eprint
  {http://arxiv.org/abs/1711.03173} {arXiv:1711.03173 [astro-ph.IM]}
  \BibitemShut {NoStop}%
\bibitem [{\citenamefont {{Nhan}}\ \emph {et~al.}(2019)\citenamefont {{Nhan}},
  \citenamefont {{Bordenave}}, \citenamefont {{Bradley}}, \citenamefont
  {{Burns}}, \citenamefont {{Tauscher}}, \citenamefont {{Rapetti}},\ and\
  \citenamefont {{Klima}}}]{2019ApJ...883..126N}%
  \BibitemOpen
  \bibfield  {author} {\bibinfo {author} {\bibfnamefont {B.~D.}\ \bibnamefont
  {{Nhan}}}, \bibinfo {author} {\bibfnamefont {D.~D.}\ \bibnamefont
  {{Bordenave}}}, \bibinfo {author} {\bibfnamefont {R.~F.}\ \bibnamefont
  {{Bradley}}}, \bibinfo {author} {\bibfnamefont {J.~O.}\ \bibnamefont
  {{Burns}}}, \bibinfo {author} {\bibfnamefont {K.}~\bibnamefont {{Tauscher}}},
  \bibinfo {author} {\bibfnamefont {D.}~\bibnamefont {{Rapetti}}}, \ and\
  \bibinfo {author} {\bibfnamefont {P.~J.}\ \bibnamefont {{Klima}}},\ }\href
  {\doibase 10.3847/1538-4357/ab391b} {\bibfield  {journal} {\bibinfo
  {journal} {\apj}\ }\textbf {\bibinfo {volume} {883}},\ \bibinfo {eid} {126}
  (\bibinfo {year} {2019})},\ \Eprint {http://arxiv.org/abs/1811.04917}
  {arXiv:1811.04917 [astro-ph.IM]} \BibitemShut {NoStop}%
\bibitem [{\citenamefont {{Tauscher}}\ \emph
  {et~al.}(2020{\natexlab{a}})\citenamefont {{Tauscher}}, \citenamefont
  {{Rapetti}},\ and\ \citenamefont {{Burns}}}]{2020ApJ...897..132T}%
  \BibitemOpen
  \bibfield  {author} {\bibinfo {author} {\bibfnamefont {K.}~\bibnamefont
  {{Tauscher}}}, \bibinfo {author} {\bibfnamefont {D.}~\bibnamefont
  {{Rapetti}}}, \ and\ \bibinfo {author} {\bibfnamefont {J.~O.}\ \bibnamefont
  {{Burns}}},\ }\href {\doibase 10.3847/1538-4357/ab9a3f} {\bibfield  {journal}
  {\bibinfo  {journal} {\apj}\ }\textbf {\bibinfo {volume} {897}},\ \bibinfo
  {eid} {132} (\bibinfo {year} {2020}{\natexlab{a}})},\ \Eprint
  {http://arxiv.org/abs/2005.00034} {arXiv:2005.00034 [astro-ph.CO]}
  \BibitemShut {NoStop}%
\bibitem [{\citenamefont {{Rapetti}}\ \emph {et~al.}(2020)\citenamefont
  {{Rapetti}}, \citenamefont {{Tauscher}}, \citenamefont {{Mirocha}},\ and\
  \citenamefont {{Burns}}}]{2020ApJ...897..174R}%
  \BibitemOpen
  \bibfield  {author} {\bibinfo {author} {\bibfnamefont {D.}~\bibnamefont
  {{Rapetti}}}, \bibinfo {author} {\bibfnamefont {K.}~\bibnamefont
  {{Tauscher}}}, \bibinfo {author} {\bibfnamefont {J.}~\bibnamefont
  {{Mirocha}}}, \ and\ \bibinfo {author} {\bibfnamefont {J.~O.}\ \bibnamefont
  {{Burns}}},\ }\href {\doibase 10.3847/1538-4357/ab9b29} {\bibfield  {journal}
  {\bibinfo  {journal} {\apj}\ }\textbf {\bibinfo {volume} {897}},\ \bibinfo
  {eid} {174} (\bibinfo {year} {2020})},\ \Eprint
  {http://arxiv.org/abs/1912.02205} {arXiv:1912.02205 [astro-ph.CO]}
  \BibitemShut {NoStop}%
\bibitem [{\citenamefont {{Tauscher}}\ \emph
  {et~al.}(2020{\natexlab{b}})\citenamefont {{Tauscher}}, \citenamefont
  {{Rapetti}},\ and\ \citenamefont {{Burns}}}]{2020ApJ...897..175T}%
  \BibitemOpen
  \bibfield  {author} {\bibinfo {author} {\bibfnamefont {K.}~\bibnamefont
  {{Tauscher}}}, \bibinfo {author} {\bibfnamefont {D.}~\bibnamefont
  {{Rapetti}}}, \ and\ \bibinfo {author} {\bibfnamefont {J.~O.}\ \bibnamefont
  {{Burns}}},\ }\href {\doibase 10.3847/1538-4357/ab9b2a} {\bibfield  {journal}
  {\bibinfo  {journal} {\apj}\ }\textbf {\bibinfo {volume} {897}},\ \bibinfo
  {eid} {175} (\bibinfo {year} {2020}{\natexlab{b}})},\ \Eprint
  {http://arxiv.org/abs/2003.05452} {arXiv:2003.05452 [astro-ph.IM]}
  \BibitemShut {NoStop}%
\bibitem [{\citenamefont {{Hibbard}}\ \emph {et~al.}(2020)\citenamefont
  {{Hibbard}}, \citenamefont {{Tauscher}}, \citenamefont {{Rapetti}},\ and\
  \citenamefont {{Burns}}}]{2020ApJ...905..113H}%
  \BibitemOpen
  \bibfield  {author} {\bibinfo {author} {\bibfnamefont {J.~J.}\ \bibnamefont
  {{Hibbard}}}, \bibinfo {author} {\bibfnamefont {K.}~\bibnamefont
  {{Tauscher}}}, \bibinfo {author} {\bibfnamefont {D.}~\bibnamefont
  {{Rapetti}}}, \ and\ \bibinfo {author} {\bibfnamefont {J.~O.}\ \bibnamefont
  {{Burns}}},\ }\href {\doibase 10.3847/1538-4357/abc3c5} {\bibfield  {journal}
  {\bibinfo  {journal} {\apj}\ }\textbf {\bibinfo {volume} {905}},\ \bibinfo
  {eid} {113} (\bibinfo {year} {2020})},\ \Eprint
  {http://arxiv.org/abs/2011.00549} {arXiv:2011.00549 [astro-ph.CO]}
  \BibitemShut {NoStop}%
\bibitem [{\citenamefont {{Bassett}}\ \emph {et~al.}(2021)\citenamefont
  {{Bassett}}, \citenamefont {{Rapetti}}, \citenamefont {{Tauscher}},
  \citenamefont {{Burns}},\ and\ \citenamefont
  {{Hibbard}}}]{2021ApJ...908..189B}%
  \BibitemOpen
  \bibfield  {author} {\bibinfo {author} {\bibfnamefont {N.}~\bibnamefont
  {{Bassett}}}, \bibinfo {author} {\bibfnamefont {D.}~\bibnamefont
  {{Rapetti}}}, \bibinfo {author} {\bibfnamefont {K.}~\bibnamefont
  {{Tauscher}}}, \bibinfo {author} {\bibfnamefont {J.~O.}\ \bibnamefont
  {{Burns}}}, \ and\ \bibinfo {author} {\bibfnamefont {J.~J.}\ \bibnamefont
  {{Hibbard}}},\ }\href {\doibase 10.3847/1538-4357/abdb29} {\bibfield
  {journal} {\bibinfo  {journal} {\apj}\ }\textbf {\bibinfo {volume} {908}},\
  \bibinfo {eid} {189} (\bibinfo {year} {2021})},\ \Eprint
  {http://arxiv.org/abs/2011.01242} {arXiv:2011.01242 [astro-ph.CO]}
  \BibitemShut {NoStop}%
\bibitem [{\citenamefont {{Tauscher}}\ \emph {et~al.}(2021)\citenamefont
  {{Tauscher}}, \citenamefont {{Rapetti}}, \citenamefont {{Nhan}},
  \citenamefont {{Handy}}, \citenamefont {{Bassett}}, \citenamefont
  {{Hibbard}}, \citenamefont {{Bordenave}}, \citenamefont {{Bradley}},\ and\
  \citenamefont {{Burns}}}]{2021ApJ...915...66T}%
  \BibitemOpen
  \bibfield  {author} {\bibinfo {author} {\bibfnamefont {K.}~\bibnamefont
  {{Tauscher}}}, \bibinfo {author} {\bibfnamefont {D.}~\bibnamefont
  {{Rapetti}}}, \bibinfo {author} {\bibfnamefont {B.~D.}\ \bibnamefont
  {{Nhan}}}, \bibinfo {author} {\bibfnamefont {A.}~\bibnamefont {{Handy}}},
  \bibinfo {author} {\bibfnamefont {N.}~\bibnamefont {{Bassett}}}, \bibinfo
  {author} {\bibfnamefont {J.}~\bibnamefont {{Hibbard}}}, \bibinfo {author}
  {\bibfnamefont {D.}~\bibnamefont {{Bordenave}}}, \bibinfo {author}
  {\bibfnamefont {R.~F.}\ \bibnamefont {{Bradley}}}, \ and\ \bibinfo {author}
  {\bibfnamefont {J.~O.}\ \bibnamefont {{Burns}}},\ }\href {\doibase
  10.3847/1538-4357/ac00af} {\bibfield  {journal} {\bibinfo  {journal} {\apj}\
  }\textbf {\bibinfo {volume} {915}},\ \bibinfo {eid} {66} (\bibinfo {year}
  {2021})},\ \Eprint {http://arxiv.org/abs/2105.01672} {arXiv:2105.01672
  [astro-ph.CO]} \BibitemShut {NoStop}%
\bibitem [{\citenamefont {{Bevins}}\ \emph {et~al.}(2021)\citenamefont
  {{Bevins}}, \citenamefont {{Handley}}, \citenamefont {{Fialkov}},
  \citenamefont {{de Lera Acedo}}, \citenamefont {{Greenhill}},\ and\
  \citenamefont {{Price}}}]{2021MNRAS.502.4405B}%
  \BibitemOpen
  \bibfield  {author} {\bibinfo {author} {\bibfnamefont {H.~T.~J.}\
  \bibnamefont {{Bevins}}}, \bibinfo {author} {\bibfnamefont {W.~J.}\
  \bibnamefont {{Handley}}}, \bibinfo {author} {\bibfnamefont {A.}~\bibnamefont
  {{Fialkov}}}, \bibinfo {author} {\bibfnamefont {E.}~\bibnamefont {{de Lera
  Acedo}}}, \bibinfo {author} {\bibfnamefont {L.~J.}\ \bibnamefont
  {{Greenhill}}}, \ and\ \bibinfo {author} {\bibfnamefont {D.~C.}\ \bibnamefont
  {{Price}}},\ }\href {\doibase 10.1093/mnras/stab152} {\bibfield  {journal}
  {\bibinfo  {journal} {\mnras}\ }\textbf {\bibinfo {volume} {502}},\ \bibinfo
  {pages} {4405} (\bibinfo {year} {2021})},\ \Eprint
  {http://arxiv.org/abs/2007.14970} {arXiv:2007.14970 [astro-ph.CO]}
  \BibitemShut {NoStop}%
\bibitem [{\citenamefont {{Anstey}}\ \emph
  {et~al.}(2021{\natexlab{a}})\citenamefont {{Anstey}}, \citenamefont
  {{Cumner}}, \citenamefont {{Acedo}},\ and\ \citenamefont
  {{Handley}}}]{2021MNRAS.tmp.2933A}%
  \BibitemOpen
  \bibfield  {author} {\bibinfo {author} {\bibfnamefont {D.}~\bibnamefont
  {{Anstey}}}, \bibinfo {author} {\bibfnamefont {J.}~\bibnamefont {{Cumner}}},
  \bibinfo {author} {\bibfnamefont {E.~d.~L.}\ \bibnamefont {{Acedo}}}, \ and\
  \bibinfo {author} {\bibfnamefont {W.}~\bibnamefont {{Handley}}},\ }\href
  {\doibase 10.1093/mnras/stab3211} {\bibfield  {journal} {\bibinfo  {journal}
  {\mnras}\ } (\bibinfo {year} {2021}{\natexlab{a}}),\
  10.1093/mnras/stab3211},\ \Eprint {http://arxiv.org/abs/2106.10193}
  {arXiv:2106.10193 [astro-ph.IM]} \BibitemShut {NoStop}%
\bibitem [{\citenamefont {{Anstey}}\ \emph
  {et~al.}(2021{\natexlab{b}})\citenamefont {{Anstey}}, \citenamefont {{de Lera
  Acedo}},\ and\ \citenamefont {{Handley}}}]{2021MNRAS.506.2041A}%
  \BibitemOpen
  \bibfield  {author} {\bibinfo {author} {\bibfnamefont {D.}~\bibnamefont
  {{Anstey}}}, \bibinfo {author} {\bibfnamefont {E.}~\bibnamefont {{de Lera
  Acedo}}}, \ and\ \bibinfo {author} {\bibfnamefont {W.}~\bibnamefont
  {{Handley}}},\ }\href {\doibase 10.1093/mnras/stab1765} {\bibfield  {journal}
  {\bibinfo  {journal} {\mnras}\ }\textbf {\bibinfo {volume} {506}},\ \bibinfo
  {pages} {2041} (\bibinfo {year} {2021}{\natexlab{b}})},\ \Eprint
  {http://arxiv.org/abs/2010.09644} {arXiv:2010.09644 [astro-ph.IM]}
  \BibitemShut {NoStop}%
\bibitem [{\citenamefont {{Bernardi}}\ \emph {et~al.}(2016)\citenamefont
  {{Bernardi}}, \citenamefont {{Zwart}}, \citenamefont {{Price}}, \citenamefont
  {{Greenhill}}, \citenamefont {{Mesinger}}, \citenamefont {{Dowell}},
  \citenamefont {{Eftekhari}}, \citenamefont {{Ellingson}}, \citenamefont
  {{Kocz}},\ and\ \citenamefont {{Schinzel}}}]{2016MNRAS.461.2847B}%
  \BibitemOpen
  \bibfield  {author} {\bibinfo {author} {\bibfnamefont {G.}~\bibnamefont
  {{Bernardi}}}, \bibinfo {author} {\bibfnamefont {J.~T.~L.}\ \bibnamefont
  {{Zwart}}}, \bibinfo {author} {\bibfnamefont {D.}~\bibnamefont {{Price}}},
  \bibinfo {author} {\bibfnamefont {L.~J.}\ \bibnamefont {{Greenhill}}},
  \bibinfo {author} {\bibfnamefont {A.}~\bibnamefont {{Mesinger}}}, \bibinfo
  {author} {\bibfnamefont {J.}~\bibnamefont {{Dowell}}}, \bibinfo {author}
  {\bibfnamefont {T.}~\bibnamefont {{Eftekhari}}}, \bibinfo {author}
  {\bibfnamefont {S.~W.}\ \bibnamefont {{Ellingson}}}, \bibinfo {author}
  {\bibfnamefont {J.}~\bibnamefont {{Kocz}}}, \ and\ \bibinfo {author}
  {\bibfnamefont {F.}~\bibnamefont {{Schinzel}}},\ }\href {\doibase
  10.1093/mnras/stw1499} {\bibfield  {journal} {\bibinfo  {journal} {\mnras}\
  }\textbf {\bibinfo {volume} {461}},\ \bibinfo {pages} {2847} (\bibinfo {year}
  {2016})},\ \Eprint {http://arxiv.org/abs/1606.06006} {arXiv:1606.06006
  [astro-ph.CO]} \BibitemShut {NoStop}%
\bibitem [{\citenamefont {{Harker}}(2015)}]{2015MNRAS.449L..21H}%
  \BibitemOpen
  \bibfield  {author} {\bibinfo {author} {\bibfnamefont {G.~J.~A.}\
  \bibnamefont {{Harker}}},\ }\href {\doibase 10.1093/mnrasl/slv011} {\bibfield
   {journal} {\bibinfo  {journal} {\mnras}\ }\textbf {\bibinfo {volume}
  {449}},\ \bibinfo {pages} {L21} (\bibinfo {year} {2015})},\ \Eprint
  {http://arxiv.org/abs/1501.05182} {arXiv:1501.05182 [astro-ph.CO]}
  \BibitemShut {NoStop}%
\bibitem [{\citenamefont {{Switzer}}\ and\ \citenamefont
  {{Liu}}(2014)}]{2014ApJ...793..102S}%
  \BibitemOpen
  \bibfield  {author} {\bibinfo {author} {\bibfnamefont {E.~R.}\ \bibnamefont
  {{Switzer}}}\ and\ \bibinfo {author} {\bibfnamefont {A.}~\bibnamefont
  {{Liu}}},\ }\href {\doibase 10.1088/0004-637X/793/2/102} {\bibfield
  {journal} {\bibinfo  {journal} {\apj}\ }\textbf {\bibinfo {volume} {793}},\
  \bibinfo {eid} {102} (\bibinfo {year} {2014})},\ \Eprint
  {http://arxiv.org/abs/1404.7561} {arXiv:1404.7561 [astro-ph.CO]} \BibitemShut
  {NoStop}%
\bibitem [{\citenamefont {{Harker}}\ \emph {et~al.}(2012)\citenamefont
  {{Harker}}, \citenamefont {{Pritchard}}, \citenamefont {{Burns}},\ and\
  \citenamefont {{Bowman}}}]{2012MNRAS.419.1070H}%
  \BibitemOpen
  \bibfield  {author} {\bibinfo {author} {\bibfnamefont {G.~J.~A.}\
  \bibnamefont {{Harker}}}, \bibinfo {author} {\bibfnamefont {J.~R.}\
  \bibnamefont {{Pritchard}}}, \bibinfo {author} {\bibfnamefont {J.~O.}\
  \bibnamefont {{Burns}}}, \ and\ \bibinfo {author} {\bibfnamefont {J.~D.}\
  \bibnamefont {{Bowman}}},\ }\href {\doibase 10.1111/j.1365-2966.2011.19766.x}
  {\bibfield  {journal} {\bibinfo  {journal} {\mnras}\ }\textbf {\bibinfo
  {volume} {419}},\ \bibinfo {pages} {1070} (\bibinfo {year} {2012})},\ \Eprint
  {http://arxiv.org/abs/1107.3154} {arXiv:1107.3154 [astro-ph.CO]} \BibitemShut
  {NoStop}%
\bibitem [{\citenamefont {Presley}\ \emph {et~al.}(2015)\citenamefont
  {Presley}, \citenamefont {Liu},\ and\ \citenamefont {Parsons}}]{Presley2015}%
  \BibitemOpen
  \bibfield  {author} {\bibinfo {author} {\bibfnamefont {M.~E.}\ \bibnamefont
  {Presley}}, \bibinfo {author} {\bibfnamefont {A.}~\bibnamefont {Liu}}, \ and\
  \bibinfo {author} {\bibfnamefont {A.~R.}\ \bibnamefont {Parsons}},\ }\href
  {\doibase 10.1088/0004-637x/809/1/18} {\bibfield  {journal} {\bibinfo
  {journal} {The Astrophysical Journal}\ }\textbf {\bibinfo {volume} {809}},\
  \bibinfo {pages} {18} (\bibinfo {year} {2015})}\BibitemShut {NoStop}%
\bibitem [{\citenamefont {Collaboration}\ and\ \citenamefont
  {Collaboration}(2016)}]{Ligo_2016}%
  \BibitemOpen
  \bibfield  {author} {\bibinfo {author} {\bibfnamefont {L.~S.}\ \bibnamefont
  {Collaboration}}\ and\ \bibinfo {author} {\bibfnamefont {V.}~\bibnamefont
  {Collaboration}},\ }\href {\doibase 10.1103/PhysRevLett.116.061102}
  {\bibfield  {journal} {\bibinfo  {journal} {Phys. Rev. Lett.}\ }\textbf
  {\bibinfo {volume} {116}},\ \bibinfo {pages} {061102} (\bibinfo {year}
  {2016})}\BibitemShut {NoStop}%
\bibitem [{\citenamefont {{Hogg}}\ \emph {et~al.}(2010)\citenamefont {{Hogg}},
  \citenamefont {{Bovy}},\ and\ \citenamefont {{Lang}}}]{Hogg2010fitting}%
  \BibitemOpen
  \bibfield  {author} {\bibinfo {author} {\bibfnamefont {D.~W.}\ \bibnamefont
  {{Hogg}}}, \bibinfo {author} {\bibfnamefont {J.}~\bibnamefont {{Bovy}}}, \
  and\ \bibinfo {author} {\bibfnamefont {D.}~\bibnamefont {{Lang}}},\
  }\href@noop {} {\bibfield  {journal} {\bibinfo  {journal} {arXiv e-prints}\
  ,\ \bibinfo {eid} {arXiv:1008.4686}} (\bibinfo {year} {2010})},\ \Eprint
  {http://arxiv.org/abs/1008.4686} {arXiv:1008.4686 [astro-ph.IM]} \BibitemShut
  {NoStop}%
\bibitem [{\citenamefont {{Mu{\~n}oz}}\ and\ \citenamefont
  {{Cyr-Racine}}(2021)}]{2021PhRvD.103b3512M}%
  \BibitemOpen
  \bibfield  {author} {\bibinfo {author} {\bibfnamefont {J.~B.}\ \bibnamefont
  {{Mu{\~n}oz}}}\ and\ \bibinfo {author} {\bibfnamefont {F.-Y.}\ \bibnamefont
  {{Cyr-Racine}}},\ }\href {\doibase 10.1103/PhysRevD.103.023512} {\bibfield
  {journal} {\bibinfo  {journal} {\prd}\ }\textbf {\bibinfo {volume} {103}},\
  \bibinfo {eid} {023512} (\bibinfo {year} {2021})},\ \Eprint
  {http://arxiv.org/abs/2005.03664} {arXiv:2005.03664 [astro-ph.CO]}
  \BibitemShut {NoStop}%
\bibitem [{\citenamefont {{Basak}}\ \emph {et~al.}(2006)\citenamefont
  {{Basak}}, \citenamefont {{Hajian}},\ and\ \citenamefont
  {{Souradeep}}}]{2006PhRvD..74b1301B}%
  \BibitemOpen
  \bibfield  {author} {\bibinfo {author} {\bibfnamefont {S.}~\bibnamefont
  {{Basak}}}, \bibinfo {author} {\bibfnamefont {A.}~\bibnamefont {{Hajian}}}, \
  and\ \bibinfo {author} {\bibfnamefont {T.}~\bibnamefont {{Souradeep}}},\
  }\href {\doibase 10.1103/PhysRevD.74.021301} {\bibfield  {journal} {\bibinfo
  {journal} {\prd}\ }\textbf {\bibinfo {volume} {74}},\ \bibinfo {eid} {021301}
  (\bibinfo {year} {2006})},\ \Eprint {http://arxiv.org/abs/astro-ph/0603406}
  {arXiv:astro-ph/0603406 [astro-ph]} \BibitemShut {NoStop}%
\bibitem [{\citenamefont {{Rotti}}\ and\ \citenamefont
  {{Huffenberger}}(2016)}]{2016JCAP...09..034R}%
  \BibitemOpen
  \bibfield  {author} {\bibinfo {author} {\bibfnamefont {A.}~\bibnamefont
  {{Rotti}}}\ and\ \bibinfo {author} {\bibfnamefont {K.}~\bibnamefont
  {{Huffenberger}}},\ }\href {\doibase 10.1088/1475-7516/2016/09/034}
  {\bibfield  {journal} {\bibinfo  {journal} {\jcap}\ }\textbf {\bibinfo
  {volume} {2016}},\ \bibinfo {eid} {034} (\bibinfo {year} {2016})},\ \Eprint
  {http://arxiv.org/abs/1604.08946} {arXiv:1604.08946 [astro-ph.CO]}
  \BibitemShut {NoStop}%
\bibitem [{\citenamefont {{Philcox}}\ \emph {et~al.}(2018)\citenamefont
  {{Philcox}}, \citenamefont {{Sherwin}},\ and\ \citenamefont {{van
  Engelen}}}]{2018MNRAS.479.5577P}%
  \BibitemOpen
  \bibfield  {author} {\bibinfo {author} {\bibfnamefont {O.~H.~E.}\
  \bibnamefont {{Philcox}}}, \bibinfo {author} {\bibfnamefont {B.~D.}\
  \bibnamefont {{Sherwin}}}, \ and\ \bibinfo {author} {\bibfnamefont
  {A.}~\bibnamefont {{van Engelen}}},\ }\href {\doibase 10.1093/mnras/sty1769}
  {\bibfield  {journal} {\bibinfo  {journal} {\mnras}\ }\textbf {\bibinfo
  {volume} {479}},\ \bibinfo {pages} {5577} (\bibinfo {year} {2018})},\ \Eprint
  {http://arxiv.org/abs/1805.09177} {arXiv:1805.09177 [astro-ph.CO]}
  \BibitemShut {NoStop}%
\bibitem [{\citenamefont {{Gorce}}\ \emph {et~al.}(2022)\citenamefont
  {{Gorce}}, \citenamefont {{Douspis}},\ and\ \citenamefont
  {{Salvati}}}]{2022arXiv220208698G}%
  \BibitemOpen
  \bibfield  {author} {\bibinfo {author} {\bibfnamefont {A.}~\bibnamefont
  {{Gorce}}}, \bibinfo {author} {\bibfnamefont {M.}~\bibnamefont {{Douspis}}},
  \ and\ \bibinfo {author} {\bibfnamefont {L.}~\bibnamefont {{Salvati}}},\
  }\href@noop {} {\bibfield  {journal} {\bibinfo  {journal} {arXiv e-prints}\
  ,\ \bibinfo {eid} {arXiv:2202.08698}} (\bibinfo {year} {2022})},\ \Eprint
  {http://arxiv.org/abs/2202.08698} {arXiv:2202.08698 [astro-ph.CO]}
  \BibitemShut {NoStop}%
\bibitem [{\citenamefont {{Astropy Collaboration}}\ \emph
  {et~al.}(2013)\citenamefont {{Astropy Collaboration}}, \citenamefont
  {{Robitaille}}, \citenamefont {{Tollerud}}, \citenamefont {{Greenfield}},
  \citenamefont {{Droettboom}}, \citenamefont {{Bray}}, \citenamefont
  {{Aldcroft}}, \citenamefont {{Davis}}, \citenamefont {{Ginsburg}},
  \citenamefont {{Price-Whelan}}, \citenamefont {{Kerzendorf}}, \citenamefont
  {{Conley}}, \citenamefont {{Crighton}}, \citenamefont {{Barbary}},
  \citenamefont {{Muna}}, \citenamefont {{Ferguson}}, \citenamefont
  {{Grollier}}, \citenamefont {{Parikh}}, \citenamefont {{Nair}}, \citenamefont
  {{Unther}}, \citenamefont {{Deil}}, \citenamefont {{Woillez}}, \citenamefont
  {{Conseil}}, \citenamefont {{Kramer}}, \citenamefont {{Turner}},
  \citenamefont {{Singer}}, \citenamefont {{Fox}}, \citenamefont {{Weaver}},
  \citenamefont {{Zabalza}}, \citenamefont {{Edwards}}, \citenamefont {{Azalee
  Bostroem}}, \citenamefont {{Burke}}, \citenamefont {{Casey}}, \citenamefont
  {{Crawford}}, \citenamefont {{Dencheva}}, \citenamefont {{Ely}},
  \citenamefont {{Jenness}}, \citenamefont {{Labrie}}, \citenamefont {{Lim}},
  \citenamefont {{Pierfederici}}, \citenamefont {{Pontzen}}, \citenamefont
  {{Ptak}}, \citenamefont {{Refsdal}}, \citenamefont {{Servillat}},\ and\
  \citenamefont {{Streicher}}}]{astropy}%
  \BibitemOpen
  \bibfield  {author} {\bibinfo {author} {\bibnamefont {{Astropy
  Collaboration}}}, \bibinfo {author} {\bibfnamefont {T.~P.}\ \bibnamefont
  {{Robitaille}}}, \bibinfo {author} {\bibfnamefont {E.~J.}\ \bibnamefont
  {{Tollerud}}}, \bibinfo {author} {\bibfnamefont {P.}~\bibnamefont
  {{Greenfield}}}, \bibinfo {author} {\bibfnamefont {M.}~\bibnamefont
  {{Droettboom}}}, \bibinfo {author} {\bibfnamefont {E.}~\bibnamefont
  {{Bray}}}, \bibinfo {author} {\bibfnamefont {T.}~\bibnamefont {{Aldcroft}}},
  \bibinfo {author} {\bibfnamefont {M.}~\bibnamefont {{Davis}}}, \bibinfo
  {author} {\bibfnamefont {A.}~\bibnamefont {{Ginsburg}}}, \bibinfo {author}
  {\bibfnamefont {A.~M.}\ \bibnamefont {{Price-Whelan}}}, \bibinfo {author}
  {\bibfnamefont {W.~E.}\ \bibnamefont {{Kerzendorf}}}, \bibinfo {author}
  {\bibfnamefont {A.}~\bibnamefont {{Conley}}}, \bibinfo {author}
  {\bibfnamefont {N.}~\bibnamefont {{Crighton}}}, \bibinfo {author}
  {\bibfnamefont {K.}~\bibnamefont {{Barbary}}}, \bibinfo {author}
  {\bibfnamefont {D.}~\bibnamefont {{Muna}}}, \bibinfo {author} {\bibfnamefont
  {H.}~\bibnamefont {{Ferguson}}}, \bibinfo {author} {\bibfnamefont
  {F.}~\bibnamefont {{Grollier}}}, \bibinfo {author} {\bibfnamefont {M.~M.}\
  \bibnamefont {{Parikh}}}, \bibinfo {author} {\bibfnamefont {P.~H.}\
  \bibnamefont {{Nair}}}, \bibinfo {author} {\bibfnamefont {H.~M.}\
  \bibnamefont {{Unther}}}, \bibinfo {author} {\bibfnamefont {C.}~\bibnamefont
  {{Deil}}}, \bibinfo {author} {\bibfnamefont {J.}~\bibnamefont {{Woillez}}},
  \bibinfo {author} {\bibfnamefont {S.}~\bibnamefont {{Conseil}}}, \bibinfo
  {author} {\bibfnamefont {R.}~\bibnamefont {{Kramer}}}, \bibinfo {author}
  {\bibfnamefont {J.~E.~H.}\ \bibnamefont {{Turner}}}, \bibinfo {author}
  {\bibfnamefont {L.}~\bibnamefont {{Singer}}}, \bibinfo {author}
  {\bibfnamefont {R.}~\bibnamefont {{Fox}}}, \bibinfo {author} {\bibfnamefont
  {B.~A.}\ \bibnamefont {{Weaver}}}, \bibinfo {author} {\bibfnamefont
  {V.}~\bibnamefont {{Zabalza}}}, \bibinfo {author} {\bibfnamefont {Z.~I.}\
  \bibnamefont {{Edwards}}}, \bibinfo {author} {\bibfnamefont {K.}~\bibnamefont
  {{Azalee Bostroem}}}, \bibinfo {author} {\bibfnamefont {D.~J.}\ \bibnamefont
  {{Burke}}}, \bibinfo {author} {\bibfnamefont {A.~R.}\ \bibnamefont
  {{Casey}}}, \bibinfo {author} {\bibfnamefont {S.~M.}\ \bibnamefont
  {{Crawford}}}, \bibinfo {author} {\bibfnamefont {N.}~\bibnamefont
  {{Dencheva}}}, \bibinfo {author} {\bibfnamefont {J.}~\bibnamefont {{Ely}}},
  \bibinfo {author} {\bibfnamefont {T.}~\bibnamefont {{Jenness}}}, \bibinfo
  {author} {\bibfnamefont {K.}~\bibnamefont {{Labrie}}}, \bibinfo {author}
  {\bibfnamefont {P.~L.}\ \bibnamefont {{Lim}}}, \bibinfo {author}
  {\bibfnamefont {F.}~\bibnamefont {{Pierfederici}}}, \bibinfo {author}
  {\bibfnamefont {A.}~\bibnamefont {{Pontzen}}}, \bibinfo {author}
  {\bibfnamefont {A.}~\bibnamefont {{Ptak}}}, \bibinfo {author} {\bibfnamefont
  {B.}~\bibnamefont {{Refsdal}}}, \bibinfo {author} {\bibfnamefont
  {M.}~\bibnamefont {{Servillat}}}, \ and\ \bibinfo {author} {\bibfnamefont
  {O.}~\bibnamefont {{Streicher}}},\ }\href {\doibase
  10.1051/0004-6361/201322068} {\bibfield  {journal} {\bibinfo  {journal}
  {\aap}\ }\textbf {\bibinfo {volume} {558}},\ \bibinfo {eid} {A33} (\bibinfo
  {year} {2013})},\ \Eprint {http://arxiv.org/abs/1307.6212} {arXiv:1307.6212
  [astro-ph.IM]} \BibitemShut {NoStop}%
\bibitem [{\citenamefont {{Astropy Collaboration}}\ \emph
  {et~al.}(2018)\citenamefont {{Astropy Collaboration}}, \citenamefont
  {{Price-Whelan}}, \citenamefont {{Sip{\H o}cz}}, \citenamefont
  {{G{\"u}nther}}, \citenamefont {{Lim}} \emph {et~al.}}]{astropy2}%
  \BibitemOpen
  \bibfield  {author} {\bibinfo {author} {\bibnamefont {{Astropy
  Collaboration}}}, \bibinfo {author} {\bibfnamefont {A.~M.}\ \bibnamefont
  {{Price-Whelan}}}, \bibinfo {author} {\bibfnamefont {B.~M.}\ \bibnamefont
  {{Sip{\H o}cz}}}, \bibinfo {author} {\bibfnamefont {H.~M.}\ \bibnamefont
  {{G{\"u}nther}}}, \bibinfo {author} {\bibfnamefont {P.~L.}\ \bibnamefont
  {{Lim}}},  \emph {et~al.},\ }\href@noop {} {\bibfield  {journal} {\bibinfo
  {journal} {ArXiv e-prints}\ } (\bibinfo {year} {2018})},\ \Eprint
  {http://arxiv.org/abs/1801.02634} {arXiv:1801.02634 [astro-ph.IM]}
  \BibitemShut {NoStop}%
\bibitem [{\citenamefont {Hunter}(2007)}]{matplotlib}%
  \BibitemOpen
  \bibfield  {author} {\bibinfo {author} {\bibfnamefont {J.~D.}\ \bibnamefont
  {Hunter}},\ }\href {\doibase 10.1109/MCSE.2007.55} {\bibfield  {journal}
  {\bibinfo  {journal} {Computing In Science \& Engineering}\ }\textbf
  {\bibinfo {volume} {9}},\ \bibinfo {pages} {90} (\bibinfo {year}
  {2007})}\BibitemShut {NoStop}%
\bibitem [{\citenamefont {{Jones}}\ \emph {et~al.}(2001)\citenamefont
  {{Jones}}, \citenamefont {{Oliphant}}, \citenamefont {{Peterson}} \emph
  {et~al.}}]{scipy}%
  \BibitemOpen
  \bibfield  {author} {\bibinfo {author} {\bibfnamefont {E.}~\bibnamefont
  {{Jones}}}, \bibinfo {author} {\bibfnamefont {T.}~\bibnamefont {{Oliphant}}},
  \bibinfo {author} {\bibfnamefont {P.}~\bibnamefont {{Peterson}}},  \emph
  {et~al.},\ }\href {http://www.scipy.org/} {\enquote {\bibinfo {title}
  {{SciPy}: Open source scientific tools for {Python}},}\ } (\bibinfo {year}
  {2001})\BibitemShut {NoStop}%
\bibitem [{\citenamefont {Oliphant}(2006)}]{numpy}%
  \BibitemOpen
  \bibfield  {author} {\bibinfo {author} {\bibfnamefont {T.}~\bibnamefont
  {Oliphant}},\ }\href {http://www.numpy.org/} {\enquote {\bibinfo {title}
  {{NumPy}: A guide to {NumPy}},}\ }\bibinfo {howpublished} {USA: Trelgol
  Publishing} (\bibinfo {year} {2006})\BibitemShut {NoStop}%
\bibitem [{\citenamefont {{Foreman-Mackey}}\ \emph {et~al.}(2013)\citenamefont
  {{Foreman-Mackey}}, \citenamefont {{Hogg}}, \citenamefont {{Lang}},\ and\
  \citenamefont {{Goodman}}}]{emcee}%
  \BibitemOpen
  \bibfield  {author} {\bibinfo {author} {\bibfnamefont {D.}~\bibnamefont
  {{Foreman-Mackey}}}, \bibinfo {author} {\bibfnamefont {D.~W.}\ \bibnamefont
  {{Hogg}}}, \bibinfo {author} {\bibfnamefont {D.}~\bibnamefont {{Lang}}}, \
  and\ \bibinfo {author} {\bibfnamefont {J.}~\bibnamefont {{Goodman}}},\ }\href
  {\doibase 10.1086/670067} {\bibfield  {journal} {\bibinfo  {journal} {PASP}\
  }\textbf {\bibinfo {volume} {125}},\ \bibinfo {pages} {306} (\bibinfo {year}
  {2013})},\ \Eprint {http://arxiv.org/abs/1202.3665} {arXiv:1202.3665
  [astro-ph.IM]} \BibitemShut {NoStop}%
\end{thebibliography}%

\end{document}